\documentclass[structabstract]{aa}
\usepackage{txfonts}
\usepackage{natbib}
\usepackage{graphicx}
\bibpunct{(}{)}{;}{a}{}{,} 
\usepackage{subfigure}
\usepackage[margin=10pt,font=small,labelfont=bf,labelsep=endash]{caption}%

\def\tex {\ifmmode{{T}_{\rm ex}}\else{$T_{\rm ex}$}\fi}
\def\tmb {\ifmmode{{T}_{\rm mb}}\else{$T_{\rm mb}$}\fi}
\def\hi     {\ifmmode{{\rm H}{\rm \small I}}\else{H\ts {\scriptsize I}}\fi}
\def\hh     {\ifmmode{{\rm H}_2}\else{H$_2$}\fi}

\def\ts     {\thinspace}
\def\kms    {\ifmmode{{\rm \ts km\ts s}^{-1}}\else{\ts km\ts s$^{-1}$}\fi}
\def\msol   {\ifmmode{{\rm M}_{\odot}}\else{M$_{\odot}$}\fi}
\def\lsol   {\ifmmode{{\rm L}_{\odot}}\else{L$_{\odot}$}\fi}
\def\zsol   {\ifmmode{{\rm Z}_{\odot}}\else{Z$_{\odot}$}\fi}

\begin{document}

\title{\object{ALMA backed NIR high resolution integral field spectroscopy of the NUGA galaxy NGC 1433}
	\thanks{Based on the ESO-VLT proposal ID: 090.B-0657(A) and on observations carried out with ALMA in cycle 0.}
 }


\author{Semir Smaji\'c\inst{1,}\inst{2}
	\and Lydia Moser\inst{1}
	\and Andreas Eckart\inst{1,}\inst{2}
	\and M\'onica Valencia-S.\inst{1}
	\and Francoise Combes\inst{3}
	\and Matthew Horrobin\inst{1}
	\and Santiago Garc\'ia-Burillo\inst{4}
	\and Macarena Garc\'ia-Mar\'in\inst{1}
	\and Sebastian Fischer\inst{5}
	\and Jens Zuther\inst{1}
	}

\institute{I. Physikalisches Institut, Universit\"at zu K\"oln, Z\"ulpicher Str. 77, 50937 K\"oln, Germany\\email: smajic@ph1.uni-koeln.de
	\and Max-Planck-Institut f\"ur Radioastronomie, Auf dem H\"ugel 69, 53121 Bonn, Germany
	\and Observatoire de Paris, LERMA, CNRS: UMR8112, 61 Av. de l'Observatoire, 75014, Paris, France
	\and Observatorio Astronómico Nacional (OAN)-Observatorio de Madrid, Alfonso XII 3, 28014, Madrid, Spain
	\and Deutsches Zentrum für Luft- und Raumfahrt (DLR), Königswinterer Str. 522-524, 53227 Bonn, Germany}

\date{Received ???/ Accepted ???}

\abstract {}
{We present the results of near-infrared (NIR) H- and K-band European Southern Observatory SINFONI integral field spectroscopy (IFS) of the Seyfert 2 galaxy \object{NGC 1433}. We investigate the central 500 pc of this nearby galaxy, concentrating on excitation conditions, morphology, and stellar content. NGC 1433 was selected from our extended NUGA(-south) sample which was additionally observed with the Atacama Large Millimeter/submillimeter Array (ALMA). NGC 1433 is a ringed, spiral galaxy with a main stellar bar in roughly east--west direction (PA 94\degr) and a secondary bar in the nuclear region (PA 31\degr). Several dusty filaments are detected in the nuclear region with the Hubble Space Telescope. ALMA detects molecular CO emission coinciding with these filaments. The active galactic nucleus is not strong and the galaxy is also classified as a LINER.}
{The NIR is less affected by dust extinction than optical light and is sensitive to the mass-dominating stellar populations. SINFONI integral field spectroscopy combines NIR imaging and spectroscopy allowing us to analyze several emission and absorption lines, to investigate the stellar populations and ionization mechanisms over the $10\arcsec\times10\arcsec$ field of view (FOV).}
{We present emission and absorption line measurements in the central kpc of NGC 1433. We detect a narrow Balmer line and several H$_2$ lines. We find that the stellar continuum peaks in the optical and NIR in the same position, indicating that there is no covering of the center by a nuclear dust lane.
A strong velocity gradient is detected in all emission lines at that position.
The position angle of this gradient is at 155$\degr$ whereas the galactic rotation is at a position angle of 201$\degr$.
Our measures of the molecular hydrogen lines, hydrogen recombination lines, and [\ion{Fe}{ii}] indicate that the excitation at the nucleus is caused by thermal excitation, i.e. shocks which can be associated with active galactic nuclei emission, supernovae or outflows. The line ratios [\ion{Fe}{ii}]/Pa$\beta$ and H$_2$/Br$\gamma$ show a Seyfert to LINER identification of the nucleus. We do not detect high star formation rates in our FOV. The stellar continuum is dominated by spectral signatures of red-giant M stars. The stellar line-of-sight velocity follows the galactic field whereas the light continuum follows the nuclear bar.
}
{
The dynamical center of NGC 1433 coincides with the optical and NIR center of the galaxy and the black hole position. Within the central arcsecond, the molecular hydrogen and the $^{12}$CO(3-2) emissions -- observed in the NIR and in the sub-millimeter with SINFONI and ALMA, respectively -- are indicative for a nuclear outflow originating from the galaxy's SMBH. A small circum nuclear disk cannot be fully excluded. Derived gravitational torques show that the nuclear bar is able to drive gas inwards to scales where viscosity torques and dynamical friction become important. The black hole mass derived using stellar velocity dispersion is $\sim10^7$~M$_{\odot}$.
}

\label{abstract}


\maketitle

\section{Introduction}

The NUclei of GAlaxies (NUGA) project \citep{garcia-burillo_molecular_2003} started with the IRAM Plateau de Bure Interferometer (PdBI) and 30 m single-dish survey of nearby low-luminosity active galactic nuclei (LLAGN) in the northern hemisphere. The aim is to map the distribution and dynamics of molecular (cool) gas in the inner kpc of LLAGN and to study the possible mechanisms for gas fueling at high angular resolution ($\approx$ 0\farcs5 -- 2$\arcsec$) and high sensitivity. The step by step implementation of the Atacama Large Millimeter/submillimeter Array (ALMA) in the Atacama desert in Chile finally allows the NUGA project to expand to the southern hemisphere at an even higher angular resolution ($\approx$ 6 -- 37 mas, assuming the use of the full array) and sensitivity. The Spectrograph for INtegral Field Observations in the Near Infrared \citep[SINFONI;][]{eisenhauer_sinfoni_2003,bonnet_first_2004-1} adds complementary information to the NUGA goal in the near-infrared (NIR). By mapping (hot-) gas and the mass dominating stellar population the fueling of nearby LLAGN can be investigated in the NIR.

\subsection{Feeding and feedback in AGN}

An active galactic nucleus (AGN), 
as a highly ionizing source, 
needs to show powerful highly ionizing emission coming from an unresolved region. This powerful emission ionizes the surrounding gas on scales from several lightdays (e.g. broad line region (BLR)) to several hundreds of parsecs (e.g. narrow line region (NLR)) and up to even larger scales via outflows and jets. Usually, a dust mantle surrounds the AGN on parsec to tens of parsec scales, inferred from high column densities towards Seyfert 2 AGN and dust black body emission in Seyfert 1 AGN with temperatures up to the sublimation temperature of dust ($\sim1300$ K). The source of dust heating must be highly ionizing emission, probably originated by the accretion of 
gas onto a supermassive black hole (SMBH).

Tight correlations between the SMBH and its host galaxy, especially the bulge stellar velocity dispersion and luminosity have been found \citep[e.g.][]{ferrarese_fundamental_2000,magorrian_demography_1998}. These correlations connect the mass, luminosity and kinematics of the galactic bulge with the mass of the central SMBH. To feed a SMBH, gas within the bulge needs to be transported towards the center. There are two ways to remove angular momentum in gas and consequently produce its infall. One way is through gravitational mechanisms that exert gravitational torques such as galaxy-galaxy interactions (e.g. galaxy merger) or non-axisymmetries within the galaxy potential (e.g. spiral density waves or stellar bars on large scales). The other way are hydrodynamical mechanisms such as shocks and viscosity torques introduced by turbulences in the interstellar medium (ISM). The NUGA project has already studied the gaseous distribution in more than ten nearby galaxies ($\sim4$ -- 40 Mpc). They show a variety of morphologies in nuclear regions, including bars and spirals \citep{garcia-burillo_molecular_2005,boone_molecular_2007,hunt_molecular_2008,lindt-krieg_molecular_2008,garcia-burillo_molecular_2009}, rings \citep{combes_molecular_2004,casasola_molecular_2008,combes_molecular_2009} and lopsided disks \citep{garcia-burillo_molecular_2003,krips_molecular_2005,casasola_molecular_2010}. 
Gravitational torques, when present, are the strongest mechanism to successfully transport gas close to the nucleus, on smaller scales (< 200 pc) viscosity torques can take over \citep[e.g.][]{combes_molecular_2004,van_der_laan_molecular_2011}. Hence, large-scale stellar bars are an important agent to transport gas towards the inner Lindblad resonance \citep[ILR; e.g.][]{sheth_secular_2005} where it can induce the formation of nuclear spirals and rings.
 
As a complement to the studies of cold gas in the NUGA sample, SINFONI can detect hot molecular and ionized gas in the NIR at similar angular resolution. Comparing the distributions of the different gas-emission lines enables us to identify ongoing star formation sites and regions ionized by shocks (i.e. supernovae (SNe) or outflows). Furthermore, this can provide information
on feeding and feedback of the AGN through its ambient gas reservoir. We will analyze the relation of nuclear star formation sites and the AGN with regard to fueling and feedback. Using the differently excited H$_2$ lines in the K-band (e.g. H$_2$ $\lambda$2.12 $\mu$m, 1.957 $\mu$m, 2.247 $\mu$m) and [\ion{Fe}{ii}] in the H-band we are able to constrain the origin of excitation (e.g. thermal, non-thermal) of the warm gas and its excitation temperature \citep{rodriguez-ardila_molecular_2004,zuther_mrk_2007}. The cold gas information is comparable to our results.

Recent or ongoing star formation on scales of 0.1 -- 1 kpc around the nucleus is frequently found in all types of AGN in contrast to quiescent galaxies \citep[e.g.][]{cidfernandes_star_2004,davies_star-forming_2006}. Whether outflows from the AGN quench or initiate star formation is still a matter of debate, but it is most probable that outflows can show both effects. Stellar absorption features (e.g. \ion{Si}{i}, $^{12}$CO(6--3), \ion{Mg}{i}, \ion{Na}{i}, $^{12}$CO(2--0)) are used as a probe of the star formation history in the nuclear region \citep[e.g.][]{davies_close_2007}. Strong star formation with young bright hot stars is able to blow enough material from its gas cloud and initiate a series of accretion events onto the AGN.

\subsection{NGC 1433}

NGC 1433 is a barred, ringed, spiral SB(rs)ab galaxy in the Dorado group \citep{buta_structure_1986,Kilborn_wide-field_2005,buta_mid-infrared_2010} at a redshift of $z\approx 0.003586$ \citep[see table \ref{tab:basic},][]{koribalski_1000_2004}. \citet{veron-cetty_miscellaneous_1986} classify this galaxy as Seyfert like, but \citet{cid_fernandes_stellar_1998} and \citet{sosa-brito_integral_2001} refined the classification in a low ionization narrow emission line region (LINER) galaxy using a stellar-continuum subtracted spectrum and extinction corrected optical emission lines. The position of the center of the galaxy coincides with the X-ray ROSAT source \citep{liu_ultraluminous_2005}.

\begin{figure}[htbp]
\centering
\includegraphics[width=0.45\textwidth]{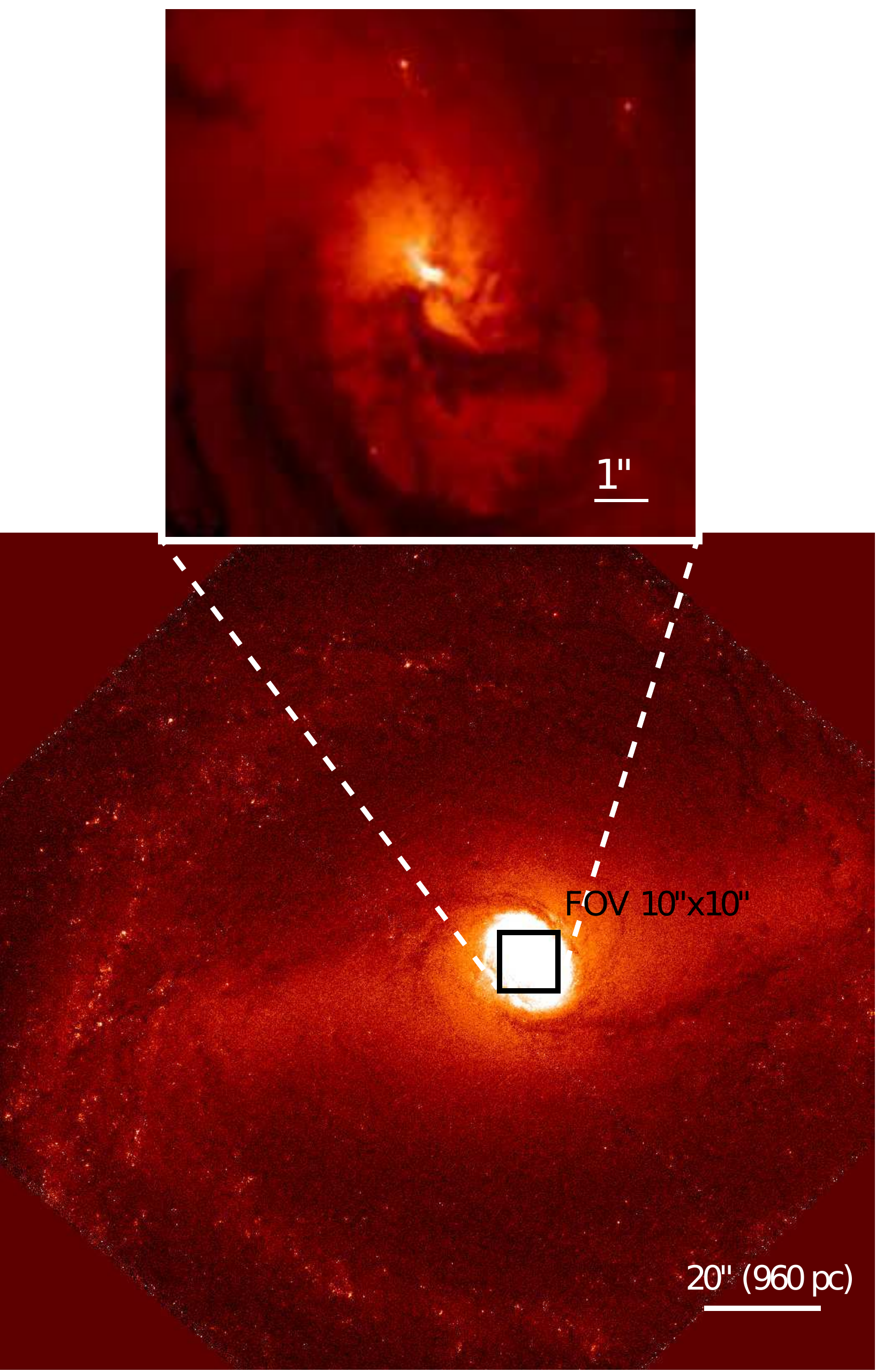}
\caption{HST F438 wide field image of NGC 1433 from the ESA Hubble Science Archive. 
The FOV of 10$\arcsec$$\times$10$\arcsec$ of our SINFONI observations is marked by a square. 
The color scale in the lower image was chosen to reflect the large scale structure of the host, e.g. main stellar bar. The top image shows the prominent dust lanes in the center of NGC 1433.
}
\label{fig:ngc1433}
\end{figure}

The galaxy shows a primary stellar bar with a radius of about 4 kpc roughly in the east-west direction (PA 94$\degr$). Two rings, an inner ring with a radius of about 5.2 kpc and a PA of about 95$\degr$ (see Fig. \ref{fig:ngc1433}) and an outer ring at a PA of 15$\degr$ and radius of 9.1 kpc, can be identified. The secondary nuclear bar has a radius of 430 pc at PA 31$\degr$ and is surrounded by a ring at the same PA and 460 pc radius \citep[see Fig. \ref{fig:ngc1433};][]{wozniak_disc_1995,buta_dynamics_2001}. The nuclear region shows no signs of massive star formation except for the nuclear ring which is the site of a starburst \citep{cid_fernandes_stellar_1998,sanchez-blazquez_star_2011}. \ion{H}{i} 21 cm measurements show atomic gas in the inner and outer rings and a depletion in the nuclear region \citep{ryder_neutral_1996}, whereas strong molecular CO emission is detected in the nuclear region \citep{bajaja_observations_1995}. The \ion{H}{i} measurement reveals a line of nodes at a PA of 201$\degr$ and an inclination of 33$\degr$. The nuclear region is filled with dusty spiral arms visible in Hubble Space Telescope (HST, see Fig. \ref{fig:ngc1433}) images \citep{maoz_atlas_1996,peeples_connection_2006}. These arms or filaments are very well traced by the molecular $^{12}$CO(3--2) emission observed with ALMA \citep{combes_ALMA_2013}. The CO shows a highly redshifted component just south of the nucleus at about 100 km s$^{-1}$ indicating a possible outflow. Spitzer data show that NGC 1433 harbors a pseudo-bulge and show prominent nuclear spirals \citep{fisher_bulges_2010}.

Because of its very complex dynamical structure, NGC 1433 is a good case to search for the connection between cold molecular gas, detected at a few hundred GHz, and warm molecular H$_2$ gas detected in the NIR. Are the cold and warm gas following the same dusty spiral arms? Are the bright emission spots of cold and warm gas coinciding? What evidences of an outflow can be seen in the warm gas?

This paper is structured as follows: In section 2 we describe the observations and the data reduction. Section 3 presents the results of our study, focusing on the emission lines in K-band and the continuum analysis. Section 4 discusses the results and compares them with the literature. In section 5 we sum up our results and phrase the conclusions we take from our study.

\section{Observation and Data Reduction}
\label{sec:obs_red}
\begin{figure}[htbp]
\centering
\includegraphics[height=0.45\textwidth,angle=90]{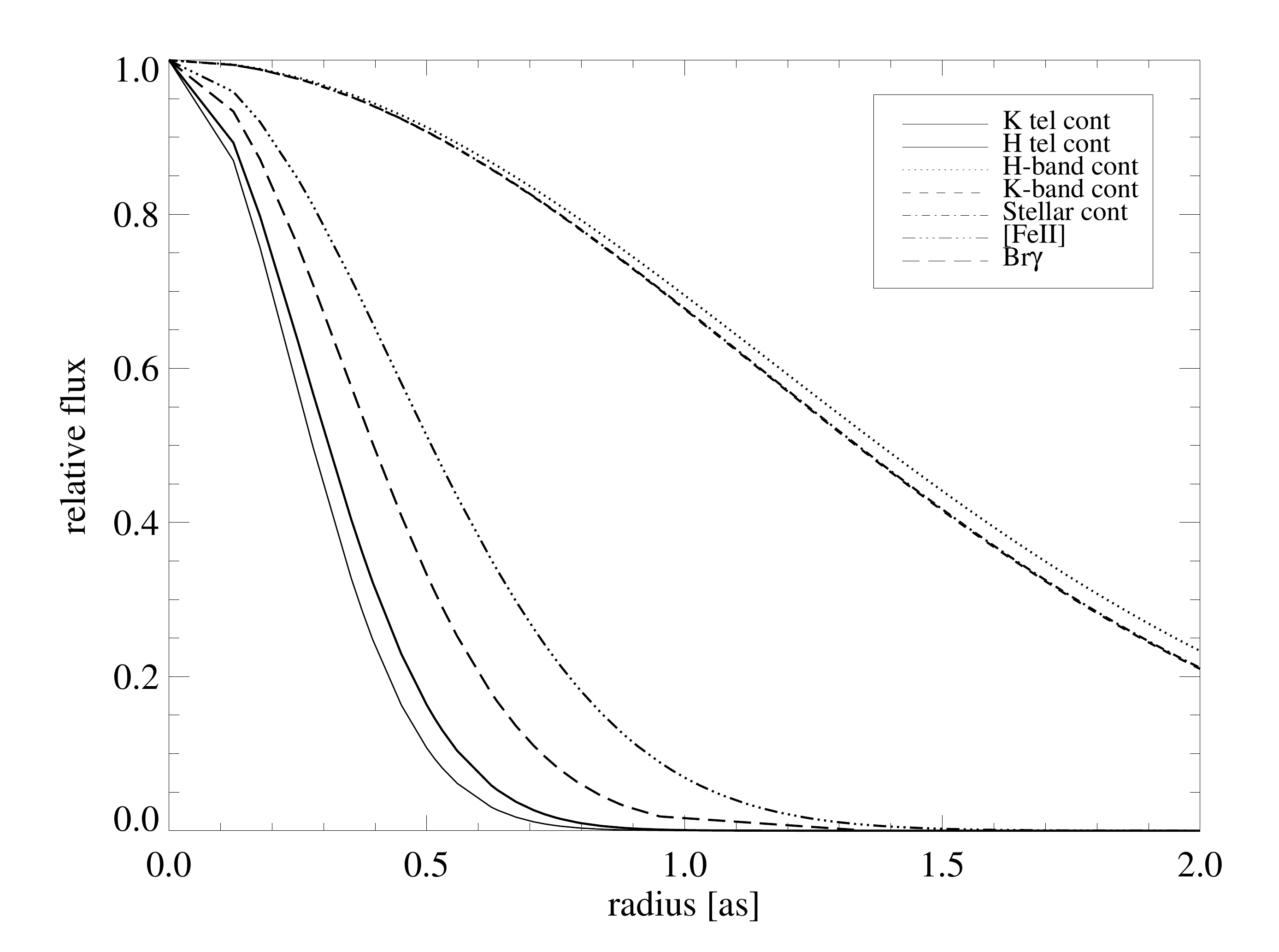}
\caption{Shown are Gaussian fits to radial luminosity profiles of the K-band standard star continuum, the H-band standard star continuum, the H- and K-bands galaxy continuum, the fitted stellar continuum in the K-band, \ion{Fe}{ii} $\lambda$1.644 $\mu$m and Br$\gamma$. Note that K-band galaxy continuum and the fitted stellar continuum lie on top of each other. The Gaussian fit was centered on the galactic center.}
\label{fig:PSF}
\end{figure}

\begin{figure*}[htbp]
\centering
\subfigure[Flux $\mbox{[}$\ion{Fe}{ii}$\mbox{]}$]{\includegraphics[width=0.33\textwidth]{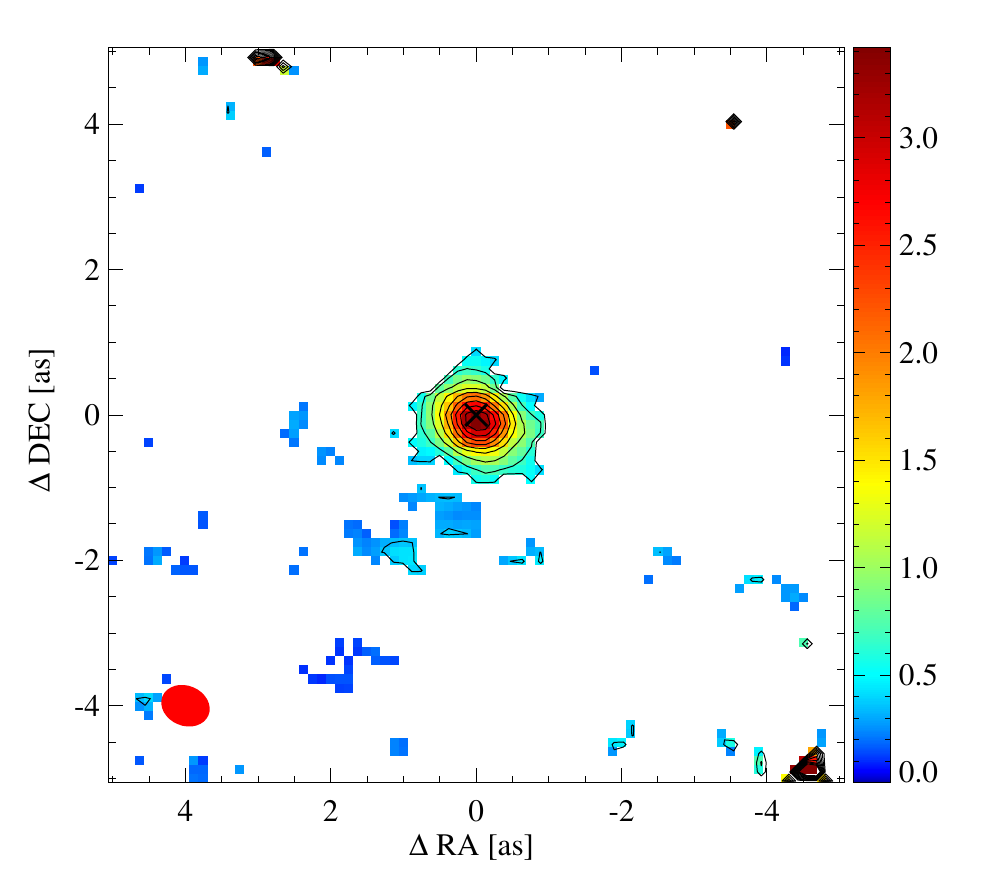}\label{fig:feiiflux}}
\subfigure[FWHM $\mbox{[}$\ion{Fe}{ii}$\mbox{]}$]{\includegraphics[width=0.33\textwidth]{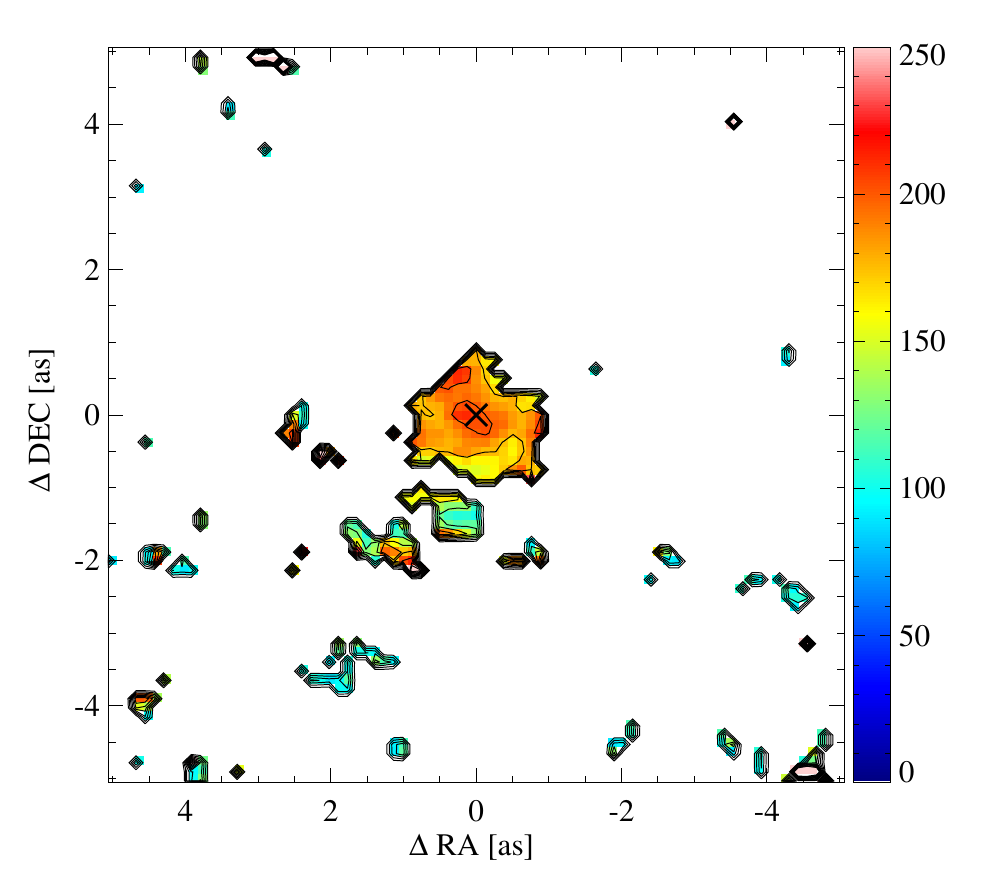}\label{fig:feiifwhm}}
\subfigure[LOSV $\mbox{[}$\ion{Fe}{ii}$\mbox{]}$]{\includegraphics[width=0.33\textwidth]{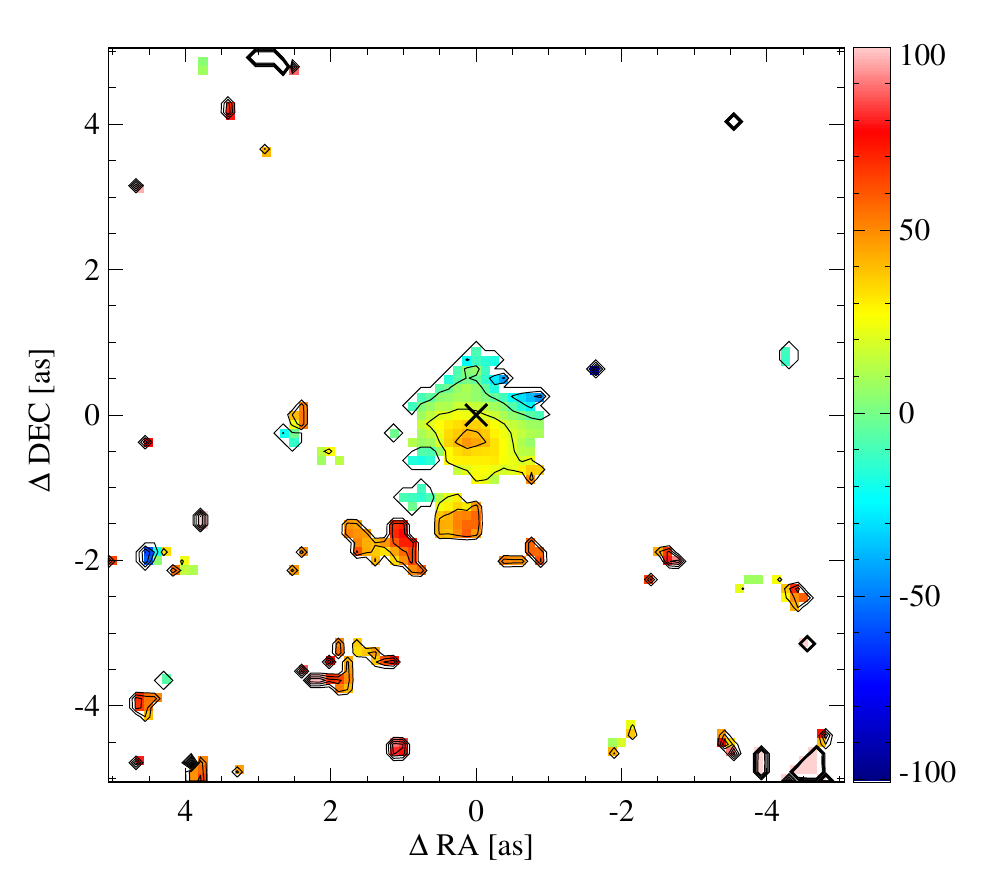}\label{fig:losvfeii}}
\subfigure[Flux Br$\gamma$]{\includegraphics[width=0.33\textwidth]{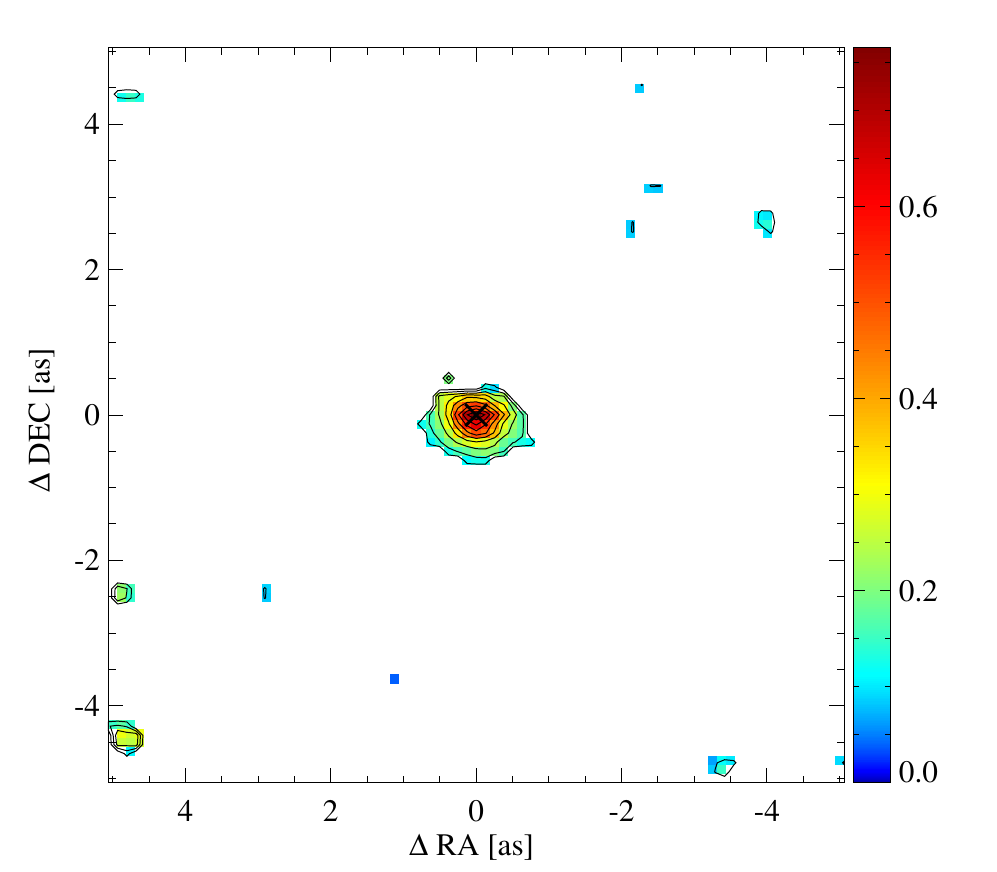}\label{fig:brgflux}}
\subfigure[FWHM Br$\gamma$]{\includegraphics[width=0.33\textwidth]{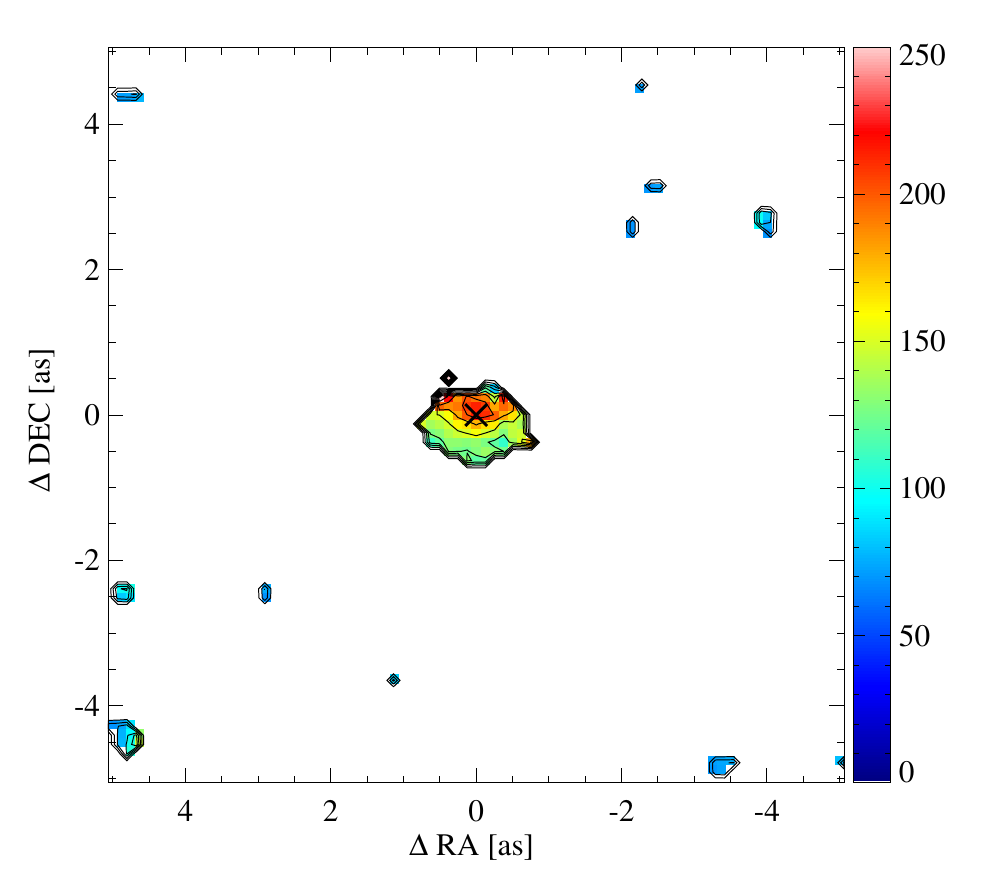}\label{fig:brgfwhm}}
\subfigure[LOSV Br$\gamma$]{\includegraphics[width=0.33\textwidth]{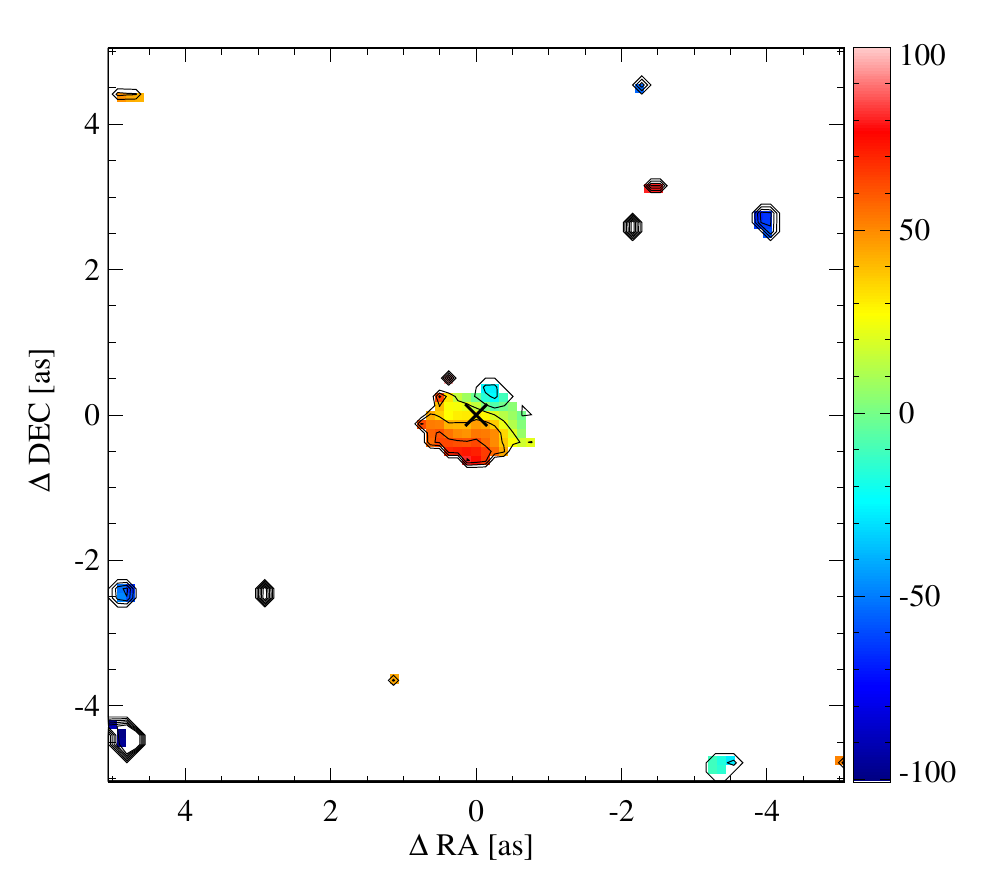}\label{fig:losvbrg}}
\subfigure[Flux H$_2$(1-0)S(1)]{\includegraphics[width=0.33\textwidth]{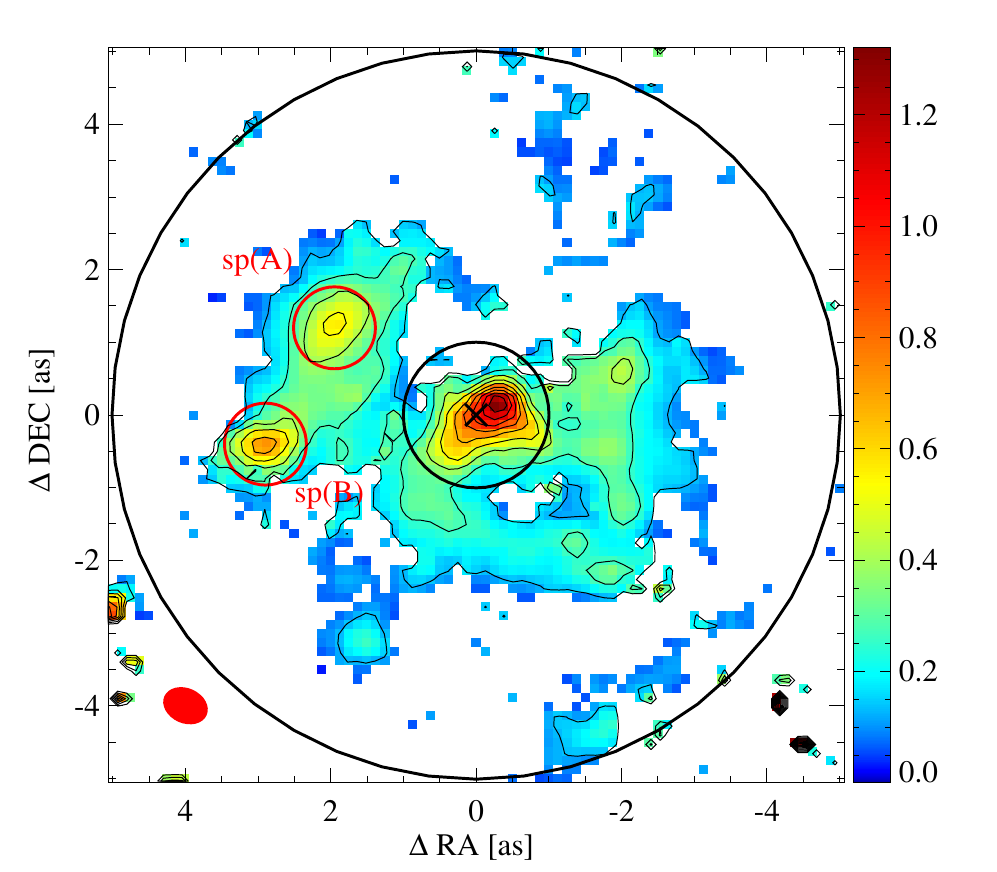}\label{fig:h212flux}}
\subfigure[FWHM H$_2$(1-0)S(1)]{\includegraphics[width=0.33\textwidth]{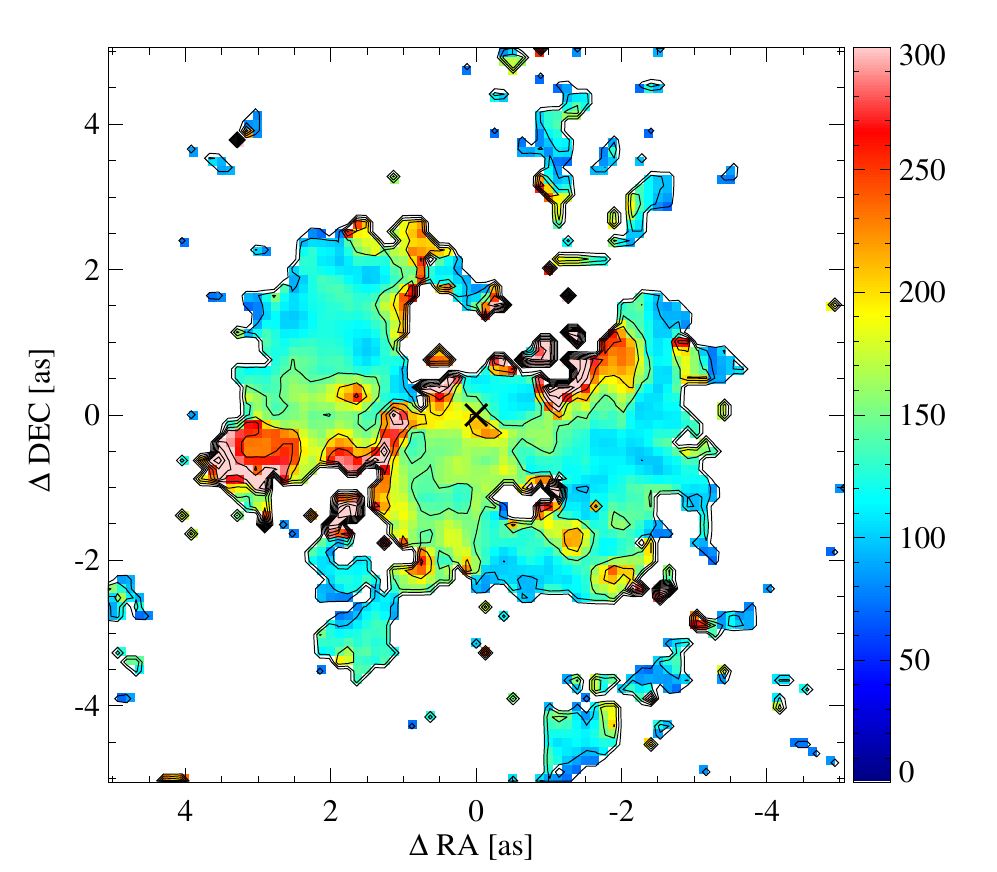}\label{fig:h212fwhm}}
\subfigure[LOSV H$_2$(1-0)S(1)]{\includegraphics[width=0.33\textwidth]{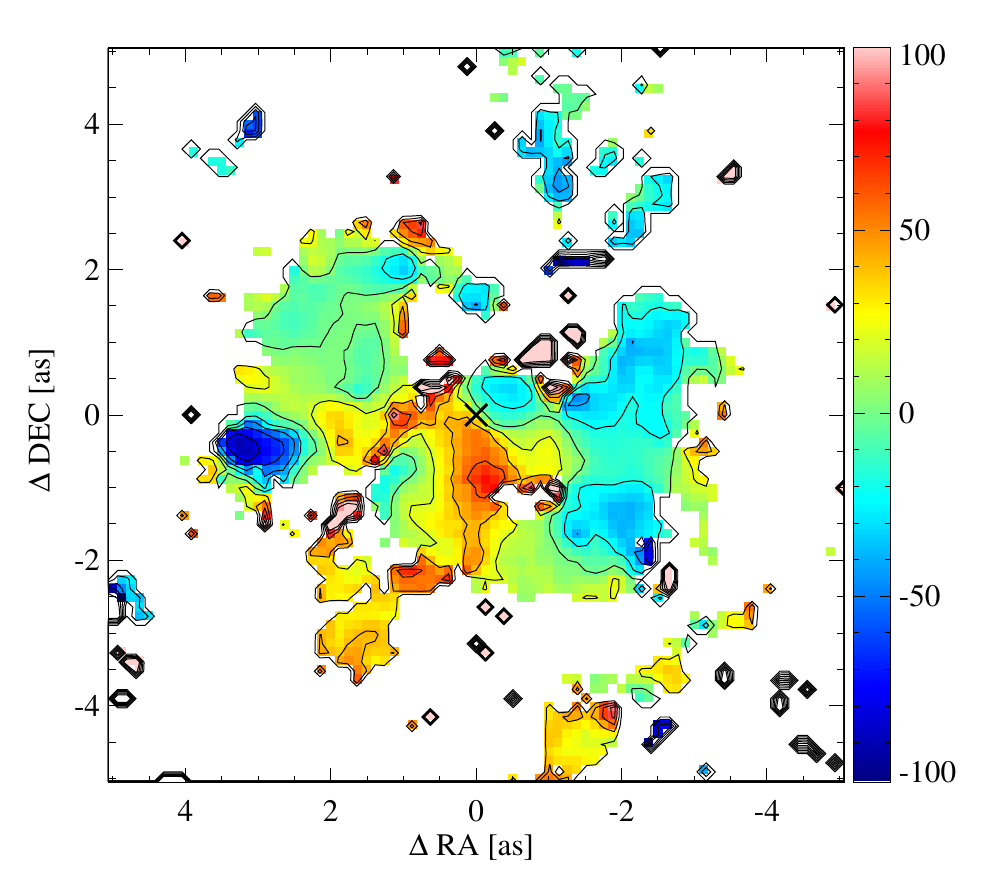}\label{fig:losvh212}}
\caption{Shown are the flux [$10^{-20}$ W m$^{-2}$], FWHM [km s$^{-1}$] and LOSV [km s$^{-1}$] maps (left to right) of the [\ion{Fe}{ii}], Br$\gamma$ and H$_2$ $\lambda$(1-0)S(1) lines (top to bottom). \subref{fig:feiiflux} and \subref{fig:h212flux} show the beam size in the lower left corner for H- and K-band respectively. \subref{fig:h212flux} shows the aperture sizes taken for our measurements. One angular resolution element (beam) corresponds to a PSF$_{\mbox{\tiny FWHM}}$ of 0$\farcs$62 in H-band and 0$\farcs$56 in K-band (see Fig. \ref{fig:PSF}). The centered circles are of radius 1$\arcsec$ and 5$\arcsec$ respectively. The red circles marking the emission regions sp(A) and sp(B) are of radius 0\farcs56. The black cross marks the position of the adopted center.}
\label{fig:lines1}
\end{figure*}

In this paper we present the results of our ESO SINFONI observations of NGC 1433 with the Unit Telescope 4 of the Very Large Telescope in Chile. The 0\farcs25 plate scale was used, which results in an 8$\arcsec\times8\arcsec$ FOV without adaptive optics assistance. The average seeing was 0\farcs5. To increase the FOV 
and minimize the overlap of dead pixels in critical areas, a $\pm2\arcsec$ dithering sequence was used. Nine dithering positions were used in which the central 4$\arcsec\times4\arcsec$ were observed at the full integration time. The resulting FOV is $12\times12\arcsec$, but for the analysis a FOV of $10\times10\arcsec$ is used. NGC 1433 was observed using the H-band grating with a spectral resolution of $R \approx 3000$ and the K-band grating with a resolution of $R \approx 4000$. The digital integration time of 150 seconds was used in both bands with a ...TST... nodding sequence (T: target, S: sky), to increase on-source time. The overall integration time on the target source in H-band was 2400 seconds and in K-band 4500 seconds with additional 1200 seconds in H-band and 2250 seconds in K-band on sky.
The G2V star HIP 017751 was observed in the H- and in K-bands within the respective science target observation. Two object-sky pairs in H-band and 4 object-sky pairs in K-band were taken.
The standard star was used to correct for telluric absorption of the atmosphere and to perform the flux calibration of the target. 
A high S/N solar spectrum was used to correct for the black body and intrinsic spectral features of the G2V star \citep{maiolino_correction_1996}. The solar spectrum was convolved with a Gaussian to adapt its resolution to the resolution of the standard star spectrum. The solar spectrum edges had to be interpolated by a black body with a temperature of $T = 5800$~K. The standard star spectrum was extracted by taking the total of all pixels within the radius of 3 $\times$ FWHM$_{\mbox{\tiny PSF}}$ of the point spread function (PSF) \citep{howell_handbook_2000}, centered on the peak of a two-dimensional Gaussian fit. The flux calibration of the target source was performed during the telluric correction procedure. We calibrated the standard star counts at $\lambda$1.662 $\mu$m and $\lambda$2.159 $\mu$m, in H- and K-band respectively, to the flux given by the 2MASS All-sky Point Source Catalog.
A PSF FWHM of 0\farcs62 in H-band and 0\farcs56 in K-band was measured from the observed standard star (see Fig. \ref{fig:PSF}). The PSF shows an elongation in the north--east to south--west direction at a PA of 69$\degr$ (see Fig. \ref{fig:lines1}).

Detector specific problems, which occurred during our observation period, were analyzed and corrected for and OH sky emission line corrections were applied (see Appx. \ref{sec:dsp}).

The data reduction was performed by the SINFONI pipeline up to single-exposure-cube reconstruction. The wavelength calibration had to be refined by hand due to a constant offset introduced to the wavelength axis. The error introduced by this shift is $\sim3$~\kms. The final cube realignment was done using our own idl routine.

We conducted the investigation and correction of the detector specific problems and the OH lines as well as the linemap and spectra extraction using our own IDL routines.

For comparison of our NIR results we use calibrated 350 GHz ALMA data \citep[][observation and calibration details therein from the ALMA archive]{combes_ALMA_2013}. For cleaning we used a mask at 50mJy, 30 mJy and 10mJy emission level. The map sizes are $720\times720$ pixels with a pixel size of 0$\farcs$05. The line cube comprises 72 channels of 5 km/s width, centered on the systemic velocity, i.e. $1076\pm180$ km s$^{-1}$. The integrated intensity and continuum maps (line-free regions of all 4 spectral windows, i.e. 7 GHz) were corrected for primary beam attenuation. Imaging, cleaning and parts of the analysis have been conducted with the CASA software \citep[v3.3][]{mcmullin_casa_2007}.

\section{Results}
\label{sec:res}

With the seeing limited imaging spectroscopy of VLT SINFONI we resolve the central 500~pc of NGC 1433 at a spatial resolution of 27~pc in K-band and 30~pc in H-band. We identify several molecular ro-vibrational hydrogen lines and the narrow hydrogen recombination line Br$\gamma$ $\lambda$2.166 $\mu$m. Several stellar absorption features in K-band (e.g. CO(2-0) $\lambda$2.29 $\mu$m, NaD $\lambda$2.207, CaT $\lambda$2.266) are detected. The H-band stellar absorption features (e.g. \ion{Si}{i} $\lambda$ 1.59 $\mu$m, CO(6-3) $\lambda$1.62 $\mu$m) and the [\ion{Fe}{ii}] $\lambda$1.644 $\mu$m line, which is important for NIR diagnostics are also detected. The emission lines properties were determined from Gaussian profile fits. A FOV of $10\arcsec\times10\arcsec$ instead of the available $12\arcsec\times12\arcsec$ is used because the signal to noise in our utmost outer regions is not good enough for a thorough analysis. 

The new galaxy center is adopted in all figures shown (see table \ref{tab:basic} and Sec. \ref{sec:nucleus}). The black $\times$ in all shown maps gives the position of the new adopted center of the galaxy, i.e. the location of the SMBH. The FWHM maps and values are not corrected for instrumental broadening which is $\sim100$~km~s$^{-1}$ in H-band and $\sim75$~km~s$^{-1}$ in K-band. Emission line properties, i.e flux and FWHM, of all discussed emission lines are summarized in table \ref{tab:region}.

\subsection{Ionized gas}
\label{sec:iongas}

The forbidden line transition [\ion{Fe}{ii}] in H-band is detected in a resolved region at the center of about 38~pc in diameter (see Fig. \ref{fig:lines1}). The flux peak is slightly offset by one spaxel (0\farcs125) to the south. The shape of the emission looks slightly elongated in the north--east to south--west direction, similar to the PSF elongation. 
The line-of-sight velocity (LOSV) shows that the line is redshifted with respect to the systemic velocity by $\sim 20$ km s$^{-1}$. The red velocity maximum is $\sim 18$~pc south of the center where the velocity is $-45 km$ s$^{-1}$. The blue counterpart with a velocity of $\sim 20$ km s$^{-1}$ lies north of the center. The line width is rather uniform with an FWHM of about 185 km s$^{-1}$ and a faint broadening with an FWHM of 210 km s$^{-1}$ in the center. The equivalent width (EW) peaks at the same position as the flux with $\sim2.2$ $\AA$ indicating that the ionization of the [\ion{Fe}{ii}] originates from the very center of the galaxy (see Fig. \ref{fig:ew}).

Br$\gamma$ is detected only at the center as well and is marginally resolved in a region with a size of about 30~pc in diameter (see Fig. \ref{fig:lines1}). Only a narrow component of this recombination line can be detected. The emission peaks at the center and is roughly elongated in the east--west direction. The line width at the very center is with an FWHM $\sim 200$ km s$^{-1}$ similar to that of the [\ion{Fe}{ii}] emission line but otherwise its at $\sim 150$ km s$^{-1}$ in the emission region. The LOSV shows a blue end in the north-west direction and a red end in the south-west. One might assume a rotation with the line of nodes in this direction (PA $\sim$ 114$\degr$), however, the region Br$\gamma$ is detected in is too small to say much more about the velocity field in the central region. The velocity at the very center is $\sim 30$ km s$^{-1}$ redshifted. The EW is small with a maximum of about 0.8 $\AA$ on the nucleus hinting at a nuclear ionization (see Fig. \ref{fig:ew}).

The line ratio log([\ion{Fe}{ii}]/Br$\gamma$) can be used to determine the main excitation mechanism \citep{alonso-herrero_using_1997} in the center. In the central emission region with a radius of $1\arcsec$ a line ratio of 0.8 is measured. This value is rather typical of a mixed ionization by X-ray photons from an AGN (log([\ion{Fe}{ii}]/Br$\gamma$) $\lesssim$ 1.3), stellar thermal UV photons and shocks.
If the central Br$\gamma$ excitation is assumed to be originating from stars only, i.e. no excitation by AGN radiation, the luminosity of Br$\gamma$ in this region can be used to determine a star formation rate \citep[SFR;][]{panuzzo_dust_2003,valencia-s._is_2012}. We derive an upper limit for the SFR within a 1$\arcsec$ radius around the nucleus of $1.4 \times 10^{-3} M_{\odot}$yr$^{-1}$ using
\begin{equation}
\mbox{SFR}=\frac{L_{Br\gamma}}{1.585\times 10^{32}\mbox{W}}M_{\odot} \mbox{yr}^{-1}
\label{eqn:}
\end{equation}
and $L_{Br\gamma}$ from above. 
Additionally, if outflows or jets are disregarded as excitation sources in this region and [\ion{Fe}{ii}] excitation mainly by shocks from supernovae is assumed an upper limit for the supernova rate (SNR) in the central region can be estimated. Following \citet{bedregal_near-ir_2009} we use two different estimators
\begin{equation}
\mbox{SNR}_{\mbox{\tiny Cal97}}=5.38\;\frac{L_{[\ion{Fe}{ii}]}}{10^{35}\mbox{W}}\;\mbox{yr}^{-1}
\label{eqn:}
\end{equation}
after \citet{calzetti_reddening_1997} and
\begin{equation}
\mbox{SNR}_{\mbox{\tiny AlH03}}=8.08\;\frac{L_{[\ion{Fe}{ii}]}}{10^{35}\mbox{W}}\;\mbox{yr}^{-1}
\label{eqn:}
\end{equation}
after \citet{alonso-herrero_[fe_2003}. Both authors derive their relation from empirical data analysis. With an [\ion{Fe}{ii}] luminosity of $L_{[\ion{Fe}{ii}]}$ = 1.92 $\times$ 10$^{30}$ W the SNRs are estimated to be 1.03 $\times$ 10$^{-4}$ yr$^{-1}$ and 1.55 $\times$ 10$^{-4}$ yr$^{-1}$, respectively. These SFR and SNR values are low \citep[e.g.][and references therein]{rosenberg_[FeII]_2012,esquej_nuclear_2014} compared to other non-starburst galaxies, particularly with regard to the assumption that all of the central Br$\gamma$ and [\ion{Fe}{ii}] emission is attributed to star formation, which is an unlikely scenario in a Seyfert like galaxy.

\subsection{Molecular gas}
\label{sec:molgas}

At a five sigma confidence level several H$_2$ molecular emission lines are detected:
H$_2$(1-0)S(3) $\lambda$1.96 $\mu$m, H$_2$(1-0)S(2) $\lambda$2.03 $\mu$m, H$_2$(1-0)S(1) $\lambda$2.12 $\mu$m, H$_2$(1-0)S(0) $\lambda$2.22 $\mu$m, H$_2$(2--1)S(1) $\lambda$2.25 $\mu$m, H$_2$(1-0)Q(1) $\lambda$2.41 $\mu$m, H$_2$(1-0)Q(3) $\lambda$2.42 $\mu$m. The most prominent lines are H$_2$(1-0)S(3) and H$_2$(1-0)S(1).

\begin{figure*}[htbp]
\centering
\subfigure[Flux $^{12}$CO(3-2)]{\includegraphics[width=0.33\textwidth]{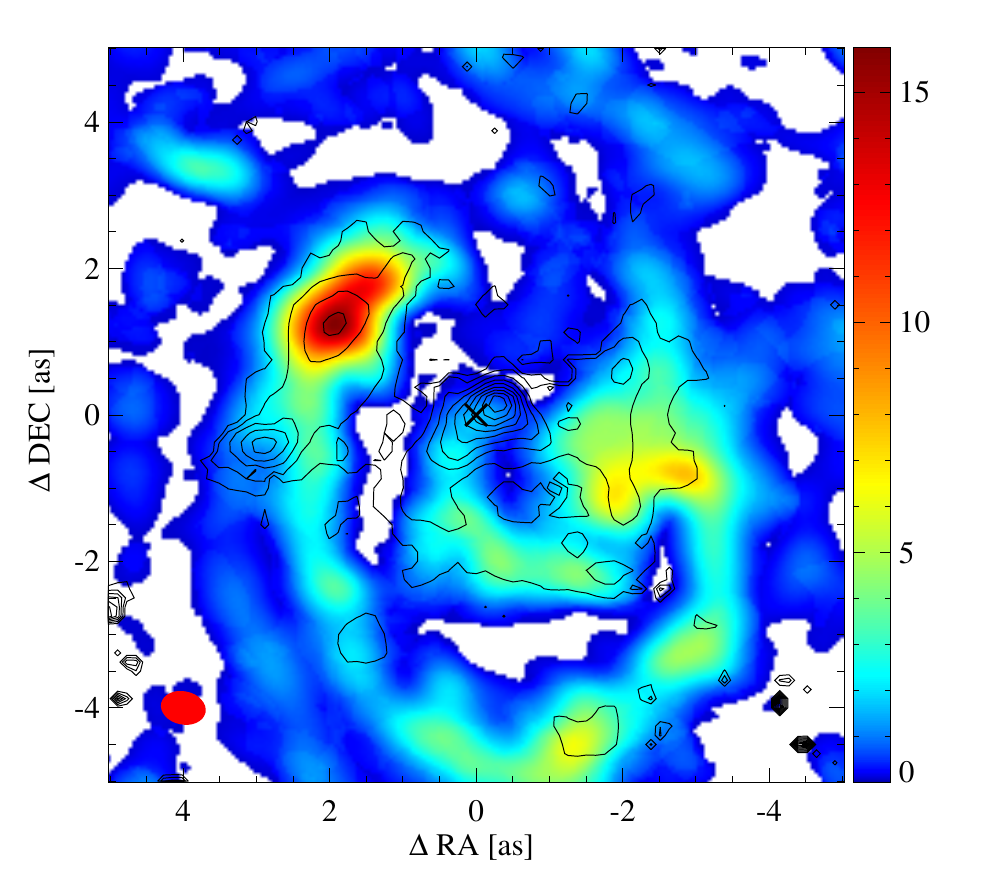}\label{fig:ALMACO32}}
\subfigure[Dispersion $^{12}$CO(3-2)]{\includegraphics[width=0.33\textwidth]{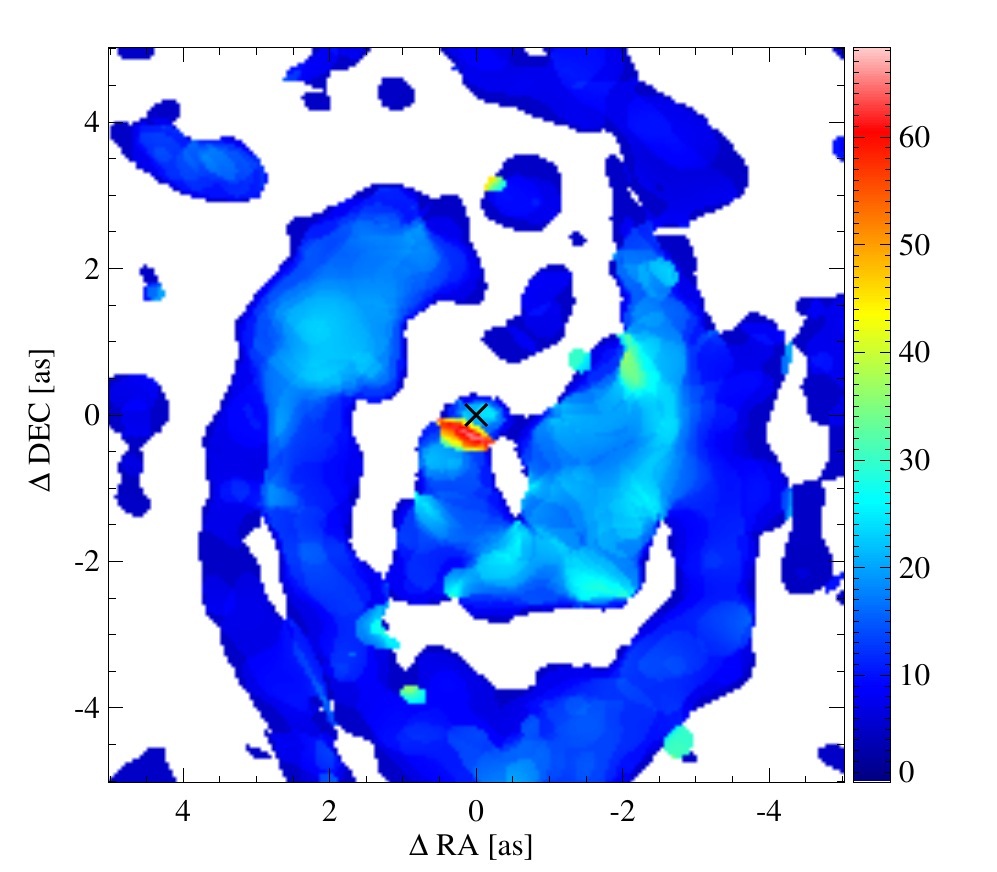}\label{fig:ALMAmom2}}
\subfigure[LOSV $^{12}$CO(3-2)]{\includegraphics[width=0.33\textwidth]{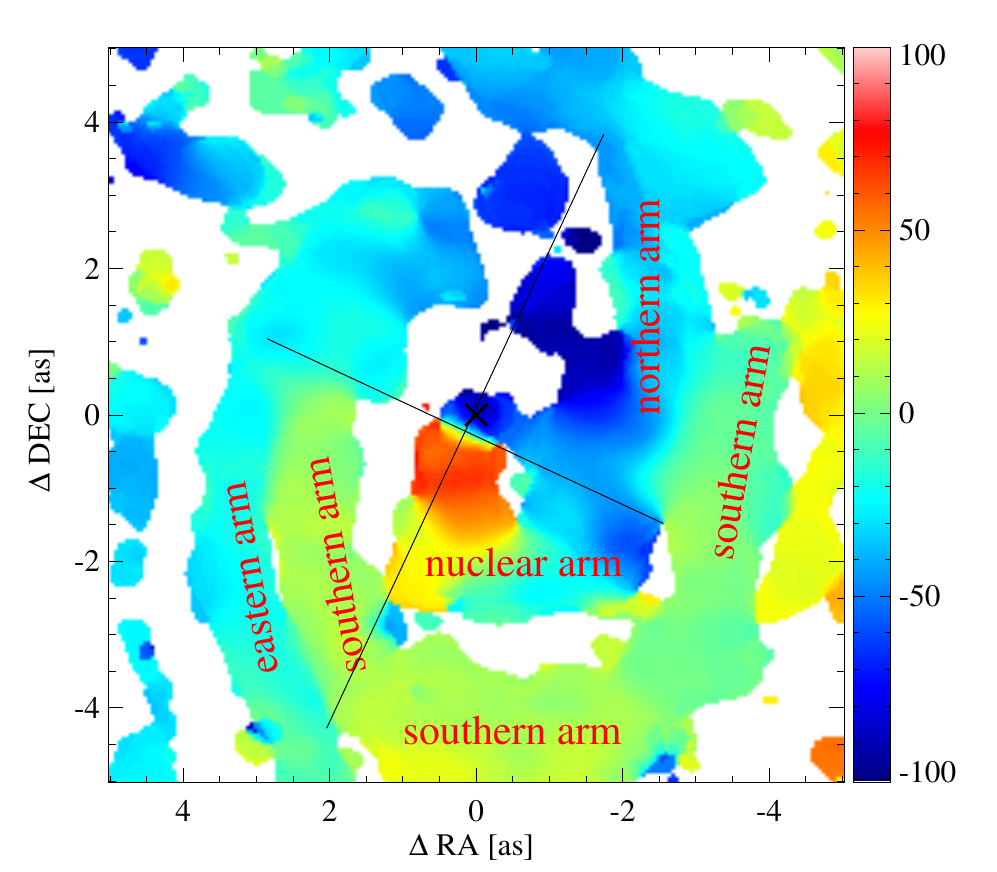}\label{fig:ALMAmom1}}
\caption{ALMA cycle 0 molecular $^{12}$CO(3-2) maps. \subref{fig:ALMACO32} shows the emission map of $^{12}$CO(3-2) [Jy beam$^{-1}$ km s$^{-1}$] overlayed with H$_2$(1-0)S(1) emission contours. The red ellipse at -4,4 describes the beam size. The dispersion (moment 2) is shown in \subref{fig:ALMAmom2} and the LOSV is shown in \subref{fig:ALMAmom1}, both maps are in [km s$^{-1}$]. \subref{fig:ALMAmom1} shows the cuts for the position velocity (PV) diagrams (see Fig. \ref{fig:pvcut}). The cuts were not set on the galactic center but along the central velocity gradient of the $^{12}$CO(3-2) 1st moment map.}
\label{fig:ALMA}
\end{figure*}

\begin{figure*}[htbp]
\centering
\subfigure[EW $\mbox{[}$\ion{Fe}{ii}$\mbox{]}$]{\includegraphics[width=0.33\textwidth]{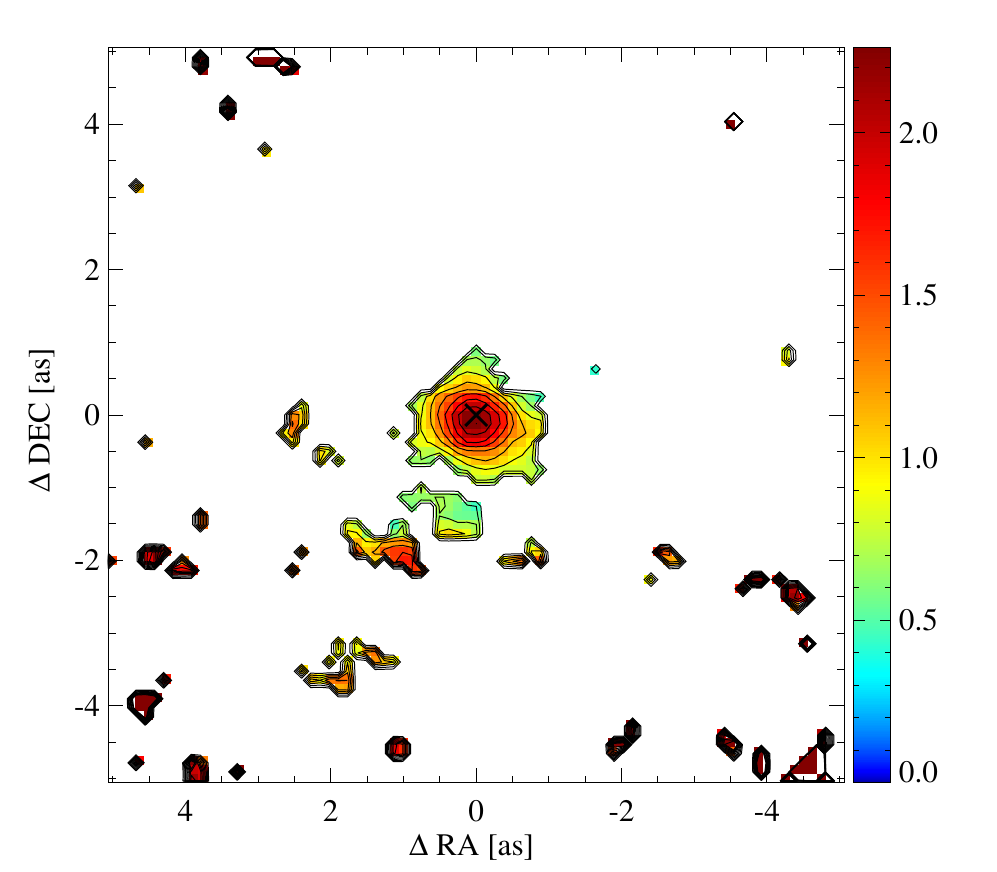}\label{fig:feiiew}}
\subfigure[EW Br$\gamma$]{\includegraphics[width=0.33\textwidth]{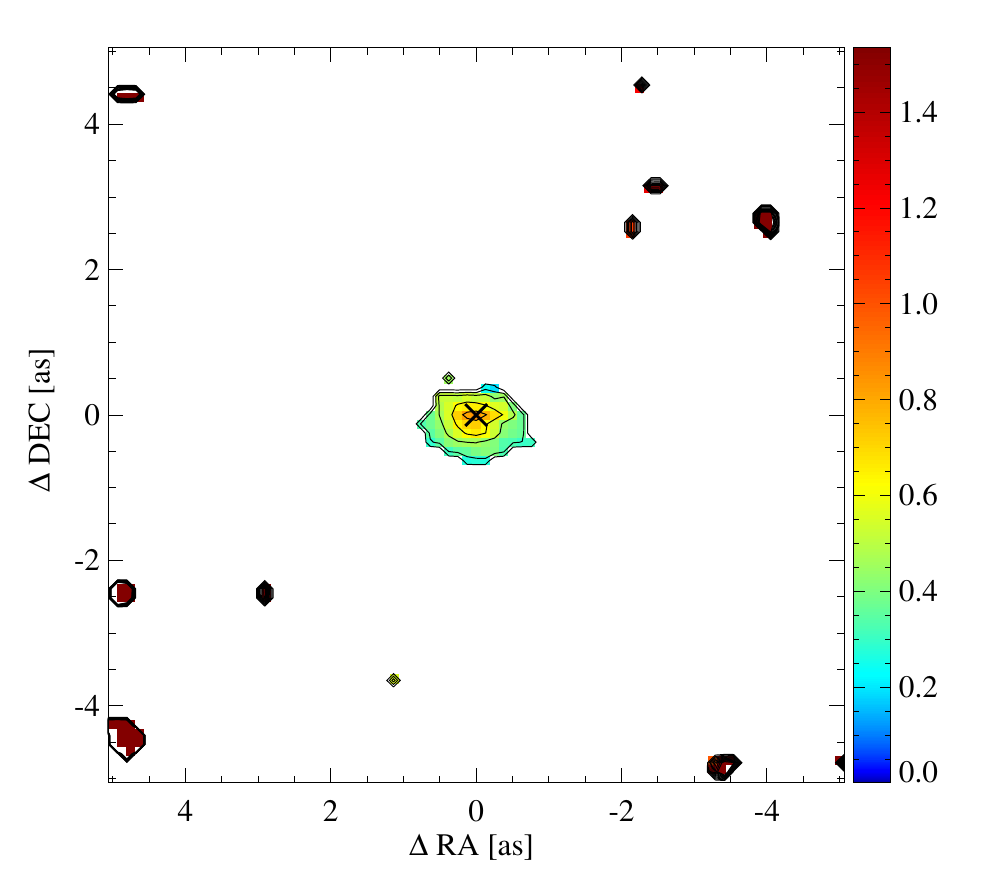}\label{fig:brgew}}
\subfigure[EW H$_2$ $\lambda$2.12 $\mu$m]{\includegraphics[width=0.33\textwidth]{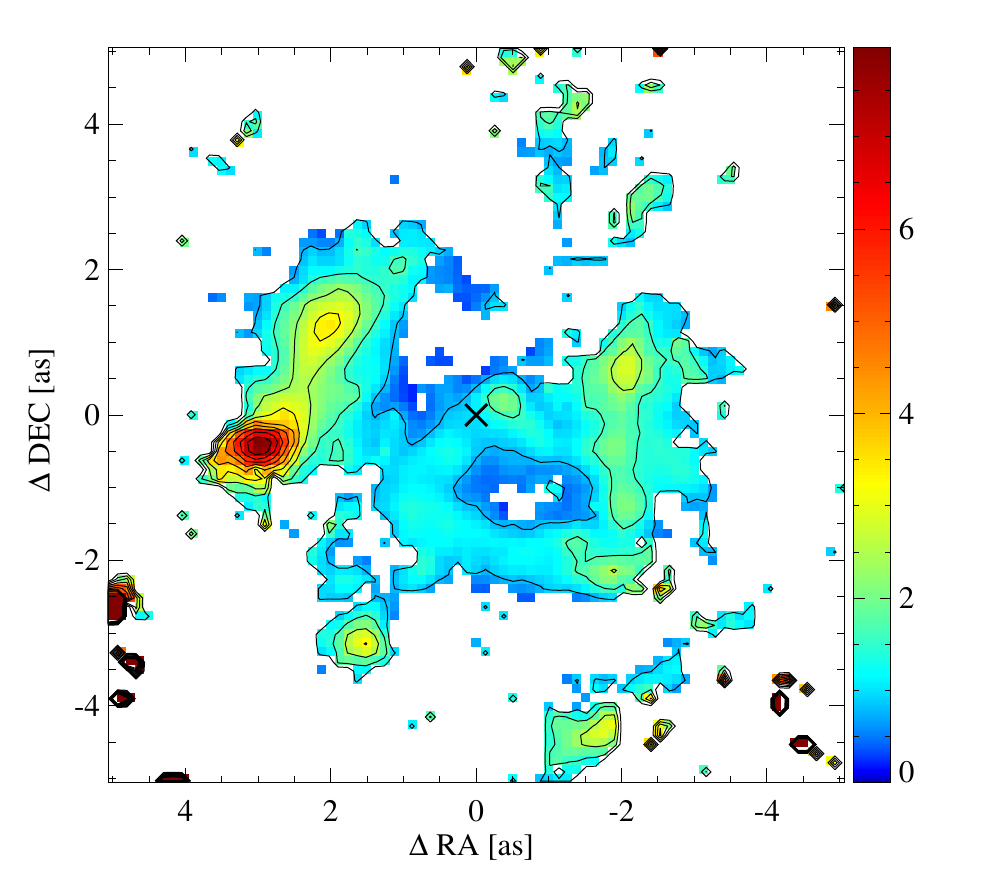}\label{fig:h212ew}}
\caption{EW maps of the narrow Br$\gamma$, [\ion{Fe}{ii}] and H$_2$ $\lambda$(1-0)S(1) lines in $\AA$. {\bf For more details see Sect. \ref{sec:res}.}}
\label{fig:ew}
\end{figure*}

The line map of H$_2$(1-0)S(1) shows an interesting structure in the nuclear region of NGC 1433 (see Fig. \ref{fig:h212flux}). First, the overall emission follows the dust lanes seen in the HST image (see Fig. \ref{fig:HSTH2}). Second, the emission peaks about 0\farcs5 north-west offset to the K-band continuum peak and forms an arc-like structure from there towards the south-east, where the continuum peak is located.
The LOSV along this region shows a strong gradient from about $-40$ km s$^{-1}$ to 40 km s$^{-1}$. There are also several bright emission spots. The most prominent are two spots 2$\arcsec$--3$\arcsec$ east--north--east of the nucleus and a less prominent emission region in the west about 2\farcs5 away from the center. There are several other emission spots distributed over the spiral dust lanes. The EW map shows that on the continuum peak the EW is $\sim 1 \AA$ while at the H$_2$(1-0)S(1) emission peak it is $\sim 2.1 \AA$. Moreover, the strongest EWs are seen in the eastern emission spots ranging from about 1.8~$\AA$ in the eastern surroundings over 3.4~$\AA$ in the north--eastern spot (sp(A)) and reaching about 8~$\AA$ in the spot (sp(B), see Fig. \ref{fig:ew}). Emission spot sp(B) shows an FWHM of more than 230 km s$^{-1}$ and a blue-shifted LOSV of 90 km s$^{-1}$ whereas the surrounding gas is rather systemic or slightly redshifted. The southern spot sp(B) belongs rather to a spiral arm that is not coupled with the inner nuclear (pseudo-) ring (see Figs. \ref{fig:lines1}, \ref{fig:ALMA}) described by \cite{combes_ALMA_2013}. 
The LOSV is rather blueshifted in the arm west of the center whereas the eastern side is rather systemic (see Fig. \ref{fig:losvh212}). 
The region south of the nucleus, where the gas is redshifted up to 70 km s$^{-1}$ in an 1$\arcsec$ distance, indicates a possible outflow. The FWHM in this region is 180 km s$^{-1}$ whereas in the center it is around 210 km s$^{-1}$. To the north--west of the nucleus there is a probable blue counter part at a velocity of only 30 km s$^{-1}$ and a distance of $0\farcs5$. An outflow scenario was thoroughly discussed in \citet{combes_ALMA_2013}. Another possible scenario that can explain this high LOSV gradient in all the emission lines is a gaseous disk in the center of NGC 1433. We will discuss this in section \ref{sec:centrarcs}.

H$_2$(1-0)S(3) is distributed similarly to the H$_2$(1-0)S(1) line. It seems noisier than H$_2$(1-0)S(3) which is due to a spectrally noisier region in which the line is situated. Nevertheless, the line peaks in the same region as H$_2$(1-0)S(1) and has a tail along the K-band continuum peak towards the south-east. The two emission peaks in the east are detected as well as other emission spots that one can identify with bright regions in the H$_2$(1-0)S(1) emission map. The EWs are similar over the whole FOV. The line width shows generally higher velocities than the H$_2$(1-0)S(1) line.

All other detected H$_2$ lines show emission around the center and/or in the north-eastern emission spot. The dust lane structure is barely traced in emission line maps other than H$_2$(1-0)S(1) and H$_2$(1-0)S(3).

\subsubsection{H$_2$ mass estimation}
The warm H$_2$ gas mass can be estimated following \citet{turner_mass--light_1977,scoville_velocity_1982,wolniewicz_quadrupole_1998}
\begin{equation}
M_{\mbox{\tiny H$_2$}}=4.243\times10^{-30}\left(\frac{L_{\mbox{\tiny H$_2$(1-0)S(1)}}}{\mbox{W}}\right)\;\mbox{M}_\odot.
\label{eqn:}
\end{equation}
A warm H$_2$ gas mass of 4.89 $M_{\odot}$ can be estimated at the central 1$\arcsec$ ($\sim$54~pc) using an H$_2$ (1-0)S(1) luminosity of $L_{\mbox{\tiny H$_2$(1-0)S(1)}}=1.15\times$10$^{30}$ W. Within $10\arcsec\times10\arcsec$ FOV a warm H$_2$ gas mass of 19.5 $M_{\odot}$ is estimated using a 5$\arcsec$ radius aperture centered on the nucleus. The conversion factor detemined by \citet{mazzalay_molecular_2013}
\begin{equation}
\frac{M_{\mbox{\tiny H}_2(\mbox{\tiny cold})}}{M_{\mbox{\tiny H}_2(\mbox{\tiny warm})}} = (0.3-1.6)\times 10^6
\label{eqn:}
\end{equation}
is used to determine the cold H$_2$ gas mass. The cold H$_2$ gas mass of (1.5 -- 7.8) $\times$ 10$^6 M_{\odot}$ is derived for the 1$\arcsec$ radius aperture and (0.6 -- 3.1)$\times$10$^7 M_{\odot}$ for the 5$\arcsec$ radius aperture. This means that one third of the H$_2$ in the central 10$\arcsec$ can be found in the central 1$\arcsec$ radius around the nucleus. \citet{combes_ALMA_2013} derive an H$_2$ gas mass for their outflow, which covers about the same region as our central 1$\arcsec$ radius, of 3.6 $\times$ 10$^6 M_{\odot}$. This is in good agreement with our measurements. However, from our measurements of the $^{12}$CO(3-2) emission we estimate an H$_2$ gas mass of $2.1\times10^5 M_{\odot}$ for a 1$\arcsec$ radius and $8.3\times10^6 M_{\odot}$ for a $5\arcsec$ radius, using the Milky Way dust-to-gas ratio \citep{solomon_CO_1991}.

\begin{table}[htbp]
\centering
\caption{Basic data on NGC1433.}
\begin{tabular}{c c c}
\hline\hline
 Parameter & Value & Reference$^c$ \\
\hline
 $\alpha_{J2000}^a$ & 03$^{\mbox{\tiny h}}$42$^{\mbox{\tiny m}}$01.49$^{\mbox{\tiny s}}$ & (1) \\
 $\delta_{J2000}^a$ & -47$\degr$13$\arcmin$20$\farcs$2 & (1) \\
 Nuclear activity & Seyfert & (2) \\
 Redshift & 0.003586 & (3) \\
 Distance & 9.9 Mpc (1$\arcsec=$~48~pc) & (1) \\
 Inclination & 33$\degr$ & (4) \\
 Position angle & $199\degr\pm1\degr$ & (4) \\
 $\alpha_{J2000}^b$ & 03$^{\mbox{\tiny h}}$42$^{\mbox{\tiny m}}$01.49$^{\mbox{\tiny s}}$ & This paper \\
 $\delta_{J2000}^b$ & -47$\degr$13$\arcmin$18$\farcs$8 & This paper \\

\hline

\end{tabular}
\caption*{\textbf{Notes}: $^{(a)}$ Galaxy center as adopted by \citet{combes_ALMA_2013}. $^{(b)}$ The new adopted center taken from the ALMA observations world coordinate system. $^{(c)}$ (1) \citet{combes_ALMA_2013}; (2) \cite{veron-cetty_miscellaneous_1986}; (3) \citet{koribalski_1000_2004}; (4) \citet{buta_dynamics_2001}.
}
\label{tab:basic}
\end{table}

\subsection{Continuum}
\label{sec:continuum}

To address the continuum emission in the nuclear region of NGC 1433 a decomposition as described in \citet{smajic_unveiling_2012} was used. 
The continuum shows only a stellar contribution with almost no extinction needed for the fit. 
Hence, the K-band continuum and the stellar continuum peak in the same position and their photon distributions look very similar. The stellar continuum was obtained by fitting the CO(2-0) bandhead with a stellar spectrum. Stellar templates from \citet{winge_gemini_2009} were used as input for the fitting routine. The template spectra were convolved with a Gaussian to adapt the resolution to our SINFONI resolution of 4000 in K-band. The red giant HD2490 -- spectral class M0III -- fitted best without the need of additional stellar spectra (e.g. asymptotic giant branch stars). What might be surprising is the fact that the NIR continuum peak coincides with the optical -- HST images from F450W, F660W and F810W filters -- and is not affected by the obvious dust lanes seen in Fig. \ref{fig:HSTH2}. Unfortunately, the spectra of these template stars have a wavelength coverage of only 2.2 $\mu$m to 2.4 $\mu$m. Therefore, the fitted spectral region is small, which may explain why no extinction component was needed for the fit, even at the spatial positions of the dust lanes.

\begin{figure}[h!]
\centering
\includegraphics[width=0.35\textwidth]{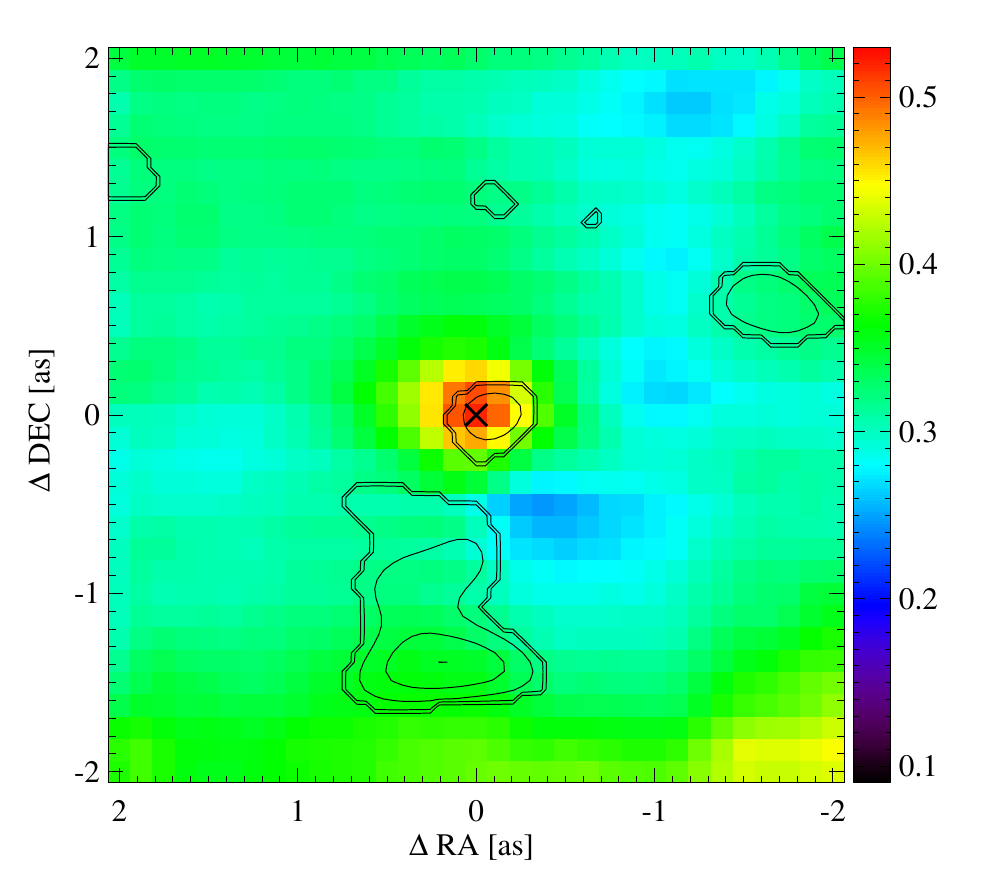}
\caption{The H-K color diagram in magnitudes showing the center of NGC 1433 measured in the SINFONI H- and K-band. The contours are from the 0.87 mm continuum emission measured by ALMA. For more details see section \ref{sec:continuum} \& \ref{sec:nucleus}.}
\label{fig:hkcont}
\end{figure}

\begin{figure}[h!]
\centering
\includegraphics[height=0.35\textwidth,angle=90]{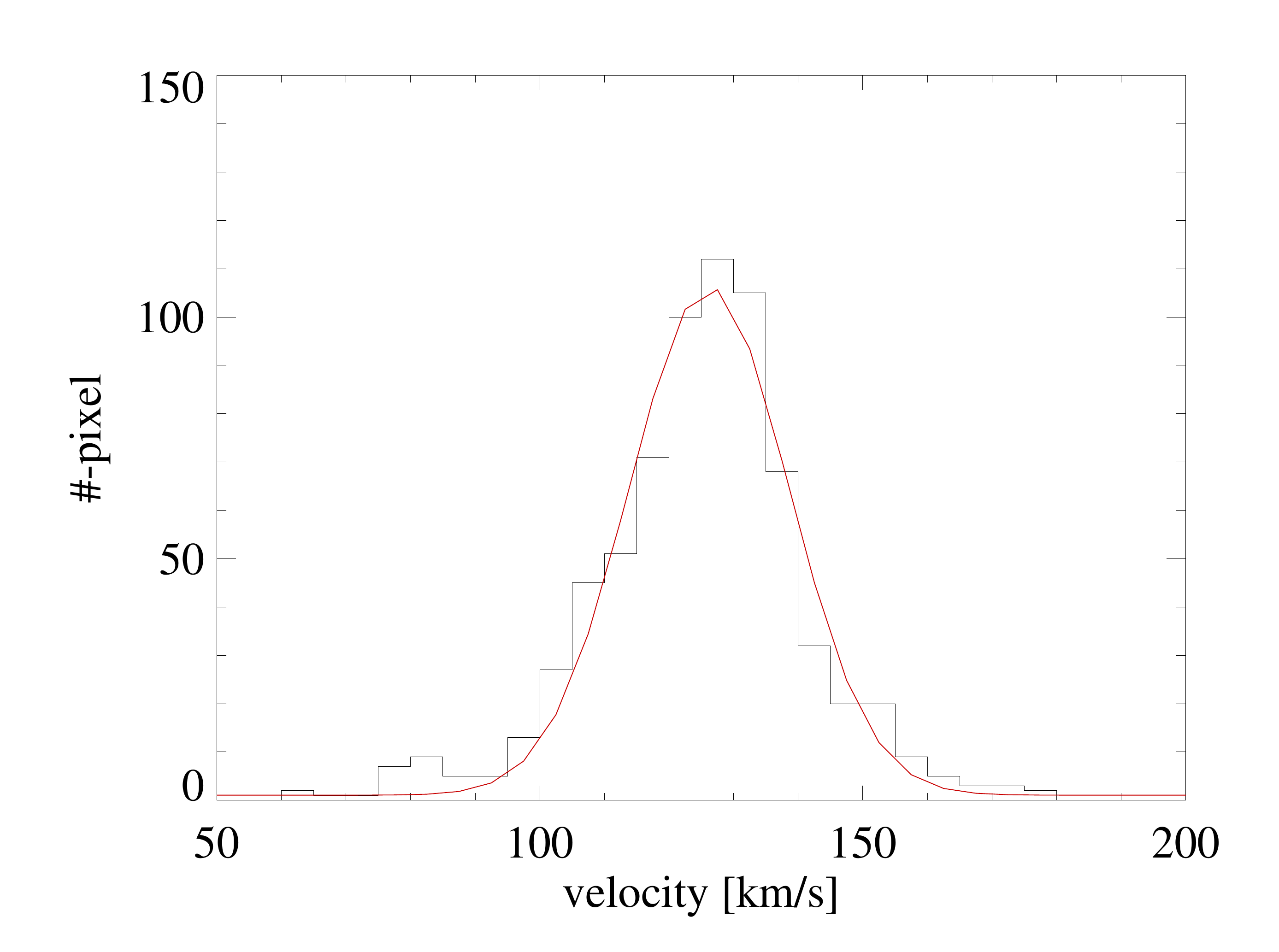}
\caption{The fit over the stellar dispersion from the continuum decomposition in 5~\kms\ bins. The X-axis shows the fitted dispersion and the Y-axis the total number of spatial pixel that correspond to this dispersion bin. The red curve represents a Gaussian fit to the distribution.}
\label{fig:CO20disp}
\end{figure}

The H-K color map (see Fig. \ref{fig:hkcont}) shows no reddening at the dust lane position,
but towards the newly adopted center a strong reddening in H-K is detected.
This can be taken as a signature of hot dust at the sublimation temperature.
In AGN this hot dust is located at the sublimation radius which 
is co-spatial with the outer edges of the accretion disk and/or the 
inner edge of the circum-nuclear torus.
From the CO(2-0) band head fitting the LOSV of the stars can be determined and a rotation direction at a PA $\sim201\degr$ is measured. The velocity field seems to be systematically redshifted by a few km s$^{-1}$. A reason may be the imperfect wavelength calibration as stated in section \ref{sec:obs_red}. The projected velocities range from about -50 to 60 km s$^{-1}$ and the 0 km s$^{-1}$ isovelocity contour lies close to the continuum center. The line of sight velocity dispersion (LOSVD) ranges from about 90 to 160 km s$^{-1}$. From the velocity dispersion we can determine a black hole mass, assuming that NGC 1433 lies on the M-$\sigma$ relation. Following \citet{gultekin_m-_2009} 
\begin{equation}
M_{\bullet}=10^{8.12\pm0.08}\times\left(\frac{\sigma_{*}}{200\;\mbox{km s}^{-1}}\right)^{4.24\pm0.41}M_{\odot}
\end{equation}
we determine a black hole mass of M$_{\bullet}$ = 7 $\times$ 10$^6 M_{\odot}$ with an assumed velocity dispersion of $\sigma_*$ = 100 km s$^{-1}$ which is measured at the very center, where a minimum in LOSV is measured. Such a minimum in the LOSVD is often found in galaxies with a nuclear disk which is not always resolved \citep{emsellem_dynamics_2001,falcon-barroso_sauron_2006}. By assuming a Gaussian distribution of the fitted LOSVD over our FOV (see Fig. \ref{fig:CO20disp}) we determine a mean velocity dispersion of $125\pm10$~\kms\ hence, a black hole mass of M$_{\bullet}=(1.8\pm0.8)\times10^7M_{\odot}$. The first estimate agrees with \citet{buta_structure_1986}, however, the latter represents the bulges LOSVD which is needed for the equation given above.

\begin{figure*}[htbp]
\centering
\subfigure[H-band continuum]{\includegraphics[width=0.33\textwidth]{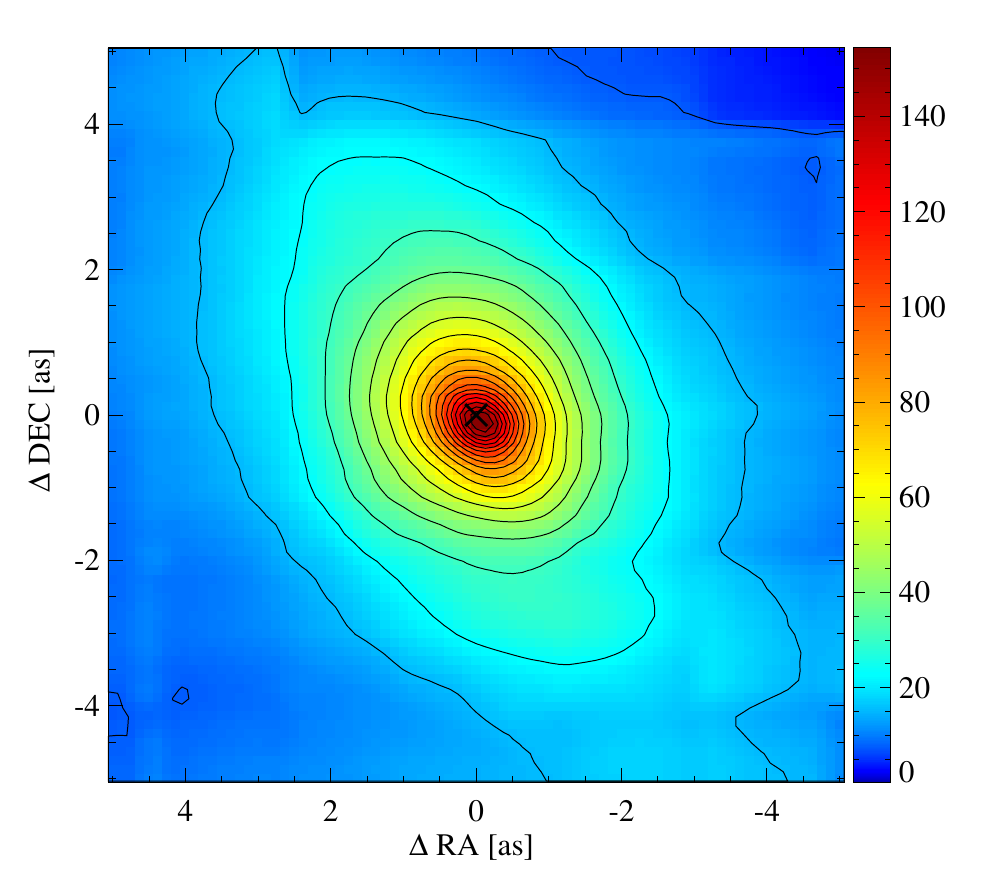}\label{fig:hcont}}
\subfigure[K-band continuum]{\includegraphics[width=0.33\textwidth]{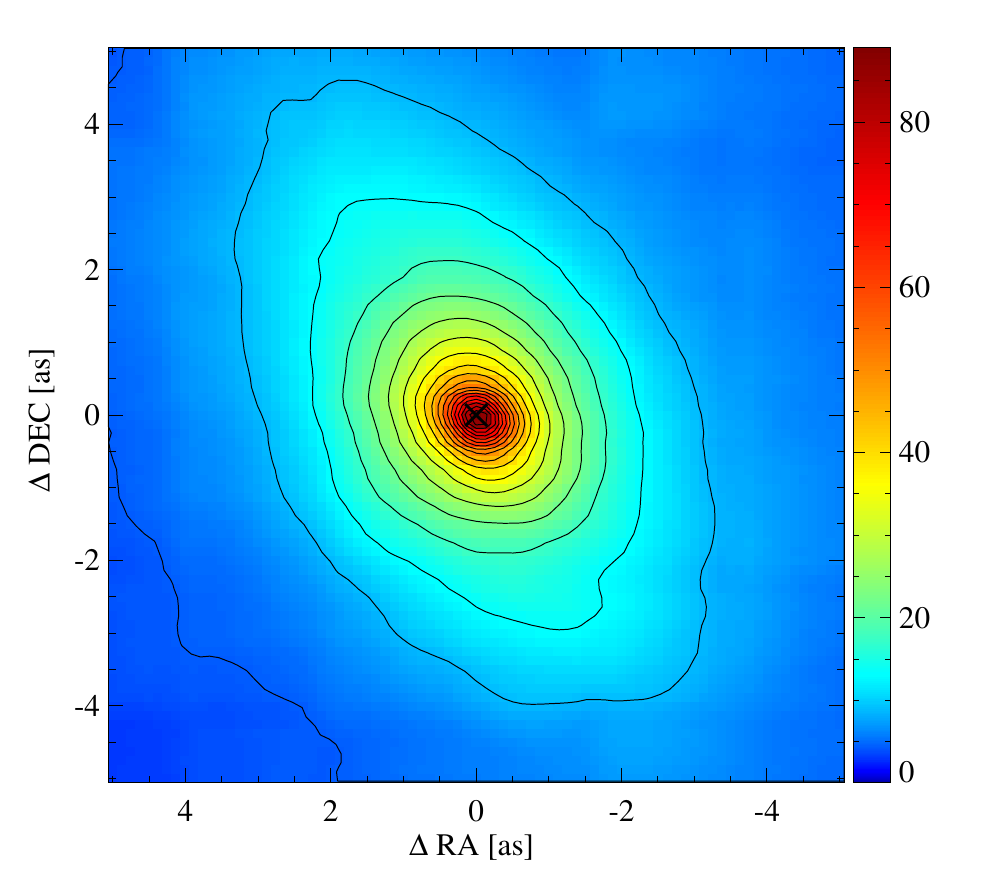}\label{fig:kcont}}
\subfigure[0.87 mm continuum]{\includegraphics[width=0.33\textwidth]{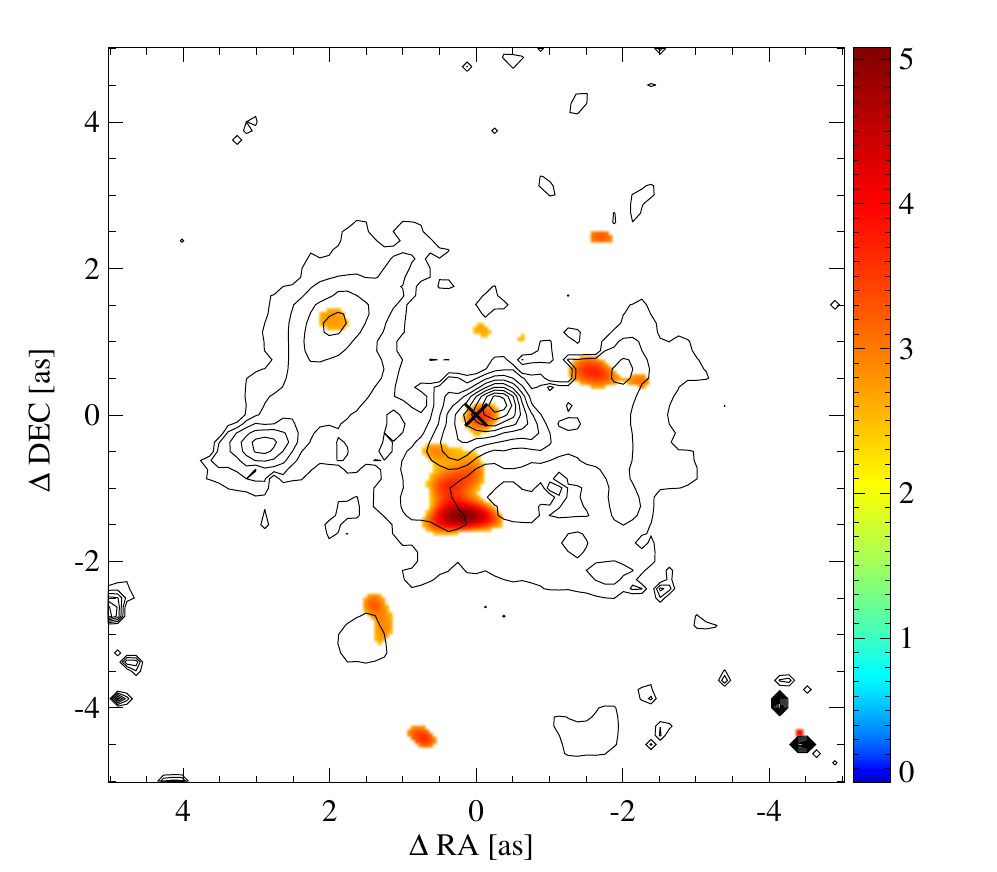}\label{fig:ALMAcont}}
\caption{From left to right H-band, K-band [$10^{-18}$ W m$^{-2}$ $\mu$m$^{-1}$] and the 0.87 mm [Jy beam$^{-1}$] continuum are shown. \subref{fig:ALMAcont} H$_2$(1-0)S(1) contour overlayed on the 3 sigma clipped 0.87 mm continuum. For more details see section \ref{sec:continuum} \& \ref{sec:nucleus}.}
\label{fig:cont}
\end{figure*}

\begin{figure*}[htbp]
\centering
\subfigure[Stellar LOSV]{\includegraphics[width=0.33\textwidth]{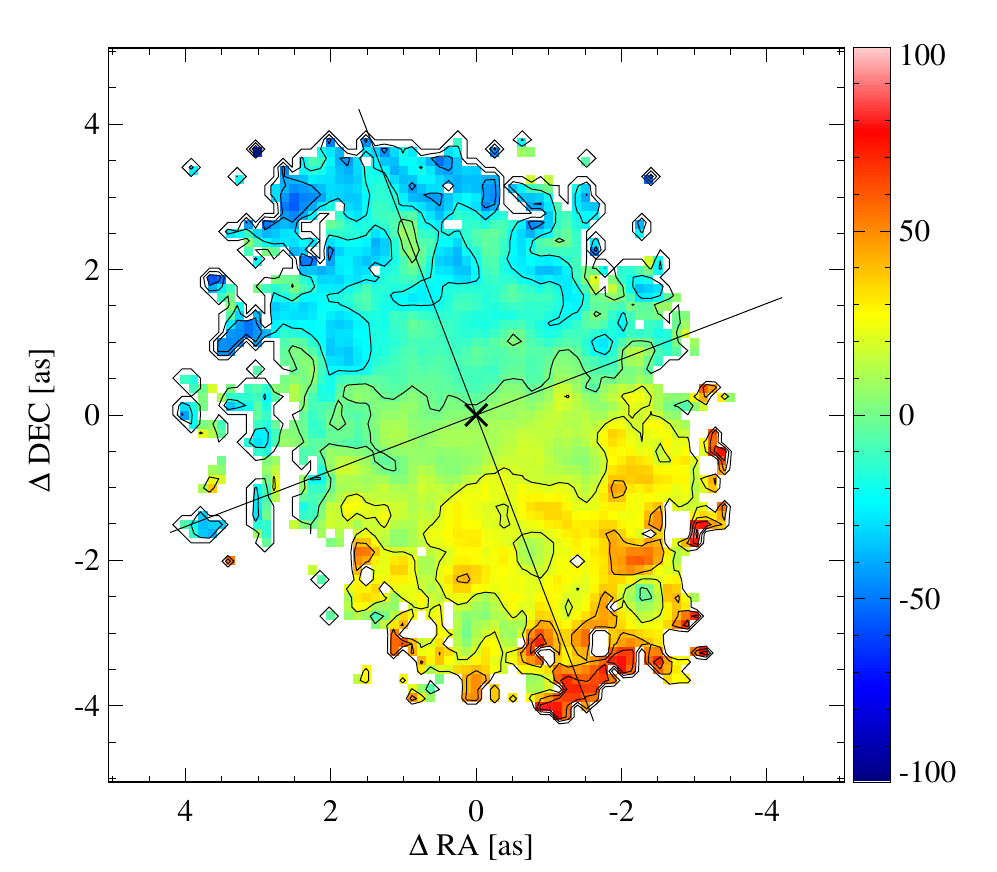}\label{fig:losvco20}}
\subfigure[Stellar dispersion]{\includegraphics[width=0.33\textwidth]{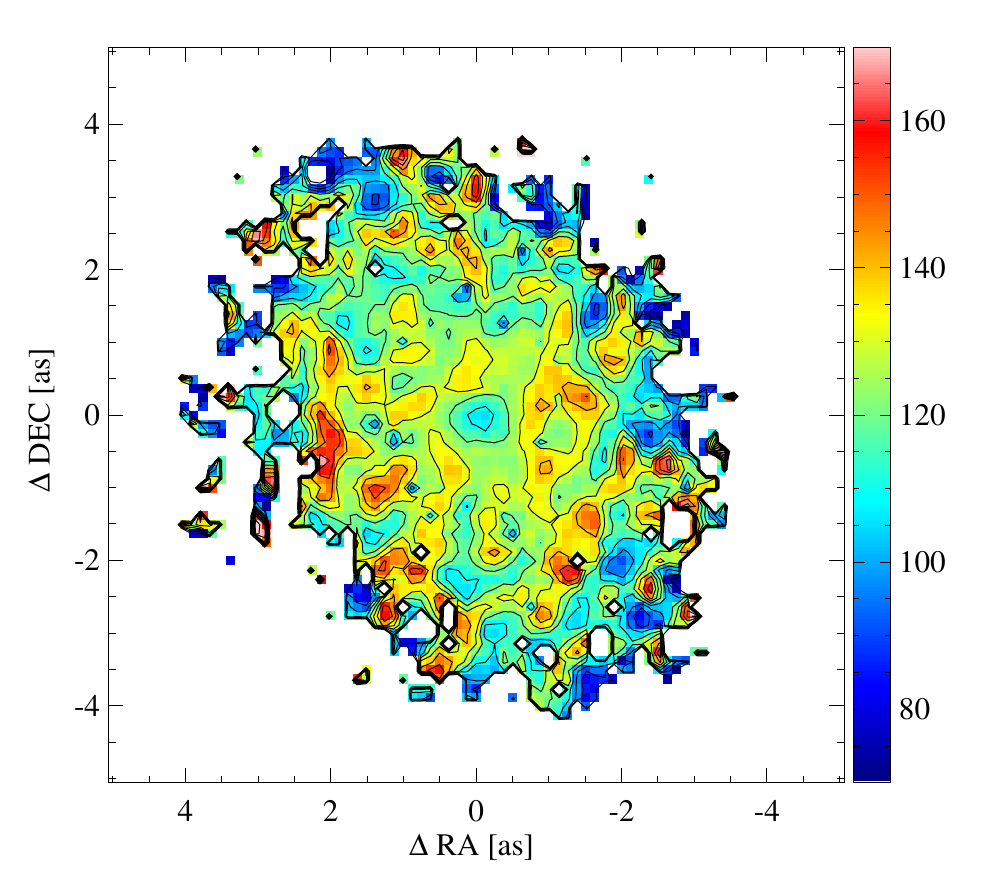}\label{fig:sigmaco20}}
\subfigure[Stellar continuum]{\includegraphics[width=0.33\textwidth]{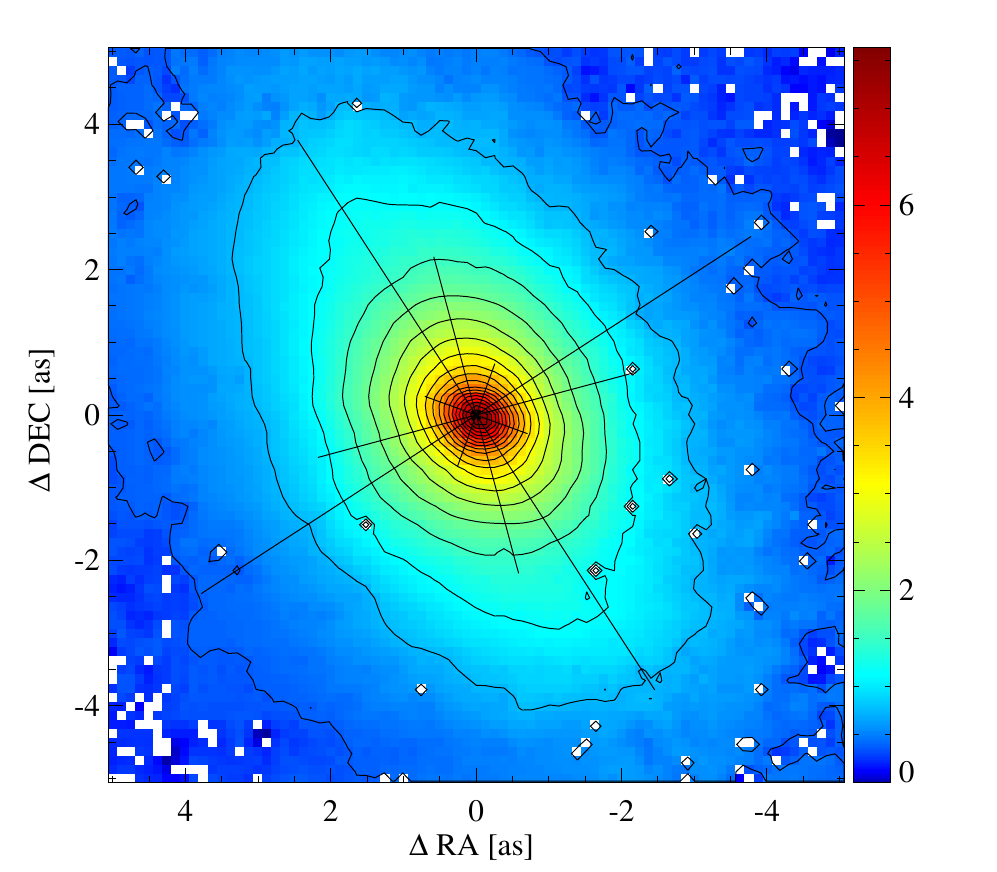}\label{fig:stellcont}}
\caption{The fitted stellar LOSV map [km s$^{-1}$], stellar disperion [km s$^{-1}$] and stellar continuum (arbitrary unit) map are shown. In fig. \subref{fig:losvco20} the LOSV PA of the galactic rotation is marked. Figure \subref{fig:stellcont} shows probable PAs of the isophotes where the largest cross marks the PA of the nuclear bar and the smallest cross the orientation of the beam. The middle cross indicates a possible twist in the isophotal lines.}
\label{fig:stellvel}
\end{figure*}

\begin{figure*}[htbp]
\centering
\subfigure[H$_2$(1-0)S(1) over F450W HST]{\includegraphics[width=0.465\textwidth]{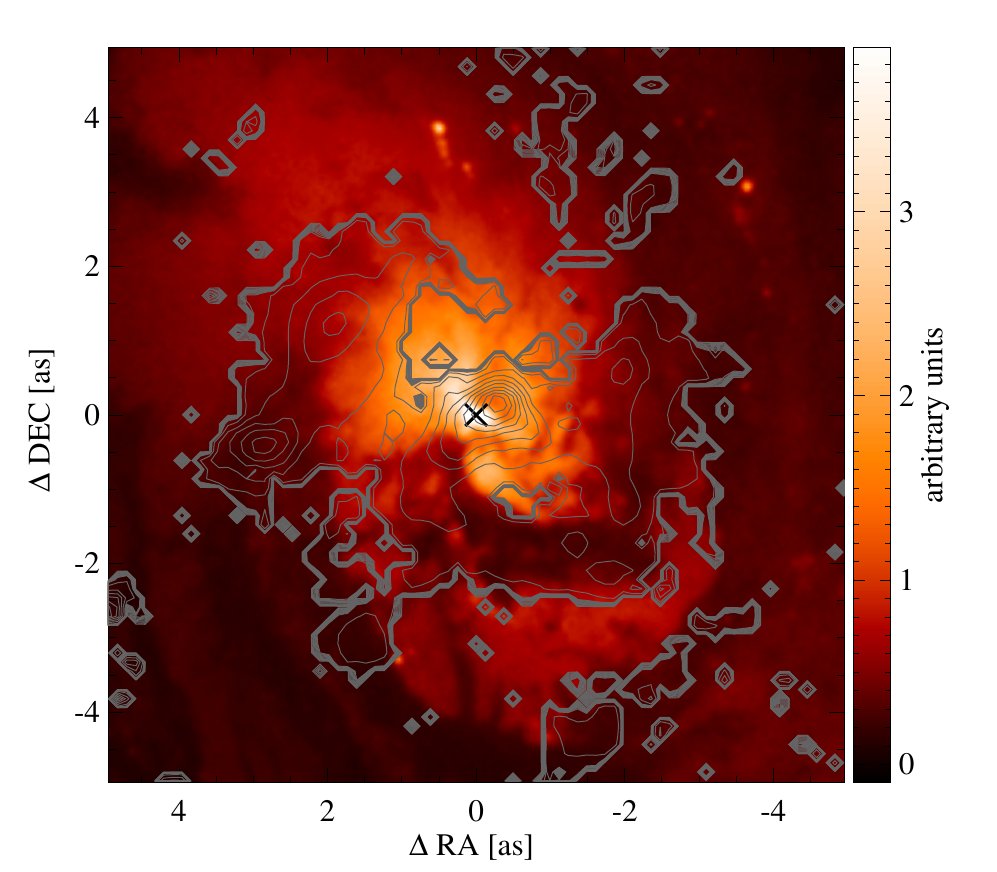}\label{fig:HSTH2}}
\subfigure[CO(3-2) over F450W HST]{\includegraphics[width=0.465\textwidth]{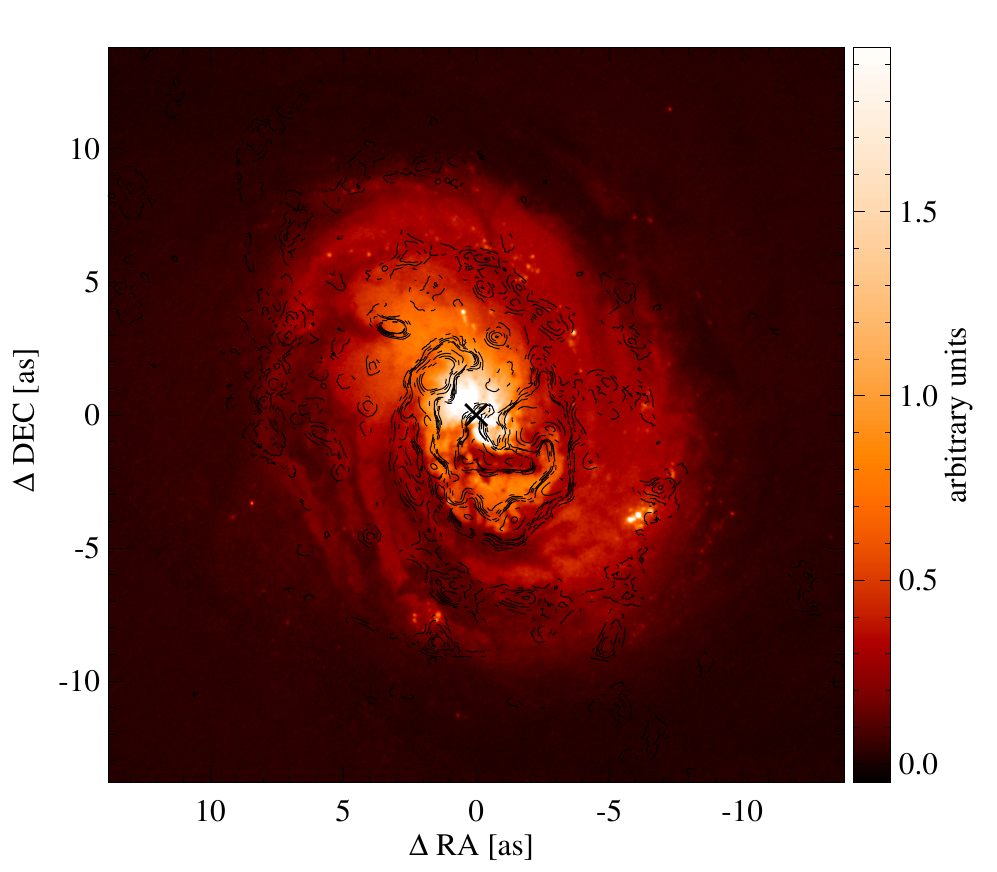}\label{fig:F450WALMA}}
\caption{Figure \subref{fig:HSTH2} shows an HST map at about 4380 $\AA$ with H$_2$(1-0)S(1) contours overlayed. Figure \subref{fig:F450WALMA} shows the same HST image in the ALMA FOV with the $^{12}$CO(3-2) line emission in contours. The HST images were taken from the archive and were observed with the WFC3 instrument and the UVIS detector using the F438W filter. The images illustrate the remarkable overlap of dust lanes and molecular hydrogen \subref{fig:HSTH2} and $^{12}$CO(3-2) emission \subref{fig:F450WALMA}, respectively.}
\label{fig:HSToverlay}
\end{figure*}

From the shape and angle of the K-band continuum isophotal ellipses we deduce that the PA of the nuclear stellar oval or bar is $\sim33\degr$ (see Fig. \ref{fig:kcont}, \ref{fig:stellcont}). For the isophotal lines that range from 25\% to 35\% of the peak luminosity the PA changes to 15$\degr$ before it shifts to a PA of about 70$\degr$ for the inner isophotes. The change in angle towards the inner isophotal lines can be attributed to the PSF which is as well elongated in the north--east to south--west direction at a PA of $\sim$70$\degr$.

\begin{figure*}[htbp]
\centering
\subfigure[H$_2$(1-0)S(1) -- stellar LOSV]{\includegraphics[width=0.33\textwidth]{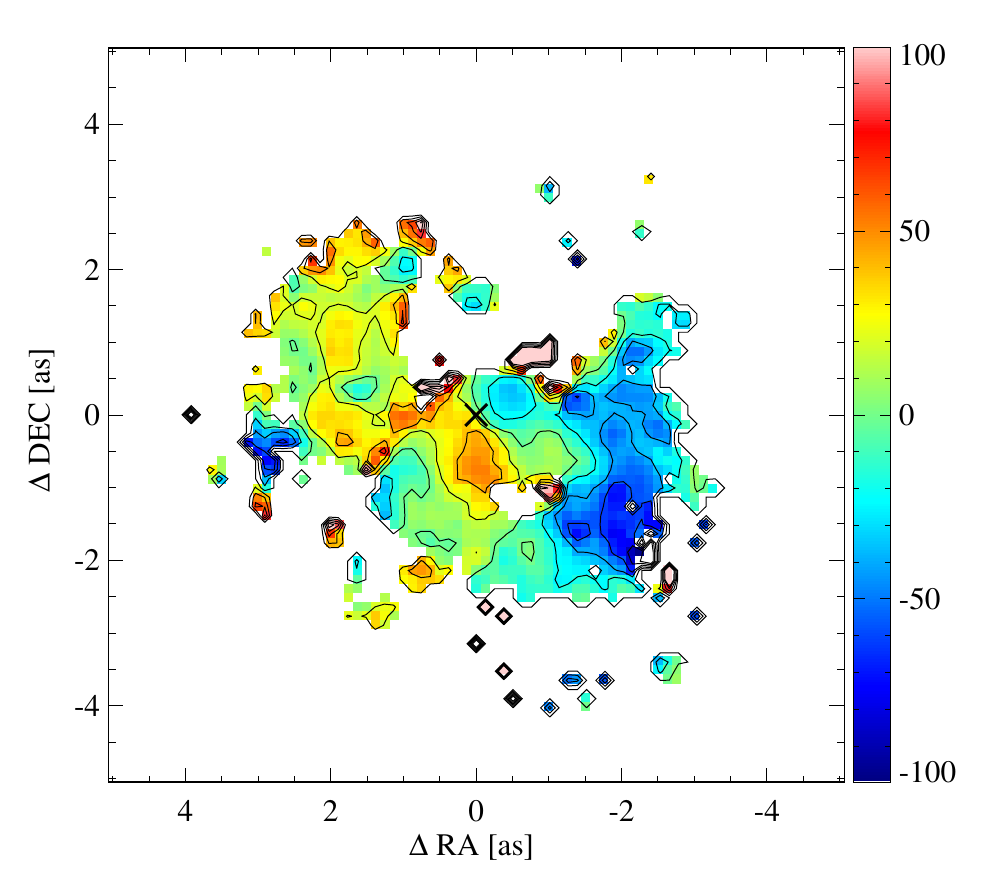}\label{fig:losvh2-co20}}
\subfigure[CO(3-2) -- stellar LOSV]{\includegraphics[width=0.33\textwidth]{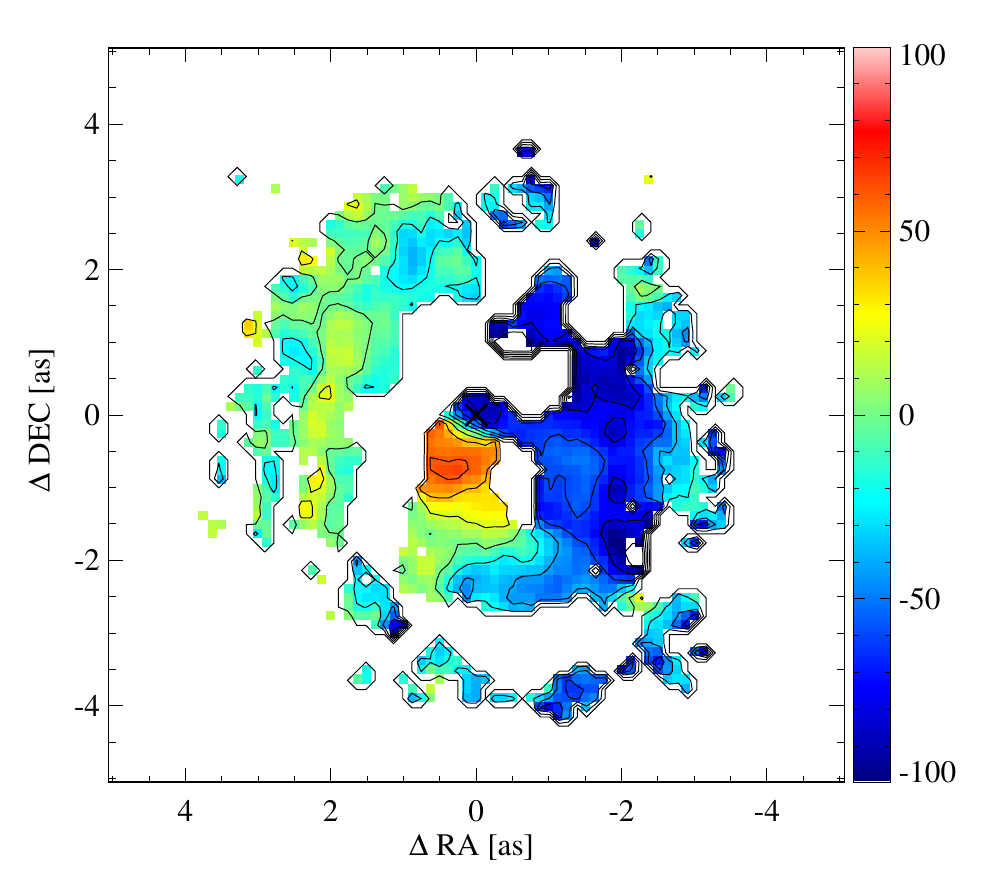}\label{fig:losvco32-co20}}
\subfigure[CO(3-2) -- stellar LOSV + 20 km s$^{-1}$]{\includegraphics[width=0.33\textwidth]{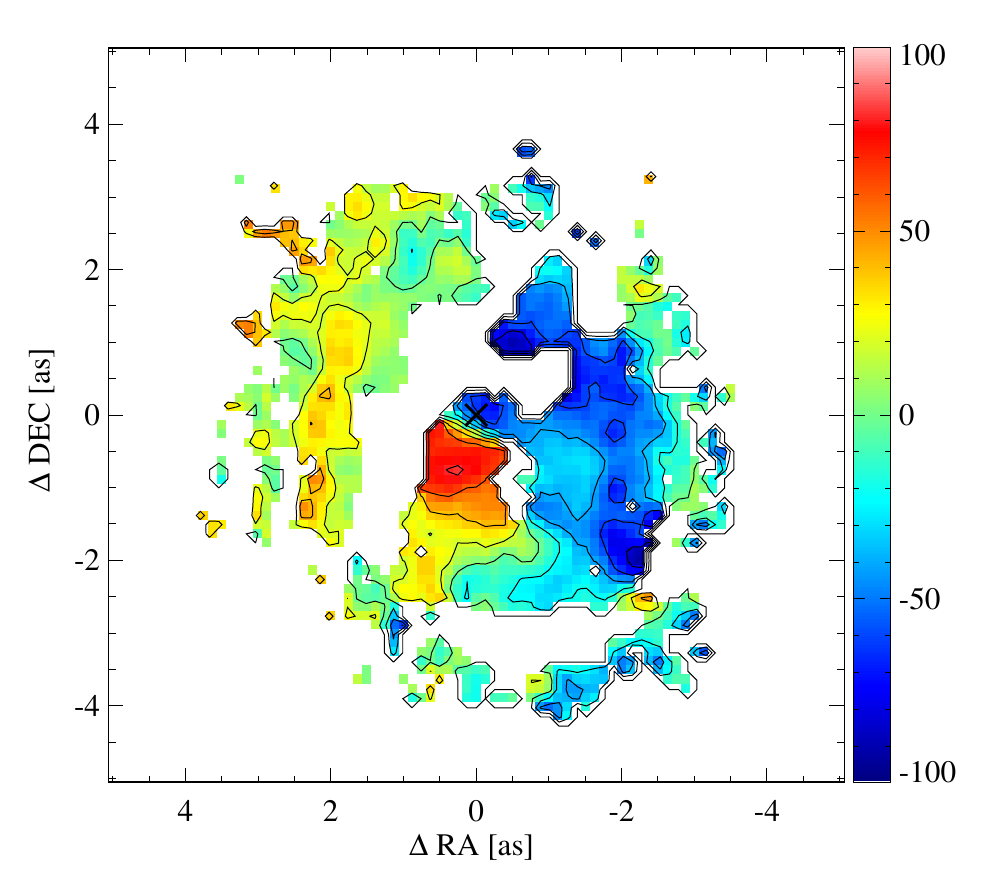}\label{fig:losvco32-co20p20}}
\caption{Stellar LOSV subtracted maps of H$_2$(1-0)S(1) and $^{12}$CO(3-2). \subref{fig:losvco32-co20p20} is the same as \subref{fig:losvco32-co20} but 20 km s$^{-1}$ were added to the residual to indicate the higher resemblance with Fig. \subref{fig:losvh2-co20}. All color-bar units are in km s$^{-1}$. For more details see Sect. \ref{sec:centrarcs}.}
\label{fig:losvlines}
\end{figure*}

\begin{table*}[htbp]
\centering
\caption{Emission lines}
\begin{tabular}{c c c c c c c}
\hline\hline
 & \multicolumn{2}{c}{center} & \multicolumn{2}{c}{sp(A)} & \multicolumn{2}{c}{sp(B)} \\
\hline
 Line & Flux & FWHM & Flux & FWHM & Flux & FWHM \\
 $\lambda$ [$\mu$m] & [$10^{-18}$ W m$^{-2}$] & [km s$^{-1}$] & [$10^{-18}$ W m$^{-2}$] & [km s$^{-1}$] & [$10^{-18}$ W m$^{-2}$] & [km s$^{-1}$] \\
\hline
 $[$\ion{Fe}{ii}$]$ & $1.94\pm0.47$ & $190\pm50$ & $...$ & $...$ & $...$ & $...$ \\
 1.644 \\
 H$_2$(1-0)S(3) & $1.33\pm0.13$ & $210\pm20$ & $0.33\pm0.02$ & $180\pm15$ & $0.29\pm0.02$ & $230\pm15$ \\
 1.957 \\
 H$_2$(1-0)S(1) & $0.86\pm0.11$ & $180\pm25$ & $0.28\pm0.02$ & $125\pm10$ & $0.28\pm0.03$ & $255\pm25$ \\
 2.122 \\
 Br$\gamma$ & $0.295\pm0.095$ & $185\pm70$ & $...$ & $...$ & $...$ & $...$ \\
 2.166 \\
 H$_2$(1-0)S(0) & $0.36\pm0.09$ & $230\pm75$ & $0.065\pm0.02$ & $190\pm70$ & $0.05\pm0.02$ & $195\pm115$ \\
 2.223 \\
 H$_2$(2-1)S(1) & $0.13\pm0.06$ & $110\pm55$ & $0.04\pm0.015$ & $140\pm60$ & $0.04\pm0.02$ & $235\pm50$ \\
 2.247 \\
 H$_2$(1-0)Q(1) & $1.08\pm0.20$ & $185\pm30$ & $0.28\pm0.03$ & $155\pm15$ & $0.21\pm0.04$ & $250\pm40$ \\
 2.408 \\
 H$_2$(1-0)Q(2) & $0.79\pm0.265$ & $290\pm75$ & $...$ & $...$ & $0.06\pm0.02$ & $115\pm35$ \\
 2.414 \\
 H$_2$(1-0)Q(3) & $1.02\pm0.20$ & $205\pm30$ & $0.23\pm0.015$ & $135\pm15$ & $0.18\pm0.02$ & $235\pm35$ \\
 2.422 \\
\hline

\end{tabular}
\caption*{For center: The lines were measured in a $1\arcsec$ radius aperture centered on the adopted galactic center and nuclear position. For sp(A) and sp(B): The lines were measured in a $0\farcs56$ radius aperture centered on the emission region. See also Fig. \ref{fig:h212flux}. The FWHM is not corrected for instrumental broadening, which is $\sim100$ km s$^{-1}$ in H-band and $\sim75$ km s$^{-1}$ in K-band.
}
\label{tab:region}
\end{table*}

\section{Discussion}
\label{sec:discussion}

\subsection{The nucleus of NGC 1433}
\label{sec:nucleus}
One result from the continuum decomposition (see Sect. \ref{sec:continuum}) is that the NIR stellar peak at the center does not fall onto the 0.87 mm continuum peak measured with ALMA (see Fig. \ref{fig:ALMAcont}). To address this issue the H$_2$(1-0)S(1) contours were overlayed on the HST F450W map, the ALMA $^{12}$CO(3-2) map and the 0.87 mm continuum map (see Figs. \ref{fig:ALMACO32}, \ref{fig:ALMAcont}, \ref{fig:HSTH2}). Fig. \ref{fig:HSTH2} shows that the dust lanes are well covered by the H$_2$(1-0)S(1) emission when the stellar NIR peak (marked with an $\times$) is placed onto the brightest pixel in the HST image. Furthermore, the H$_2$(1-0)S(1) emission contours were overlayed on the ALMA $^{12}$CO(3-2) map (see Fig. \ref{fig:ALMACO32}). The molecular CO emission fits best to the molecular H$_2$ emission when the stellar continuum center is placed on a small local CO peak about 1\farcs5 north and 0\farcs2 west from the assumed center in \citet{combes_ALMA_2013}. Finally, the H$_2$(1-0)S(1) contours were overlayed on the 0.87 mm continuum map by aligning it to the same spatial pixel that was found from the overlay onto the $^{12}$CO(3-2) map (see Fig. \ref{fig:ALMAcont}). Interestingly, the NIR stellar continuum peak is situated on a local 0.87 mm peak that is about three quarters as strong as the brightest 0.87 mm peak. The eastern side of the central arclike structure in the H$_2$(1-0)S(1) line map as described in section \ref{sec:molgas} traces a 0.87 mm emission towards the peak of the 0.87 mm emission. Furthermore, the H-K color diagram in Fig. \ref{fig:hkcont} shows that towards our adopted center the color becomes significantly redder, resulting from the warm dust emission of the SMBH surrounding torus.
Unfortunately, the bright spots in the $^{12}$CO(3-2) and the H$_2$(1-0)S(1) emission maps, which could be used to align the NIR and the sub-millimeter maps, do not coincide exactly with each other. 

As it is shown in Fig. \ref{fig:ALMACO32}, the brightest $^{12}$CO(3-2) region, at the north--east, is very well traced by the H$_2$(1-0)S(1), as well as the bright clumps in the west. Using these emission regions a proper alignment can be obtained as it is appreciable also in the correspondence of the extended flux of both emitting species. The error estimate of the center positioning, using the described method, turns out to be 0$\farcs$2. The estimated error becomes evident when studying the 0th and 1st moment maps of $^{12}$CO(3-2), overlayed with the corresponding H$_2$ maps. The error represents a shift of the ALMA data with respect to our NIR/optical center. By shifting the CO maps with respect to the H$_2$ emission line map a cross-correlation of the strongest $^{12}$CO(3-2) emission spots with probable corresponding emission in the H$_2$ map is performed.
Additional support for the newly estimated nuclear position is the strong velocity gradient across this region in all emission lines, e.g. $^{12}$CO(3-2), H$_2$(1-0)S(1), Br$\gamma$ and [\ion{Fe}{ii}]. Also, [\ion{Fe}{ii}] and Br$\gamma$ only show detectable emission at the position of the stellar continuum peak co-spatial with emission from the circum-nuclear hot dust. The exact position of the adopted new center is given in table \ref{tab:basic}. This is convincing combined evidence for the presence
of an AGN and a SMBH at the newly found position.

\subsection{The central arcsecond}
\label{sec:centrarcs}

As already mentioned in section \ref{sec:molgas} the high velocities in the central region were already discussed for the $^{12}$CO(3-2) line observed with ALMA. \citet{combes_ALMA_2013} argued mainly about an outflow scenario and its origin. Although they placed the center of NGC 1433 and therefore the AGN more than 1$\arcsec$ further south their argumentation is still solid. But, due to the change in the position of the center a rotating gaseous disk surrounding the nucleus and the center of the stellar distribution of NGC 1433 needs to be considered as an alternative solution.

\begin{figure}[htbp]
\centering
\subfigure[Along central velocity gradient]{\includegraphics[width=0.40\textwidth]{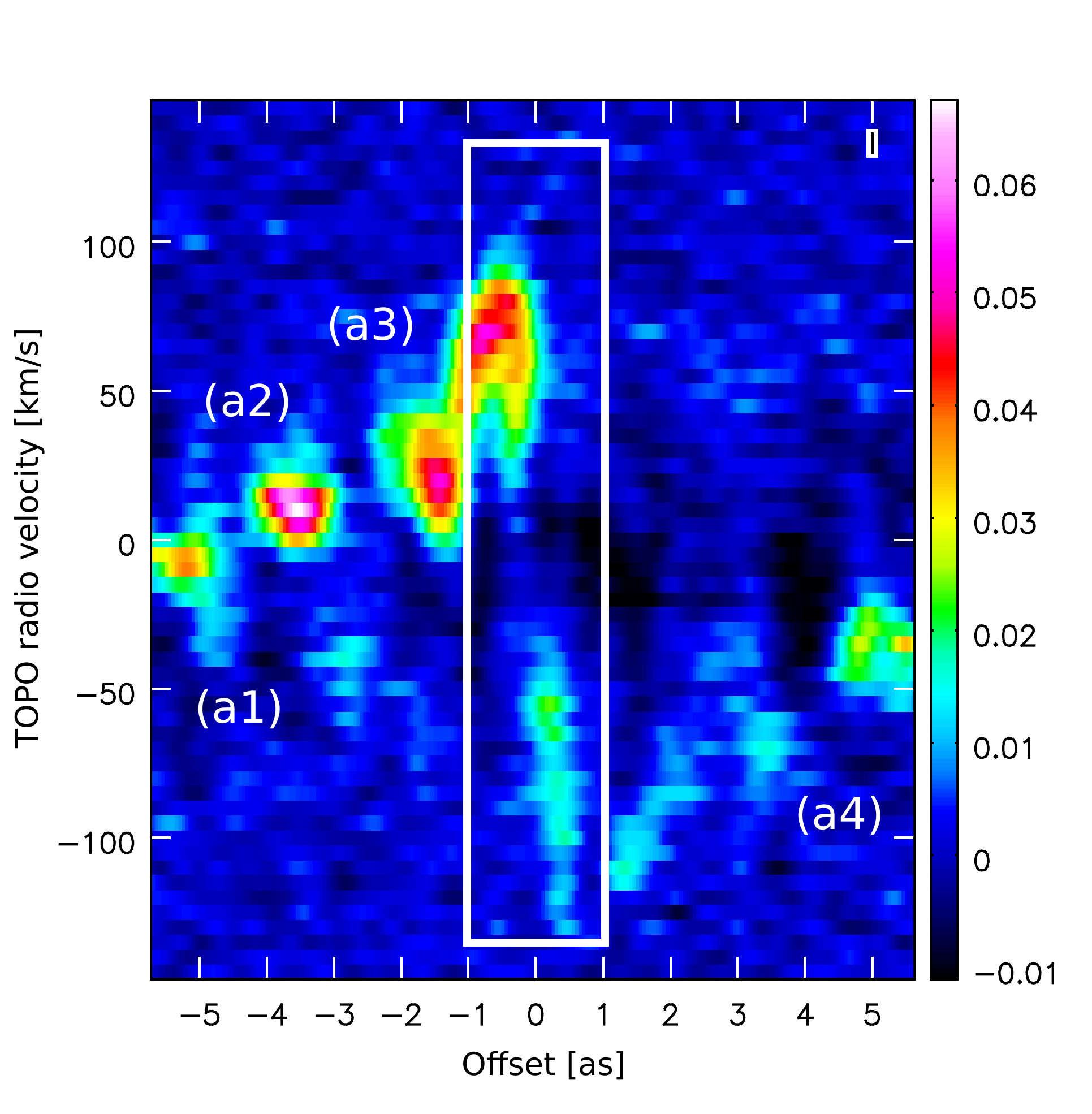}\label{fig:pvag}}
\subfigure[Perpendicular to central velocity gradient]{\includegraphics[width=0.40\textwidth]{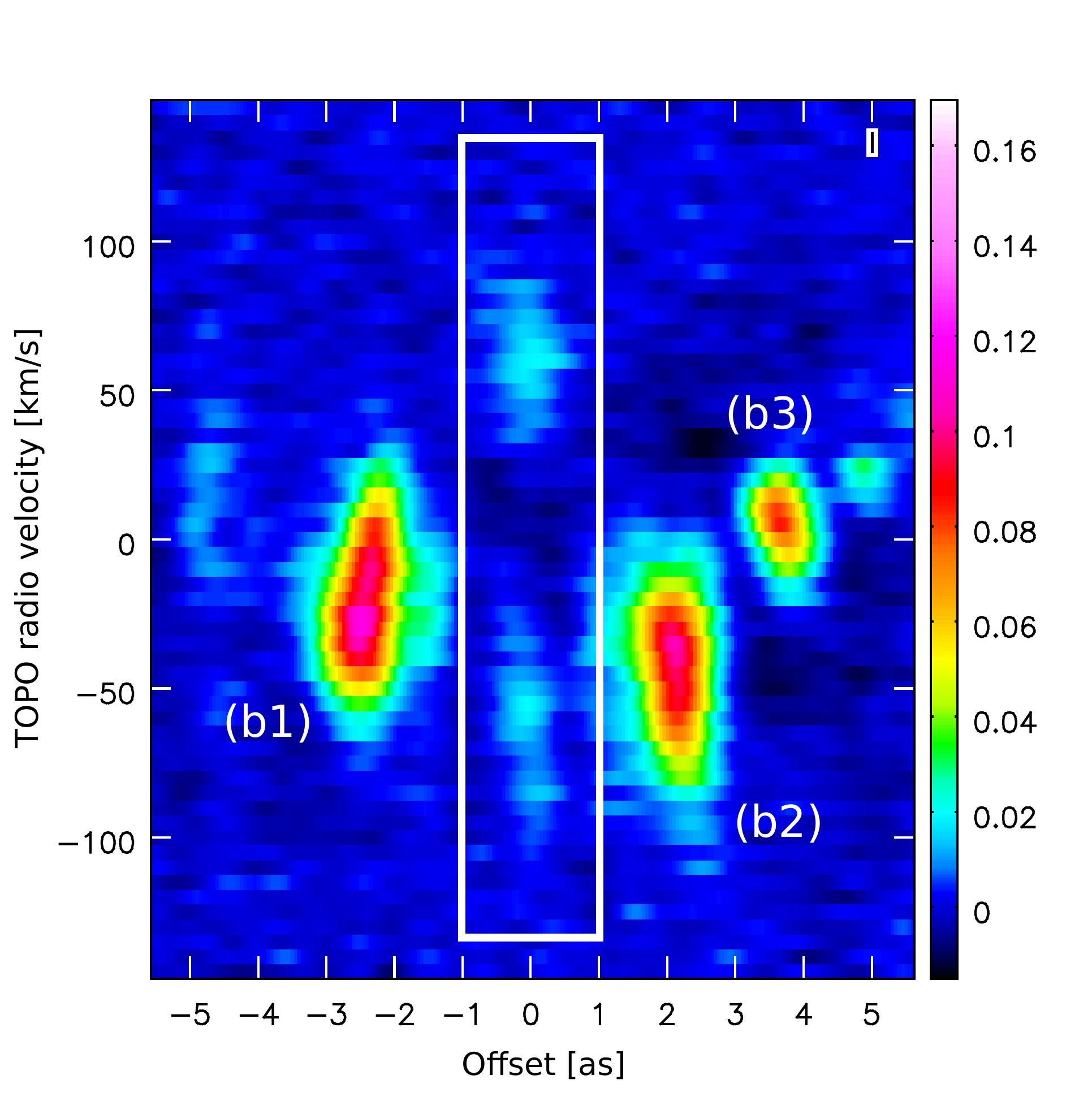}\label{fig:pvpg}}
\caption{PV diagrams from the $^{12}$CO(3-2) ALMA cube \subref{fig:pvag} along the central velocity gradient (PA$=140\degr$) and \subref{fig:pvpg} perpendicular to the central velocity gradient (PA$=50\degr$). The color-bar shows the flux density in [Jy beam$^{-1}$]. The position of the cuts is displayed in Fig. \ref{fig:ALMAmom1}. The white rectangle marks the nuclear region in which the central velocity gradient is dominent. (a1) to (b3) mark dust arm positions: (a1) eastern arm, (a2) southern arm, (a3) nuclear arm, (a4) northern arm, (b1) overlap between eastern and southern arm, (b2) nuclear arm, (b3) southern arm (west).}
\label{fig:pvcut}
\end{figure}

{\it A central nuclear disk?:} In H$_2$ a velocity gradient of -~40~km~s$^{-1}$ to 40~km~s$^{-1}$ is measured at a 1$\arcsec$ diameter distance over the nucleus (see Fig. \ref{fig:losvh212}). This gradient is higher in $^{12}$CO(3-2) reaching velocities of -~80~km~s$^{-1}$ to 120~km~s$^{-1}$ (see Figs. \ref{fig:ALMAmom1},\ref{fig:pvag}).
Using the H$_2$(1-0)S(1) velocities we determine a central mass to be $M_{cent}$ of 7.3 $\times$ 10$^{6} M_{\odot}$. This mass is in-between the two measured black hole masses determined using the M--$\sigma$ relation (7 $\times$ 10$^6$ and 1.7 $\times$ 10$^7 M_{\odot}$, see also Sect. \ref{sec:continuum}).
The velocity gradient across the center is at a PA of 140$\degr$ and is aligned with the far--side of the galaxy whereas the stellar velocity gradient is at a PA of 201$\degr$. This nuclear velocity gradient is seen in all detected NIR emission lines except for [\ion{Fe}{ii}], which can also resemble a north--south aligned gradient. 

There is a discrepancy in the line of nodes angle between H$_2$ and $^{12}$CO(3-2). For $^{12}$CO(3-2) we measure a line of nodes PA of 155$\degr$, this difference is visible in the LOSV maps of the respective lines (see Figs. \ref{fig:losvh212},\ref{fig:ALMAmom1}). The effect is probably introduced by PSF smearing since we are looking at a barely resolved region. A possible inclination (face--on to edge--on) of this gaseous nuclear disk cannot be determined from this data and since the PA differs strongly from the stellar PA a rotation in the galactic disk plane does not distinguish itself.

The position velocity (PV) diagram in Fig. \ref{fig:pvag} shows a possible central rotating disk profile that reaches the maximum velocities of $\pm100$~km~s$^{-1}$ at $\pm0\farcs4$ from the center. Here the PA of the PV diagram was chosen based on the velocity field in the central region (see Fig. \ref{fig:ALMAmom1}). Assuming an edge on disk a lower limit for the dynamical mass is estimated to be $4\times10^7 M_{\odot}$ with the measured velocity and radius. This mass is higher than our estimated BH mass values. However, we derived here the dynamical mass within an aperture of $0\farcs4$ radius which is supposed to be higher than the BH mass since it accounts for the stellar mass included in this aperture. Assuming that the inclination is $i=33^o$ this implies a higher mass by a factor of 3 \citep[see][]{combes_ALMA_2013}. But inclination and PA of a circum nuclear disk do not at all have to agree with the corresponding values of the larger scale galaxy structure.
A large stellar mass contribution for the inner 20~pc can be expected for NGC~1433.
For example at the center of the Milky Way the mass density in
the inner parsec reaches values of the order of 10$^5$~\msol~pc$^{-3}$ \citep[see Fig. 22 by][]{schodel_nuclear_2009,schodel_star_2002}.
This results in an enclosed stellar mass of 4$\times$10$^7$~\msol~pc$^{-3}$
at a radius of 10~pc and about
10$^8$~\msol~pc$^{-3}$ at a radius of 20~pc. This implies that also in the case of NGC1433
the stellar mass contribution will be substantial.

The tips of the disk are located at the slightly increased dispersion spots as seen in the 2nd moment $^{12}$CO(3-2) image (Fig. \ref{fig:ALMAmom2}).
The lack of $^{12}$CO(3-2) emission at the very center can be explained by highly excited (possibly even optically thin) molecular gas at temperatures of $>$55~K that is suppressed in its emission through the $^{12}$CO(3-2) transition. 
Molecular rings filled in by line emission of higher rotational transitions are
a commonly observed phenomenon in galactic nuclear regions.
In the case of I~Zw~1 \citet{staguhn_bima_2001,eckart_molecular_2001,eckart_molecular_2000}
have shown that the circum-nuclear ring structure, which is present in interferometric $^{12}$CO(1-0)
maps, is not evident in $^{12}$CO(2-1) maps since the center of the ring has been filled up by
$^{12}$CO(2-1) line emission enhanced through higher molecular excitation towards the nucleus.
There is also ample evidence of elevated $^{12}$CO(3-2)/(1-0) ratios towards the central positions
of galactic nuclei, as demonstrated by \citet{irwin_jcmt_2011}.
Their Fig.7 shows that at densities of 10$^{4}$~cm$^{-3}$ a variation of that ratio
by a factor of up to 4 can be obtained if the molecular gas temperature varies between
typical nuclear \citep{muraoka_aste_2007,tilanus_co_1991,israel_ci_2006}
and off-nuclear values \citep{mauersberger_dense_1999,wilson_james_2009}
of T$\sim$50~K and T$\sim$10~K.
Such a behavior of the molecular gas structure around an AGN is also expected theoretically,
based on three-dimensional non-LTE calculations of CO lines \citep{wada_molecular_2005}. 

The H$_2$ mass derived in the nuclear region is an indicator that the gas close to the nucleus is indeed highly excited since we derive from the $^{12}$CO(3-2) transition a lower mass than from the H$_2$(1-0)S(1) transition.
To resolve this issue, however, we need sub-mm observations of the central region at higher frequencies and higher angular resolution.

{\it The central nuclear outflow:} 
The newly determined nuclear position agrees well with the kinematic origin of a potential molecular outflow. Interpreting the observed nuclear [\ion{Fe}{ii}] and Br$\gamma$ line emission as tracers of star formation activity (see Sec. \ref{sec:iongas}) make star formation an unattractive source for driving such an outflow. 
In fact, it is very likely that intermittent nuclear accretion events are the driving force \citep[see also][]{combes_ALMA_2013}.

The rotation profile might as well be produced by an outflow which is then smeared out by the PSF to look like a rotation. 
We estimate the disk radius to be about 0$\farcs$4, by measuring the distance of the absolute disk velocity maxima from the center, which is about the size of the PSF. 
The PV diagram perpendicular to the velocity gradient (see Fig. \ref{fig:pvpg}) shows conspicious faint emission lobes at the $0\arcsec$ position and at a velocity of about $\pm60$~km~s$^{-1}$. Almost no emission is detected in-between these velocities on the nucleus which is expected for a molecular disk, but is not necessary (see above). The other emission lobes in the PV diagram are the molecular arms.

{\it A combined model:} A third possibility is a combination of rotating disk and outflow. This scenario would explain the rotational character of the center and the red tail south of the nucleus visible in H$_2$(1-0)S(1) and $^{12}$CO(3-2). 
Assuming a disk with a single sided outflow away from the observer this explains also the higher redshifted velocity compared to the blueshifted velocity in the SINFONI observation.
The different PAs of the central velocity gradients as measured for H$_2$ and CO might then be introduced by PSF smearing due to the higher $^{12}$CO(3-2) velocity of the outflow compared to the H$_2$ velocities.
This approach confirms the increased redshifted velocity 1$\arcsec$ south of the nucleus seen in the H$_2$ and CO LOSV images (Fig. \ref{fig:losvh212}, \ref{fig:ALMAmom1}) that follow the 0.87 mm continuum tail.

Due to the newly determined position of the center of NGC1433 the outflow's origin is the SMBH, located at the connection point of blueshifted and redshifted gas in the center. With the new central position we can now also explain the off-nuclear dust continuum emission. Fig. \ref{fig:ALMAcont} shows the 0.87 mm emission overlayed with H$_2$(1-0)S(1) emission contours. Comparing these two emission maps shows that radio continuum emission is present at least on the center and with its emission peak about 1$\farcs$5 south of it. Following the H$_2$(1-0)S(1) emission from nucleus to radio continuum peak one traces also weak radio continuum emission. This emission is a three sigma detection and the overlap of radio continuum and H$_2$ emission is remarkable. This radio continuum trail puts the outflow(+disk) scenario at least in front of the rotational disk only scenario. The trail is then explained by grey body dust emission where the dust was excited by the outflow. Interestingly, the EW of H$_2$(1-0)S(1) is almost constant along this trail and around the nucleus except for the emission peak. One might expect an increase of the EW in the outflow regions with respect to the surroundings due to higher H$_2$ fluxes, but none can be detected in any of the emission lines. The dust lane 2$\arcsec$ south of the nucleus shows the same EW as the outflow tail region.

\subsubsection{A simple model approach}
\label{sec:model}
\begin{figure*}[htbp]
\centering
\subfigure[Disk model]{\includegraphics[width=0.33\textwidth]{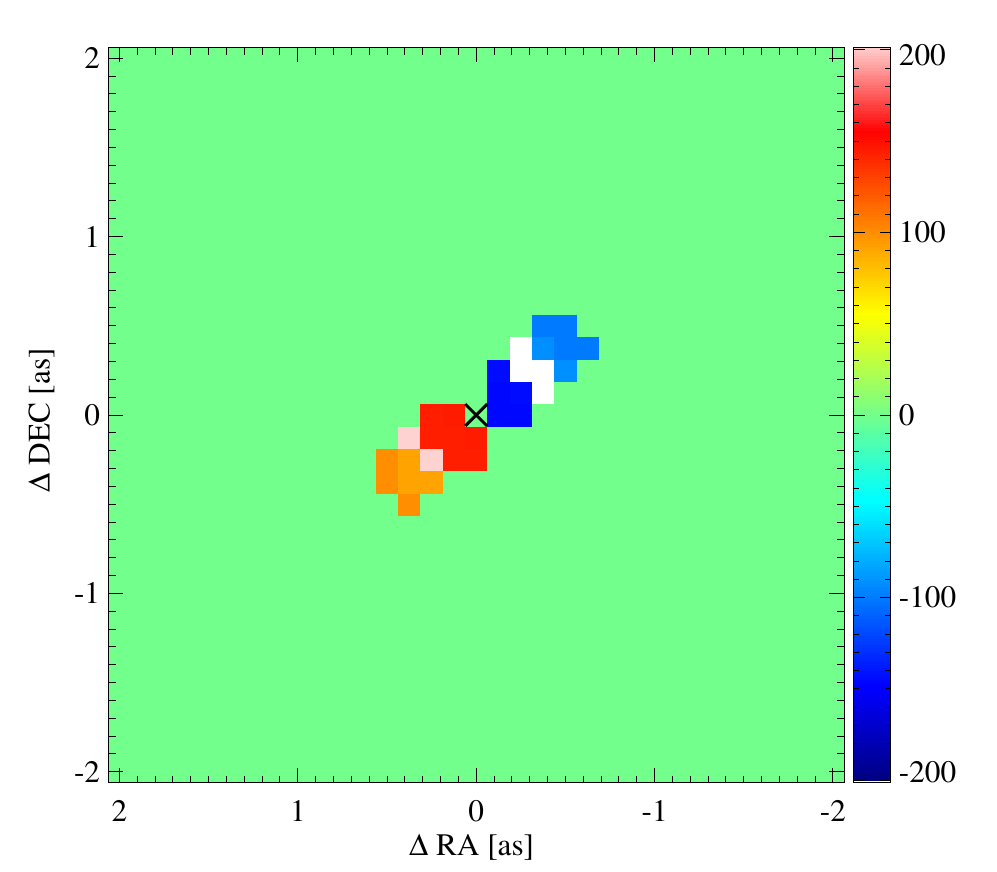}\label{fig:moddisk}}
\subfigure[Outflow model]{\includegraphics[width=0.33\textwidth]{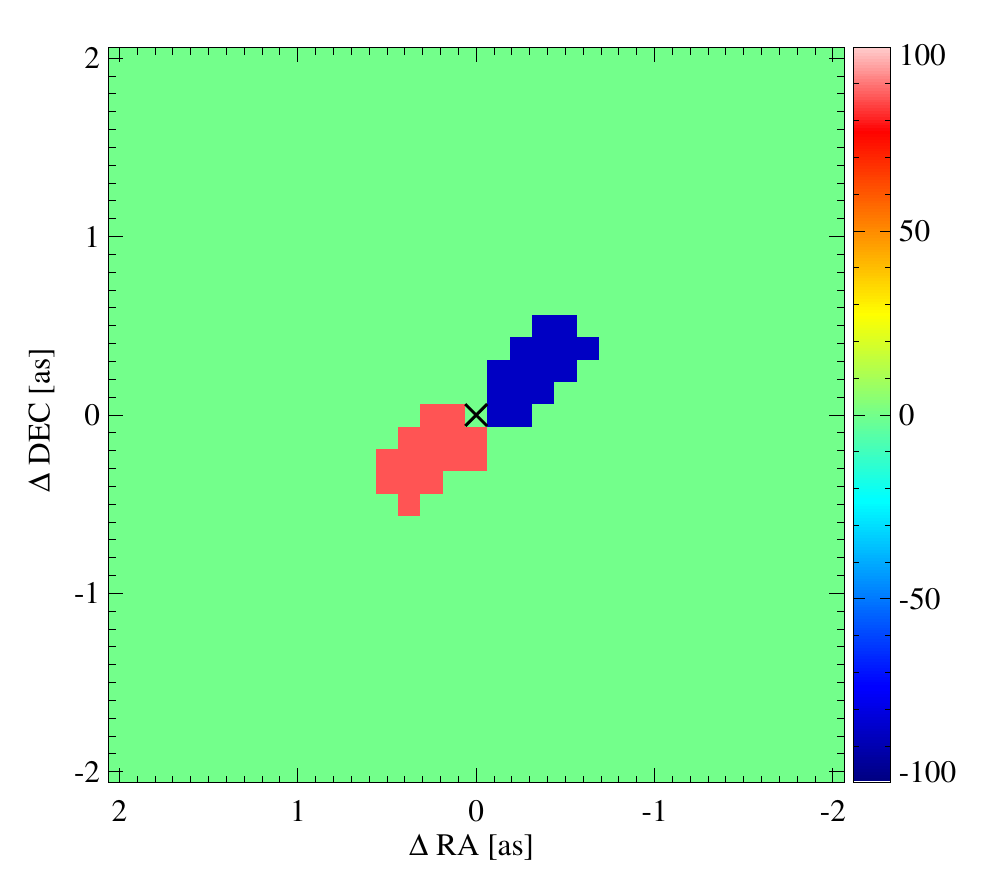}\label{fig:modoutfl}}
\subfigure[Combined model]{\includegraphics[width=0.33\textwidth]{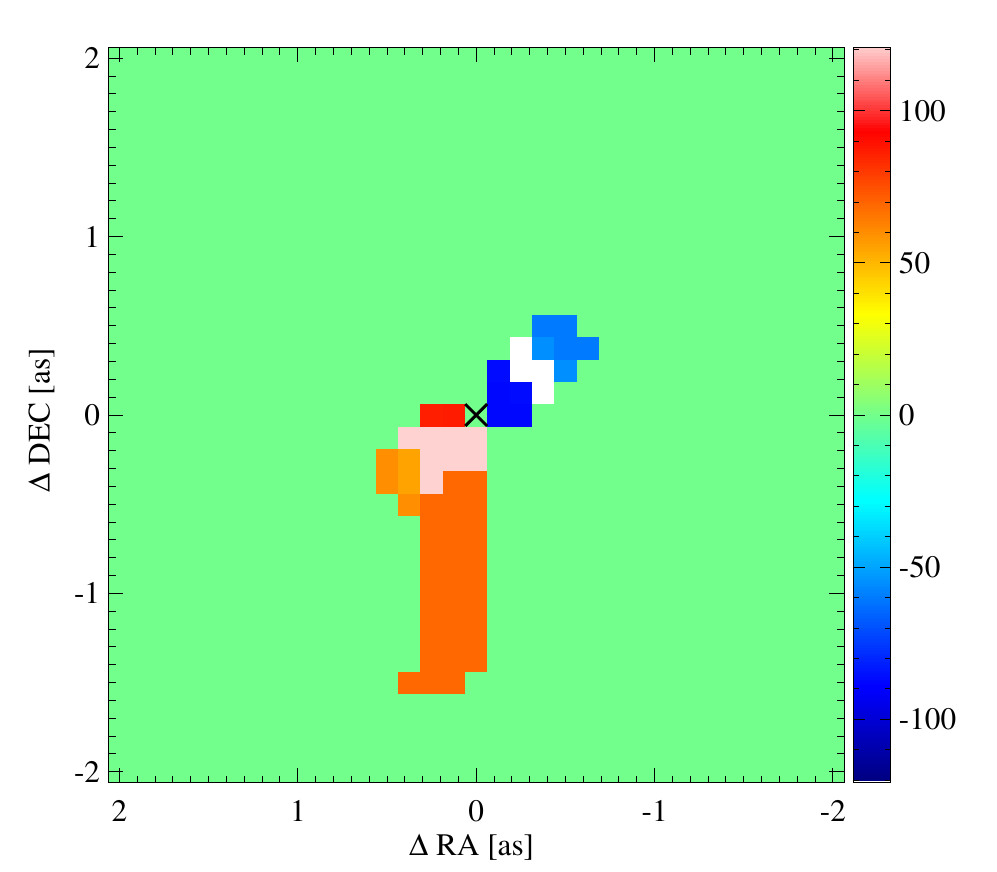}\label{fig:moddisko}}
\subfigure[Convolved disk model]{\includegraphics[width=0.33\textwidth]{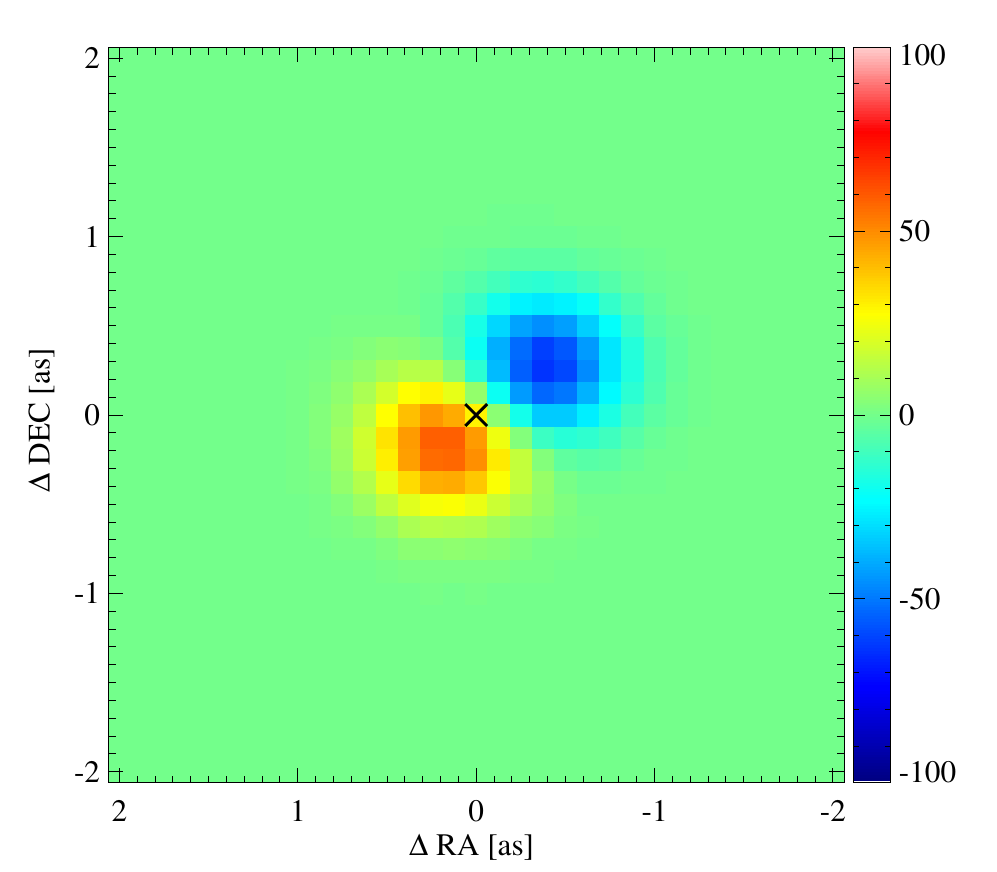}\label{fig:cmoddisk}}
\subfigure[Convolved outflow model]{\includegraphics[width=0.33\textwidth]{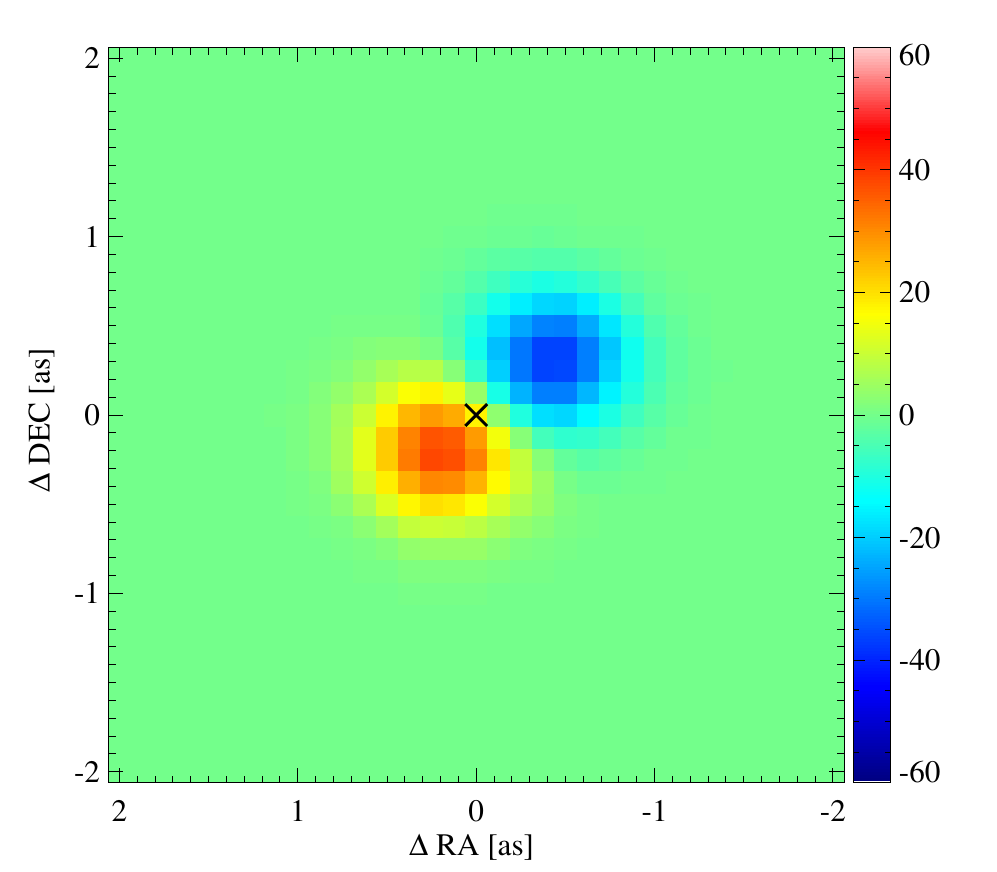}\label{fig:cmodoutfl}}
\subfigure[Convolved combined model]{\includegraphics[width=0.33\textwidth]{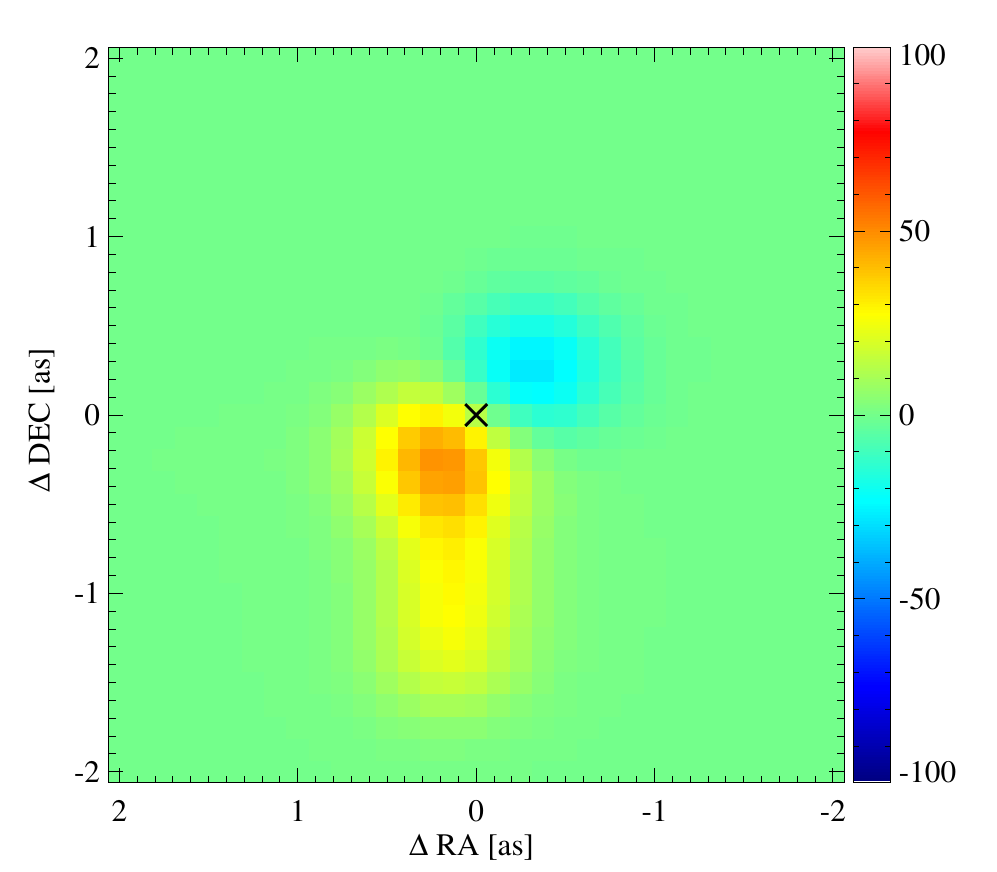}\label{fig:cmoddisko}}
\caption{The models used in section \ref{sec:model} are presented. \subref{fig:moddisk}-\subref{fig:moddisko} are the constructed models, \subref{fig:cmoddisk}-\subref{fig:cmoddisko} are the PSF convolved models. For more details see Sect. \ref{sec:model}.}
\label{fig:models}
\end{figure*}

To substantiate the discussion three simple models for the central velocity gradients were derived. A disk model based on a differentially rotating edge-on disk with a rotational velocity of $v\sim100$~km~s$^{-1}$ at the outer edge was constructed (see Fig. \ref{fig:moddisk}). An inclination parameter was not taken into account since this will mainly change the value of the LOSV of the disk. For the outflow model a constant velocity of $v\sim100$~km~s$^{-1}$ was assumed up to a radius of $0\farcs625$ (see Fig. \ref{fig:modoutfl}). A fixed inclination of $30\degr$ was chosen for the outflow. The orientation for both models is at a PA of $130\degr$. The scale height (e.g. disk thickness and width of the outflow) was chosen to be 3 spatial pixels. A change in scale height results directly in a different observed LOSV, therefore, this parameter was fixed as well. These two models were then convolved with a Gaussian PSF, using the FWHM measurements of the telluric standard star in K-band. These models were placed on an emissionless background.

As a result the convolved models of disk and outflow look very similar and cannot unambiguously be traced back to their underlying model. The models fit well in the central $1\arcsec$ diameter region.
The redshifted part, however, shows an elongated tail southwards which cannot be accounted for with these two isolated models.

Therefore, as a third model we chose a combination of the disk model as described above and a one-sided outflow model. The outflow is now at a PA of $\sim175\degr$ and only the redshifted part is taken into account. The radius of the disk is the same as above. The radius for the outflow is $1\farcs5$. The velocities chosen for this model are 60~\kms\ for the disk and 80~\kms\ for the outflow. The convolution was done with the same PSF as mentioned above. The combined model is shown in Fig. \ref{fig:moddisko}. To derive similar velocities as in our H$_2$(1-0)S(1) LOSV map we changed the velocities of disk and outflow and the PA of the outflow, only. All other parameters have the same values as described above. A one-sided outflow can be justified if an otherwise double-sided nuclear wind gets into contact with a molecular cloud at only one side.

\begin{figure}[htbp]
\centering
\subfigure[H$_2$(1-0)S(1) detail]{\includegraphics[width=0.33\textwidth]{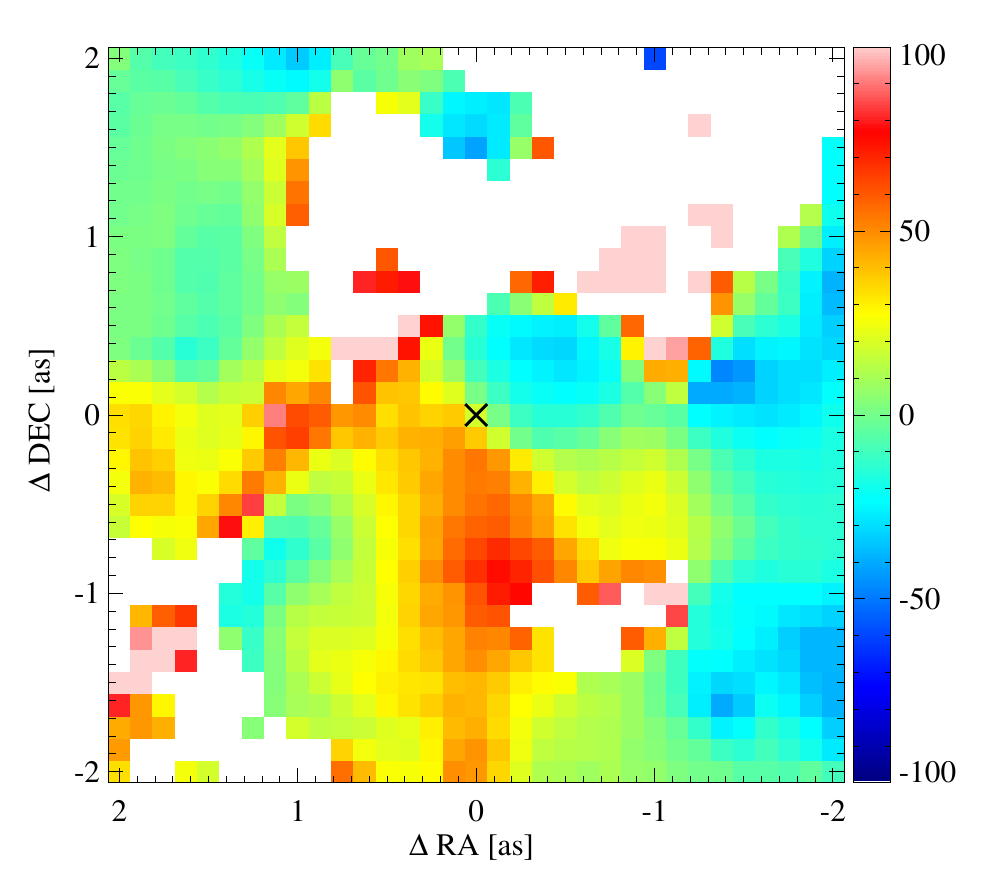}\label{fig:losvh2s}}
\subfigure[H$_2$(1-0)S(1) $-$ model]{\includegraphics[width=0.33\textwidth]{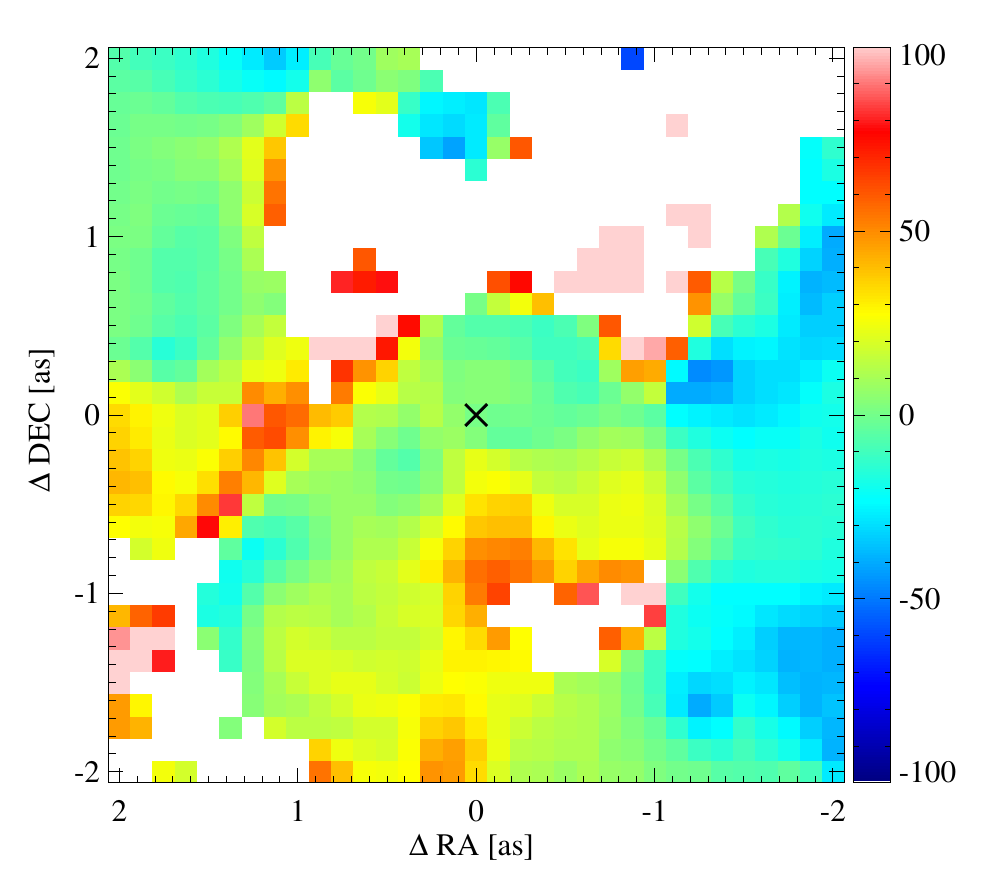}\label{fig:h2-model}}
\caption{\subref{fig:losvh2s} shows the central $4\arcsec\times4\arcsec$. \subref{fig:h2-model} shows the same detail but subtracted by the combined model (see Fig. \ref{fig:cmoddisko}). For more details see Sect. \ref{sec:model}.}
\label{fig:modres}
\end{figure}

This third approach can describe the central gradient and the southwards going redshifted tail. Fig. \ref{fig:modres} shows a $4\arcsec\times4\arcsec$ detail of the H$_2$(1-0)S(1) velocity field centered on the nucleus and the model subtracted residual. The residual is at $0\pm10$~\kms\ in the vicinity of the central gradient. The redshifted tail shows higher redshifted residuals at $\sim1\arcsec$ distance from the center, which can mainly be attributed to the underlying galactic rotation.

\subsection{The dusty nuclear spiral arms}
\label{sec:spiralarms}

The primary bar of NGC 1433 has an efficient way to pull gas towards the central region where it accumulates in a nuclear ring of about 10$\arcsec$ radius (see Fig. \ref{fig:ngc1433}). This is where the nuclear star formation ring gathers its fuel from (note the various bright spots (probably stellar clusters) in Fig. \ref{fig:F450WALMA}). From this nuclear ring several dust lanes wind towards the center forming a pseudo-ring of about 4$\arcsec$ radius right outside the inner ILR (iILR) at 3$\farcs$6 \cite{buta_dynamics_2001}. The dust arms show differential LOSV patterns as seen in Fig. \ref{fig:losvh212} and \ref{fig:ALMAmom1}.

In the $10\arcsec\times10\arcsec$ FOV we identify at least three different arms. The northern arm comes from the north along the western side towards the center and is blueshifted. The southern arm stretches from the west along the south towards the emission peak in $^{12}$CO(3-2) and is rather redshifted. The eastern arm comes from the south out of the $10\arcsec\times10\arcsec$ FOV and goes also towards the $^{12}$CO(3-2) emission peak but is blueshifted (see Fig. \ref{fig:ALMAmom1}). The western part of the inner 2$\arcsec$ is quite blueshifted and seems connected to the region of the H$_2$ nuclear peak. This nuclear arm also extends southwards and turns around from the south to the north towards the nucleus and becomes more redshifted the closer it gets.

The arms are well detected in $^{12}$CO(3-2) and the LOSVs are similar to the H$_2$ LOSVs except in one region that belongs to the eastern arm. This H$_2$(1-0)S(1) emission, sp(B) (see Sect. \ref{sec:molgas}), is the most blueshifted region in the H$_2$ LOSV map. It reaches almost --100~km~s$^{-1}$ at a distance of 3$\farcs$5 to the nucleus. The region north of it, sp(A) (see Sect. \ref{sec:molgas}), the strongest $^{12}$CO(3-2) emission does not seem to be as blue as the eastern arm and not as red as the southern arm but rather a mix of both. This region is where eastern and southern arm meet or overlap, which would explain the strong $^{12}$CO(3-2) and the H$_2$ emission there. Note that although we see a maximum in $^{12}$CO(3-2) emission the dust lane here becomes fainter as seen in the HST images. The southern spot sp(B) is stronger in H$_2$ than sp(A) and shows an FWHM of up to 250~km~s$^{-1}$. Either we see already an interaction between the two arms or we see a massive star formation region hidden behind and pushing at all the dust and gas. At least this would explain the strong blueshift and the strong H$_2$ emission due to UV excitation. The linewidth of the H$_2$ line in sp(B) speaks in favor of a collision of the two arms. Also no Br$\gamma$ emission can be detected in sp(B) which would imply that the column density of the dust and molecular gas (e.g, H$_2$) between observer and star forming region has to be very high so that the \ion{H}{ii} region cannot be detected. In addition no CO emission peak coincides with the position of sp(B). The velocity dispersion in the arms at 2$\arcsec$ distance from the nucleus is up to 50~km~s$^{-1}$ and higher (see Fig. \ref{fig:pvcut}) which may indicate an arm--arm interaction or hidden star formation activity.

The measurements of the stellar LOSV with an PA of 21$\degr$ are consistent with measurements by \citet{buta_dynamics_2001}. The stellar continuum map shows an angle of 33$\degr$ for the nuclear bar which is confirmed by \citet{combes_ALMA_2013}. There does not seem to be a greater disturbence of the stellar distribution. A minor twist from 33$\degr$ in the outer isophotes to 15$\degr$ and then to 69$\degr$ towards the central isophotes is detected (see Fig. \ref{fig:cont}). The central isophotes are dominated by the PSF structure which is also alongated along a PA of 69$\degr$. 
An ellipses fit to the continuum in optical and NIR confirms our PA measurements.

\subsection{Computation of the torques}
\label{torq}

A nuclear bar is well detected in near-infrared images of NGC 1433 \citep[see for example][]{jungwiert_near-ir_1997}. It lies inside the nuclear ring, and extends out to about 400 pc in radius, with a position angle of PA=30$^\circ$. Inside the ring, a patchy nuclear spiral structure was discovered in {\it HST} images by Peeples \& Martini (2006).
In the present HST data set the dust extinction is smallest in the I-band image that we select to trace the old stellar component, and derive the gravitational potential, with high spatial resolution. We have not separated the bulge from the disk contribution, considering that the bulge is highly flattened. \citet{buta_dynamics_2001} concludes that the inner part of the bulge is as highly flattened as the disk, from both photometry and kinematic arguments. The dark matter fraction is also negligible in the central kpc. The {\it HST}-F814W image has been rotated and deprojected according to PA=19$^\circ$ (equivalent to 199$^\circ$) and i=33$^\circ$, and then Fourier transformed to compute the gravitational potential and forces. A stellar exponential disk thickness of $\sim$1/12th of the radial scale-length of the galaxy (h$_{\rm r}$=3.9kpc) has been assumed, giving h$_{\rm z}$=328pc. This is the average scale ratio for galaxies of this type \citep[e.g.,][]{barteldrees_parameters_1994, bizyaev_photometric_2002, bizyaev_structural_2009}. The potential has been obtained assuming a constant mass-to-light ratio of M/L = 0.5 \msol/\lsol\ in the I-band over the considered portion of the image of 2~kpc in size. This value is realistic in view of what is found statistically for spiral galaxies \citep{bell_stellar_2001}. The pixel size of the map is 0.1\arcsec=4.8~pc. The stellar M/L value was fit to reproduce the observed CO rotation curve.

\begin{figure}[h!]
\centerline{
\includegraphics[angle=-90,width=7cm]{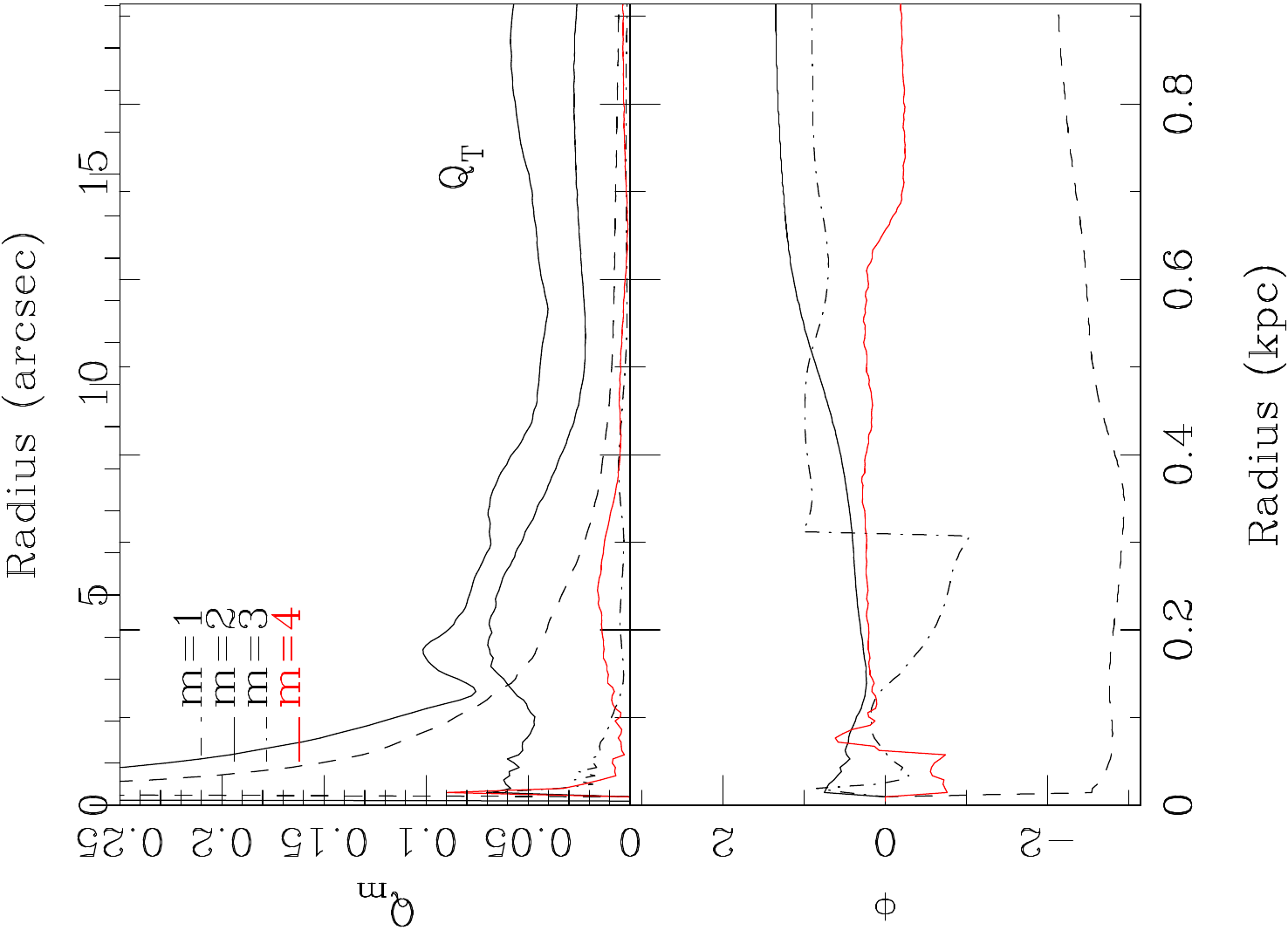}
}
\caption{{\it Top} Strengths (Q$_m$  and total Q$_T$) of  the $m=1$ to $m=4$ Fourier
  components of the stellar potential within the central kpc.  Inside a radius of 130~pc,
the $m=1$ term dominates, and the nuclear bar extends until 350~pc radius.
{\it Bottom} Corresponding phases in radians of the Fourier components, taken from the major axis, 
in the deprojected image.} 
\label{fig:pot1433}
\end{figure}

The potential has been decomposed into its different Fourier components, as in \citet{combes_ALMA_2014}. The radial distribution of their normalized amplitudes and their radial phase variations are displayed in Fig. \ref{fig:pot1433}.

\begin{figure}[h!]
\centerline{
\includegraphics[angle=0,width=8cm]{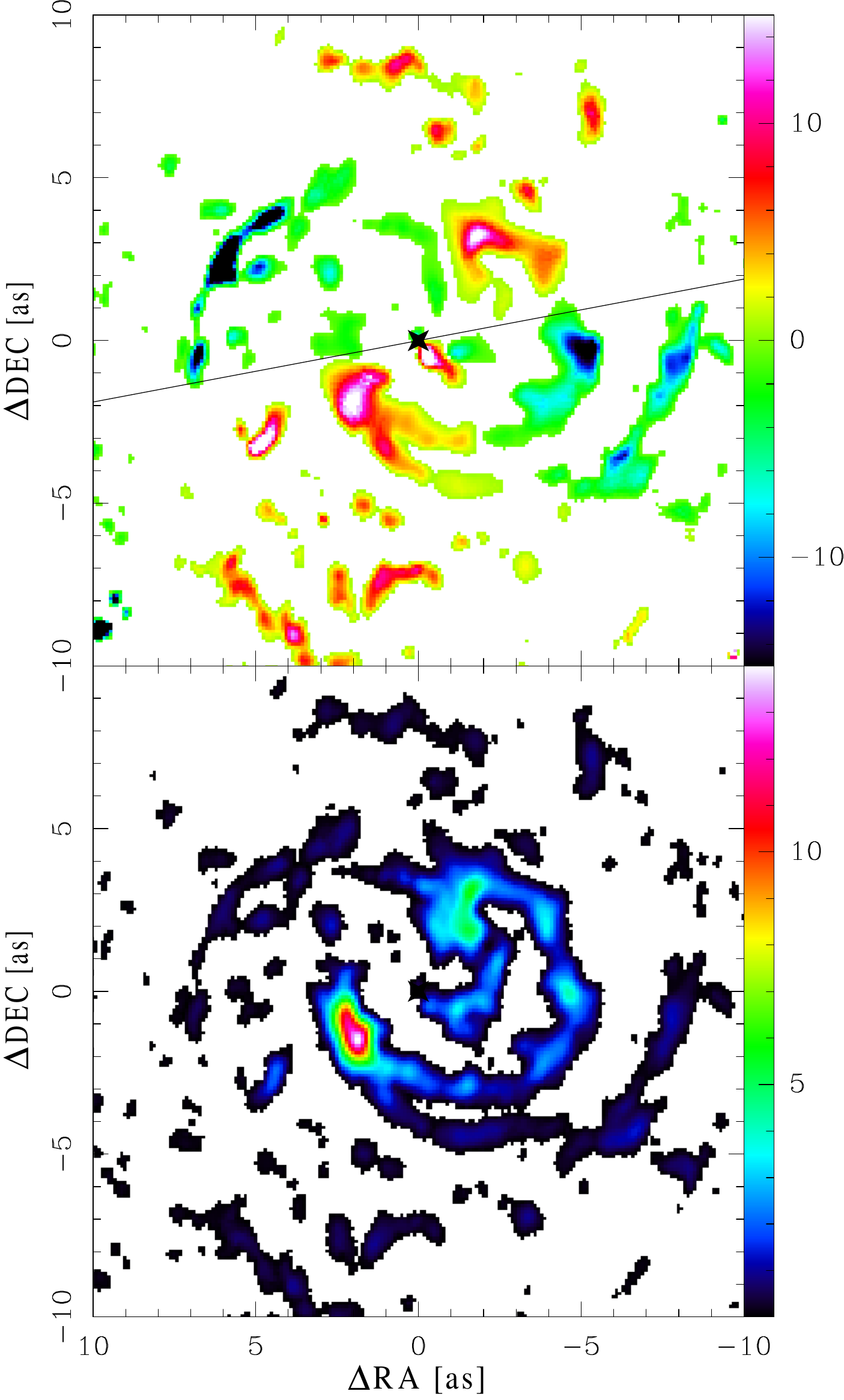}
}
\caption{{\it Top:} Map of the gravitational torque,  t(x,y)~$\times$~$\Sigma$(x,y), weighted by the 
gas surface density $\Sigma$, assumed proportional to the $^{12}$CO(3-2) emission.
The torque per unit mass is $ t(x,y) = x~F_y -y~F_x$ at each pixel (x,y), where
$F_x$ and $F_y$ are  the forces per unit mass, derived from the potential.
  The torques change sign as expected in a four-quadrant pattern
(or butterfly diagram). The orientation of the quadrants follows
  the nuclear bar's orientation. In this deprojected picture,
  the major axis of the galaxy is oriented parallel to the horizontal axis.
  The inclined line reproduces the mean orientation of the bar
  (PA = 101$^\circ$ on the deprojected image). 
{\it Bottom:} The deprojected image of the $^{12}$CO(3-2) emission, at the same scale,
and with the same orientation, for comparison. The axes are labelled in arcsecond relative to the center.
The color scales are linear, in arbitrary units.}
\label{fig:torq1}
\end{figure}

 From the potential, we derive the torques at each pixel, as described in  \citet{garcia-burillo_molecular_2005}.
The sign of the torque is determined relative to the sense of rotation in the plane of the galaxy.
  The product of the torque by the gas density $\Sigma$ is shown in Fig. \ref{fig:torq1}.

\begin{figure}[h!]
\centerline{
\includegraphics[angle=-90,width=7cm]{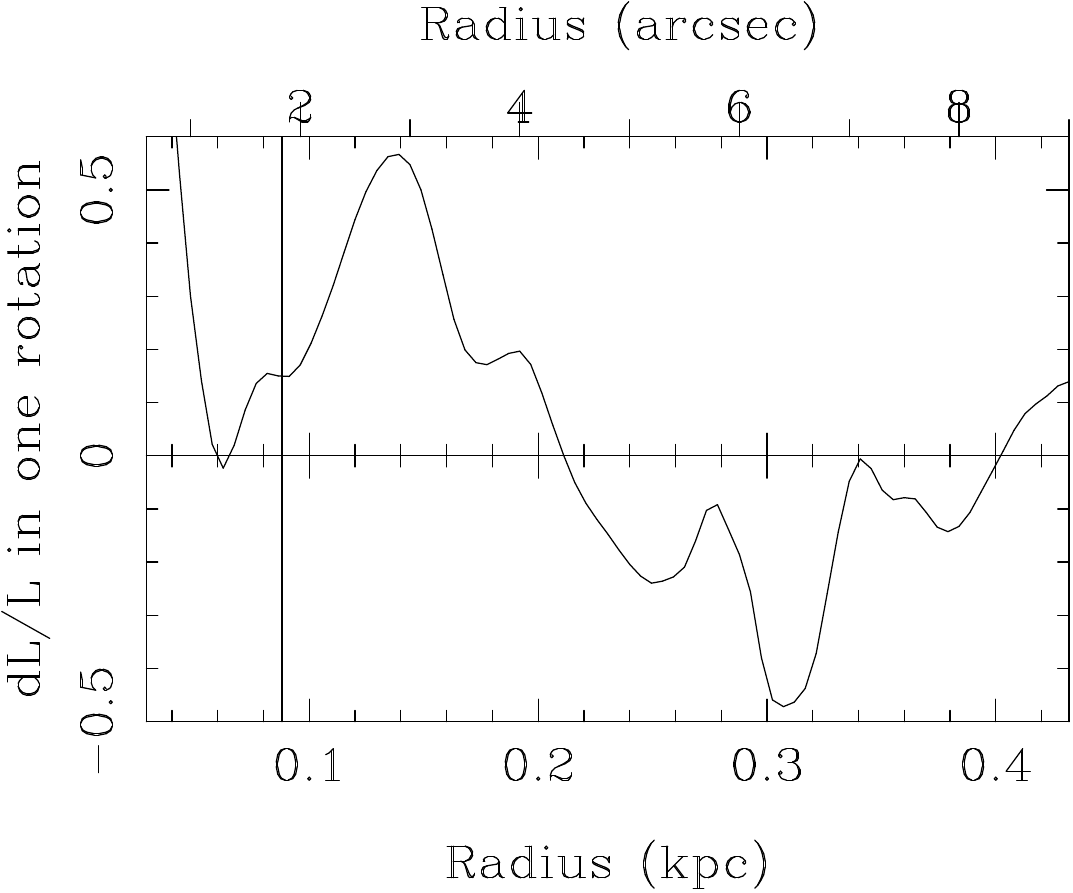}
}
\caption{
The radial distribution of the torque, quantified by the fraction of the angular momentum transferred
from the gas in one rotation--$dL/L$, estimated from the $^{12}$CO(3-2) deprojected map.
The vertical line at 88~pc radius delimitates the extent of the central gas outflow,
and the computation has no meaning here.
The torque is positive inside a 200~pc radius and then negative outside.
}
\label{fig:gastor}
\end{figure}

  The torque weighted by the gas density $\Sigma(x,y)$ is then averaged over azimuth, 
and allows us to estimate the time variations of the specific angular momentum $L$ of the gas.
The torque efficiency in driving gas flows is then obtained by
 normalizing at each radius by the angular momentum and rotation period,
as shown in   Fig.~\ref{fig:gastor}.

The definition of the exact center of the galaxy has an uncertainty
of $\pm$0.2 arcsec. To estimate the implied uncertainty on the torques, we
have computed them whith several centers, displaced by $\pm$0.2 arsec in RA and DEC.
This resulted in an uncertainty on torques of about 20\%.

Although the nuclear bar strength is diluted by the flattened bulge,  Fig.~\ref{fig:gastor} shows that its
efficiency is still high. Between 88 and 200~pc, the gas gains up to 50\% of its angular momentum in one rotation,
i.e. in $\sim$  40 Myr. Outside of this radius, the torque is negative.  The radius of 200~pc corresponds
to a pseudo-ring, which \citet{combes_ALMA_2013} interpreted as the inner ILR of the bar, i.e. the ILR of the nuclear bar.
The observed torques drive the gas towards this ring, and re-inforce it.
Inside 88~pc, the torque results have no meaning, since we observe
mostly outflowing gas, dragged by the putative AGN jets \citep{combes_ALMA_2013}. The computation cannot be interpreted
in terms of average torque here, since the gas is not in quasi-stationary orbits in rotation around
the center of the galaxy, aligned on the galaxy plane. Instead the gas is ejected at some angle from
the plane \citep{combes_ALMA_2013}, and the deprojections and torque computations do not apply here.

In summary, the gravity torques appear negative at a radius of R$>$200~pc and positive at R$<$200~pc. This leads to a accumulation of gas at the position of the R$=$200~pc ring, which is expected for the ILR of the nuclear bar. To drive
the gas further in, other mechanisms, such as viscous torques or dynamical friction,
have to be invoked to take over and fuel the nucleus.

\subsection{Line diagnostics}
\label{sec:diag}
The nuclear region and emission spots sp(A) and sp(B) are the only regions to show a variaty of line species that allow for a line diagnostic.\\
The diagnostic diagram of the logarithmic line ratios of H$_2$(1-0)S(1) over Br$\gamma$ and [\ion{Fe}{ii}] over Pa$\beta$ distinguishes three different galaxy types. Starburst galaxies, in the lower left, are photoionized and show very high \ion{H}{ii} emission. LINERs are mainly shock ionized, hence shock tracers like H$_2$(1-0)S(1) and [\ion{Fe}{ii}] $\lambda\lambda$ 1.257$\mu$m,1.644$\mu$m show stronger emission. Since both excitation mechanisms can be found in Seyfert galaxies this galaxy type is placed between the two extrema.

\begin{figure}[htbp]
\centering
\includegraphics[width=0.40\textwidth]{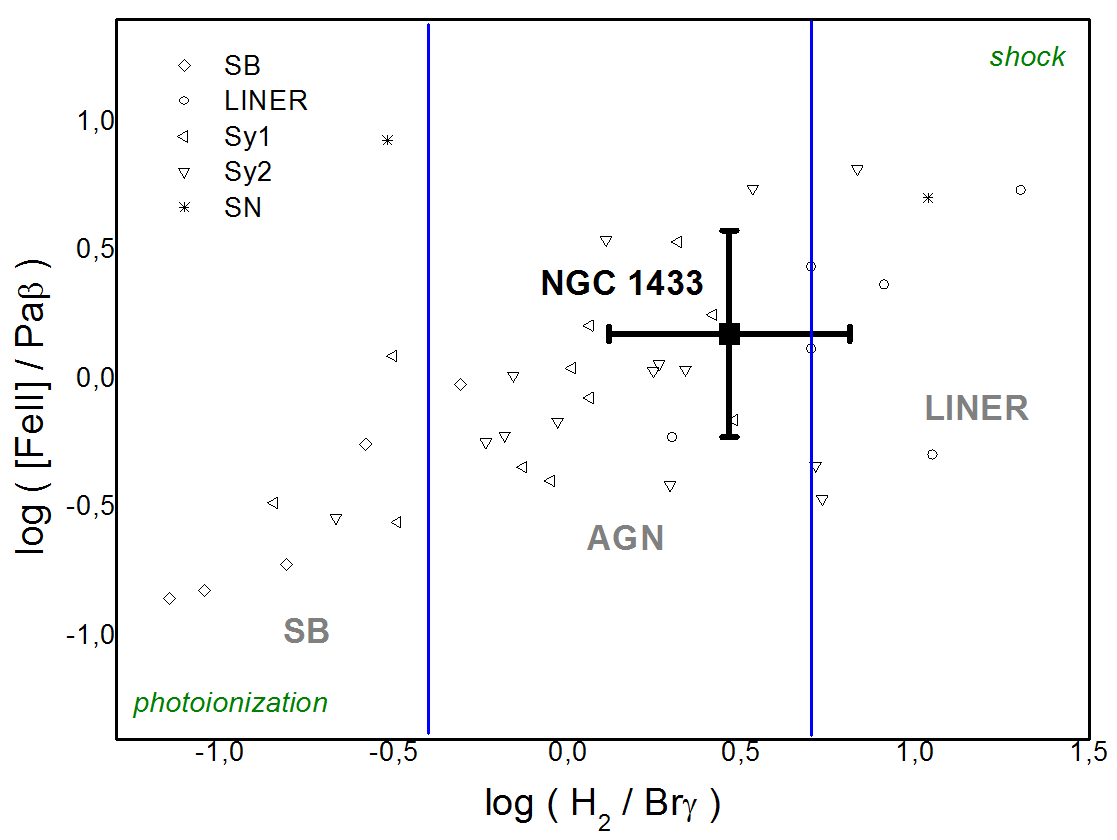}
\caption{Diagnostic diagram to classify the emission at the nucleus. The [\ion{Fe}{ii}] $\lambda$1.257 $\mu$m and Pa$\beta$ values were derived from the [\ion{Fe}{ii}] $\lambda$1.644 $\mu$m and the Br$\gamma$ line.}
\label{fig:ddiag}
\end{figure}

NGC 1433 takes its place in the AGN regime (see Fig. \ref{fig:ddiag}), but it is close to the shock ionized LINERs. Aperture effects might shift the classification into the LINER regime. This is mainly due to a deficiency of \ion{H}{ii} in the central region where we detect Br$\gamma$ only on the very center, the location of the nucleus. We confirm earlier classifications \citep[e.g.][]{veron-cetty_miscellaneous_1986,cid_fernandes_stellar_1998} that describe this galaxy as being Seyfert and LINER like. Due to the borderline classification of NGC 1433 as being a Seyfert/LINER galaxy the scenarios that include outflows are further supported.

The conversion between the [\ion{Fe}{ii}] $\lambda$1.644$\mu$m flux, obtained from our H-band observation and [\ion{Fe}{ii}] $\lambda$1.257$\mu$m was done using the factor $\lambda$1.644$\mu$m/$\lambda$1.257$\mu$m $=0.744$ \citep{nussbaumer_transition_1988}. For the conversion of Br$\gamma$ to Pa$\beta$ the case B ratio of Br$\gamma$/Pa$\beta=0.17$ was used, for a temperature of $T=10^4$ K and an electron density of $n_e=10^4$.\\

The detection of several molecular hydrogen emission lines allows for a line ratio diagnostic \citep{rodriguez-ardila_molecular_2005}. There are three main excitation mechanisms of molecular hydrogen in the NIR namely (i) UV flourescence (non-thermal); (ii) X-ray heating (thermal); (iii) shocks (thermal).

\begin{enumerate}[i]
\item {\it UV-flourescence}: High energetic UV photons from the Lyman--Werner band (912-1108 $\AA$) are absorbed by H$_2$ molecules and then re-emitted at lower frequencies. This can occur in a warm high-density gas where one finds thermal emission line ratios for the lower levels. Because of the high-density the lower levels are dominated by collisional excitation. Hence, observations of the higher level transitions are required to distinguish UV pumping (non-thermal) from collisional excitation (thermal).

\item {\it X-ray heating}: X-ray dominated regions (XRDs) in which H$_2$ is mainly ionized by X-rays are observable in regions with temperatures $<1000$ K. At higher temperatures shock excitation due to collisions populates the lower excitation levels. The ionization rate per H-atom is limited to $<10^{-15}$ cm$^{3}$ s$^{-1}$ since higher rates will destroy the H$_2$ molecules \citep[e.g.][]{tine_infrared_1997,draine_h2_1990,maloney_x-ray--irradiated_1996}. In dense, static photodissociation regions UV instead of X-ray photons can heat the molecular gas.

\item {\it Shock fronts}: The collisional thermal excitation via shock fronts in a medium populates the electronic ground levels of H$_2$ molecules. This population of the ro-vibrational transitions is described by a Boltzmann distribution. The kinetic temperatures in these shock excited regions can be higher than 2000 K \citep{draine_theory_1993}.
\end{enumerate}

\begin{figure}[htbp]
\centering
\includegraphics[height=0.5\textwidth,angle=90]{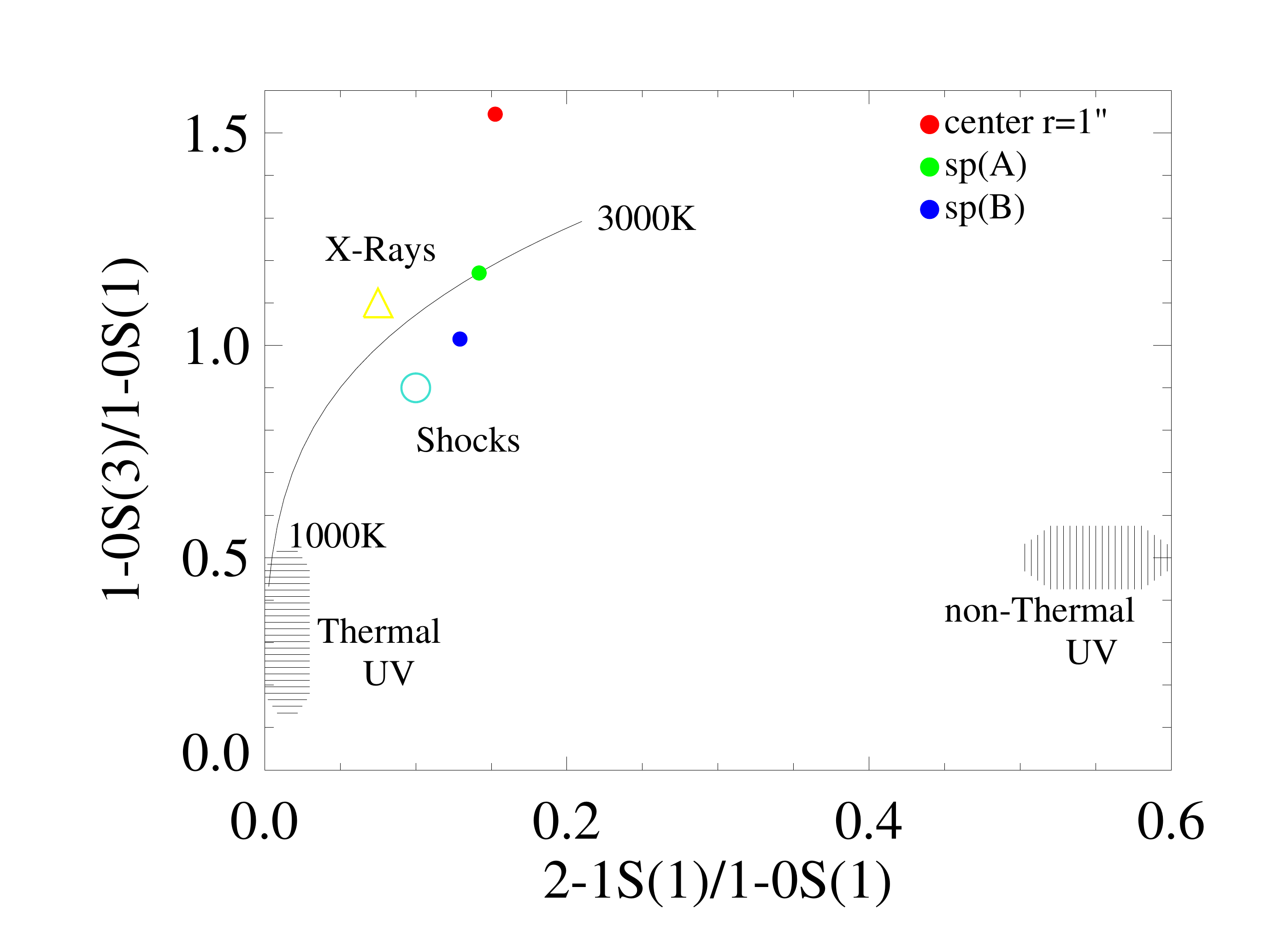}
\caption{Molecular hydrogen diagnostic diagram to classify the emission at the nucleus in red and in emission spots sp(A) in green and sp(B) in blue. Line ratios are H$_2$(2-1)S(1)/H$2$(1-0)S(1) and H$_2$(1-0)S(3)/H$_2$(1-0)S(1). The curve represents thermal emission at 1000--3000 K. Horizontal stripes are thermal UV excitation models by \citet{sternberg_infrared_1989}. Vertical stripes are non-thermal models by \citet{black_fluorescent_1987}. The area of X-ray heating models by \citet{draine_h2_1990} is marked by an open triangle (yellow). The open turquoise circle marks the region of the shock model by \citet{brand_constancy_1989}.}
\label{fig:h2diag2}
\end{figure}

Non-thermal excitation cannot be the excitation mechanism in any of our analyzed regions since all regions show (2-1)S(1)/(1-0)S(1) ratios of $<0.2$ (see Fig. \ref{fig:h2diag2}). The central region is situated above the thermal emission curve which is probably due to an overestimation of the (1-0)S(3) line that is situated in a noisy spectral region. Nevertheless, \citet{rodriguez-ardila_molecular_2005} show that the bulk of Seyfert galaxies, independent of Seyfert type, have similar (1-0)S(3)/(1-0)S(1) ratios or even higher than our nuclear region. Hence thermal excitation is the dominant mechanism in the center but further differentiation cannot be concluded.

Emission spot sp(A) lies on the thermal emission curve at an excitation temperature between 2000 K and 3000 K. This implies that the gas there is heated up, probably by a hidden star formation region. The $^{12}$CO(3-2) map shows the brightest emission in this region. The star formation region has then already formed stars, which explains that \citet{combes_ALMA_2013} report a non-detection of the dense gas tracers HCO$^+$(4--3) and HCN(4--3) in this region. The fact that we do not detect Br$\gamma$ emission hints at a high column density towards the star forming region's \ion{H}{ii} region.

Emission spot sp(B) lies slightly off the thermal excitation curve at about 2000 K and is close the predicted shock excitation models by \citet{brand_constancy_1989}. The $^{12}$CO(3-2) shows no increased flux in this region hence sp(B) seems not to be excited by the same mechanisms as sp(A). The broadness of the H$_2$ lines in sp(B) as well as the very blueshifted velocities indicate an interaction of the two arms, eastern and southern arm (see Fig. \ref{fig:ALMAmom1}). Due to the interaction of the arms shock fronts are induced in the gas which then excite the H$_2$ molecules in sp(B). Another possibility is a strong star formation that shocks the surrounding gas. As in sp(A) no Br$\gamma$ or HCO$^+$(4-3) and HCN(4-3) are detected indicating either a high column density towards the region, a not strong enough ionization source or not dense enough gas.

\section{Conclusion and summary}
We have presented the first ALMA backed SINFONI results for the nearby LINER/Seyfert 2 galaxy NGC 1433 from the extended NUGA south sample. We constrain the center of the galaxy and the black hole position to the optical and NIR stellar luminosity peaks with an error of $\pm$0$\farcs$2. The new center lies about 1$\farcs$5 north-north-west from the adopted center (see table \ref{tab:basic}) by \citet{combes_ALMA_2013}.

With the new adopted center we discuss the velocity field of the H$_2$ and CO gas in the very center and propose three interpretations of the results: gaseous disk, molecular outflow or a combination of both. No outflow characteristics were observed in X-ray, however, the resolution is with 2$\arcsec$ not good enough to trace the possible outflow that we describe above. Our simple modeling approach shows that, spatially, there is no difference between a central disk model and a nuclear double-sided outflow. Depending on the yet unknown inclination the disk might be decoupled. Furthermore, the combination of a circum nuclear disk and a one-sided outflow is a scenario that can explain the southwards reaching redshifted tail as well as the central velocity field, but it has a higher number of degrees of freedom.

Further analysis of the spiral arms and their LOSVs hints at an inflow scenario for the center of NGC 1433 (see Fig. \ref{fig:losvlines}). The dust arms seem to leave the disk which is implied by the blue LOSV 2$\arcsec$ west of the nucleus and turn around again to fall in the direction of the nucleus where they encounter the strong nuclear velocity gradient which is oriented at a PA $\sim140\degr$.
Our torque calculations show that the gas is driven towards a nuclear
ring of 200~pc radius, which could be the ILR of the nuclear R$=$430~pc bar. A possible gas
infall towards the very center requires that other mechanisms, e.g. 
viscous torques, can take over from there. Indeed we see several dust arms within a 2$\arcsec$ radius, i.e the nuclear--arm and several faint dust lanes (see Fig. \ref{fig:HSToverlay}) which could be driven inwards by viscosity torques \citep{combes_molecular_2004,van_der_laan_molecular_2011}.
The observed dip in LOSVD of the stars can then be explained by infalling gas that creates a nuclear disk in which stars are formed with a lower velocity dispersion than the bulge LOSVD \citep{emsellem_dynamics_2001,falcon-barroso_sauron_2006}.

The PV diagrams speak in favor of an outflow (see Fig. \ref{fig:pvcut}) scenario. With the new center the outflow has its origin at the position of the SMBH. The lower H$_2$ velocity with respect to the CO velocity in the nuclear region indicates a shock ionization of the the gas. However, a small circum nuclear disk cannot be excluded.

The measurement of the stellar LOSV is consistent with literature values. The isophotes in the central region follow the nuclear bar structure although a small deviation may be detected towards the center (see Fig. \ref{fig:stellvel}).

The emission lines allow a line diagnostic on the nucleus and in emission spots sp(A) and sp(B). We confirm a Seyfert-to-LINER like excitation mechanism from the diagnostic diagram in Fig. \ref{fig:ddiag}. Thermal excitation dominates emission spots sp(A) and sp(B). The thermal emission temperature is higher in sp(A) than in sp(B) whereas sp(B) is close to the shock excitation models by \cite{draine_theory_1993}. Sp(A) lies exactly on the thermal emission curve at $\sim2500$ K. An outflow inferred from strong star formation in the center can be excluded due to a very low SFR. We detect either hidden star formation in sp(A) and sp(B) or a strong dust arm interaction in sp(B) due to strong $^{12}$CO(3--2) and H$_2$ emission in sp(A) and strong and turbulent H$_2$ emission in sp(B).

We measure a stellar mean LOSVD of 124 km s$^{-1}$ and determine a black hole mass of $M_{\bullet}=1.74\times10^7 M_{\odot}$ which is by a factor 2 higher than literature values. The H$_2$ gas mass is derived from $^{12}$CO(3--2) and from the warm H$_2$ gas in the NIR. We derive in the central 1$\arcsec$ radius aperture an H$_2$ mass of $\sim10^6 M_{\odot}$ from the warm NIR gas which is a factor 2 higher than the value from the CO luminosity. For the larger central 5$\arcsec$ radius aperture we derive from both lines a mass of $\sim10^7 M_{\odot}$. This difference in the center may result from a lack of emission of the $^{12}$CO(3--2) transition due to highly excited gas with temperatures of $>55$~K.

The dust and gas arms reach towards the center. But is the center accreting this gas mass through a disk? Or is it repelling a possible infall through an outflow? Or do we see both mechanisms working without any larger interaction?
Higher spatial resolution in forthcoming ALMA cycles and SINFONI with AO assistance as well as higher excited CO transitions are needed to sufficiently resolve the center and the mechanisms that create this strong gradient which is not aligned with the stellar rotation.

\begin{acknowledgements}
This paper makes
use of the following ALMA data: ADS/JAO.ALMA\#2011.0.00208.S. ALMA
is a partnership of ESO (representing its member states), NSF (USA) and
NINS (Japan), together with NRC (Canada) and NSC and ASIAA (Taiwan),
in cooperation with the Republic of Chile. The Joint ALMA Observatory is
operated by ESO, AUI/NRAO and NAOJ. The National Radio Astronomy
Observatory is a facility of the National Science Foundation operated under co-
operative agreement by Associated Universities, Inc. 
We use data products from the Two Micron All Sky Survey, which is a joint project of the University of Massachusetts and the Infrared Processing and Analysis Center/California Institute of Technology, funded by the National Aeronautics and Space Administration and the National Science Foundation.
We used observations made with the NASA/ESA Hubble Space Telescope, obtained from the data archive at the Space Telescope Institute. STScI is operated by the association of Universities for Research in Astronomy, Inc. under the NASA contract NAS 5-26555.
We used observations made with the NASA/ESA Hubble Space Telescope, and obtained from the Hubble Legacy Archive, which is a collaboration between the Space Telescope Science Institute (STScI/NASA), the Space Telescope European Coordinating Facility (ST-ECF/ESA) and the Canadian Astronomy Data Centre (CADC/NRC/CSA).
This work was supported in part by the Deutsche Forschungsgemeinschaft
(DFG) via the Cologne Bonn Graduate School (BCGS),
and via grant SFB 956, as well as by
the Max Planck Society and the University of Cologne through
the International Max Planck Research School (IMPRS) for Astronomy and
Astrophysics and by the German federal department for education and research (BMBF) under the project number 50OS1101.
We had fruitful discussions
with members of the European Union funded COST Action MP0905: Black
Holes in a violent Universe and the
COST Action MP1104:
Polarization as a tool to study the Solar System and beyond.
We received funding from the
European Union Seventh Framework Programme (FP7/2007-2013)
under grant agreement No.312789.
\end{acknowledgements}


\bibliographystyle{aa}
\bibliography{mybib,zotero,book}

\begin{appendix}
	\section{Detector specific pattern and atmospheric emission correction}
	\label{sec:dsp}
\begin{figure}[htbp]
\centering
\subfigure[Pattern One]{\includegraphics[width=0.4\textwidth]{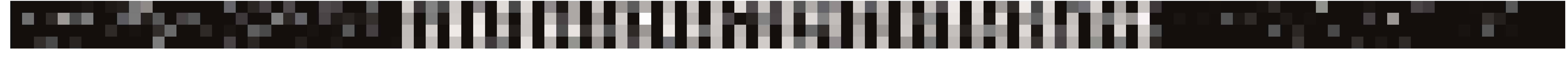}\label{fig:080_1}}
\subfigure[Pattern Two]{\includegraphics[width=0.4\textwidth]{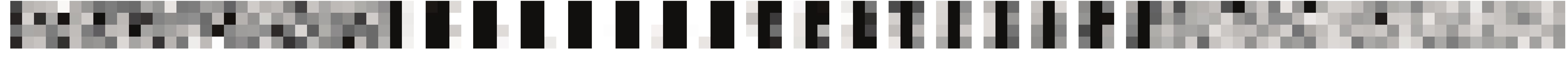}\label{fig:102}}
\caption{The figure shows the last four pixels (from bottom to top). Shown are the two types of pattern that randomly show up on the SINFONI detector. When affected, it is always at least one amplifier (64 pixels or columns).}
\label{fig:pattern}
\end{figure}

\begin{figure}[htbp]
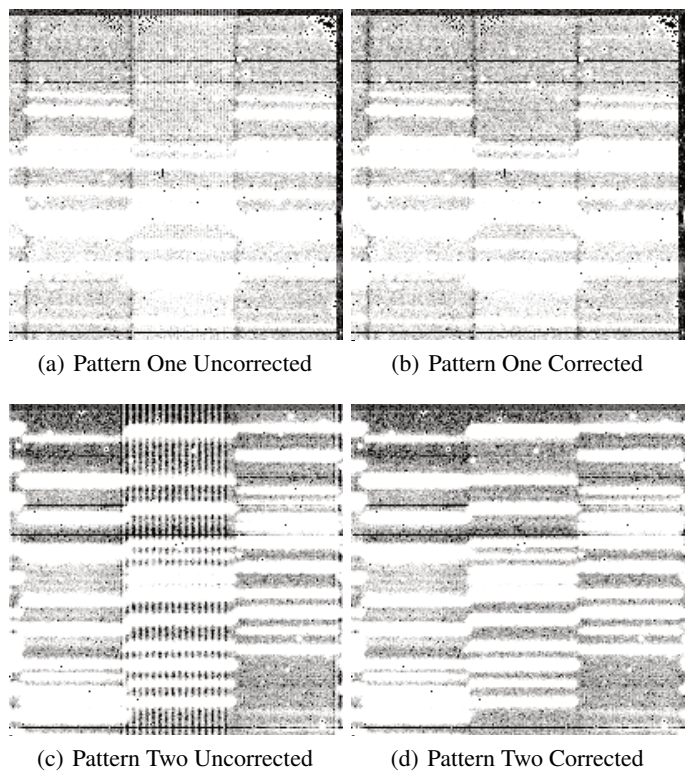

\centering
\subfigure[Pattern One Uncorrected]{\includegraphics[width=0.24\textwidth]{SINFO080.pdf}\label{fig:080}}
\subfigure[Pattern One Corrected]{\includegraphics[width=0.24\textwidth]{cor080.pdf}\label{fig:cor080}}
\subfigure[Pattern Two Uncorrected]{\includegraphics[width=0.24\textwidth]{SINFO102_1.pdf}\label{fig:102_1}}
\subfigure[Pattern Two Corrected]{\includegraphics[width=0.24\textwidth]{cor102_1.pdf}\label{fig:cor102_1}}
\caption{The figure shows parts of the detector where the pattern was detected \subref{fig:080} \& \subref{fig:102_1} and then corrected \subref{fig:cor080} \& \subref{fig:cor102_1}. Note that the dark horizontal lines are part of an already known detector problem. The bright extended horizontal lines are OH sky lines. The three darkish extended vertical lines in \subref{fig:080} \& \subref{fig:cor080} are the slit borders of the 32 slitlets on the detector.}
\label{fig:patternfull}
\end{figure}

During the data reduction we had to address a problem with the SINFONI detector that occurred during the observing period. The detector amplifiers introduce two different types of variation in the arbitrary digital units (ADU) (see Fig. \ref{fig:pattern}). These variations would come and go from exposure to exposure and were also found in the calibration data provided to the observation data. From our investigation of the affected files we have recognized this problem in particular amplifiers of the 32 amplifiers of SINFONI with random occurrence (from file to file) in these amplifiers, but we also identify fixed patterns. There are two different noise patterns within the amplifiers:

Pattern one is a constant offset in every second pixel column of an amplifier (see first row of Figs. \ref{fig:pattern} \& \ref{fig:patternfull}). But it is neither constant between two amplifiers nor between different exposures. It seems that this pattern occurs in several but not in all amplifiers and that there are amplifiers where it is usually stronger than in others.\footnote{We did not investigate in which amplifiers it occurred exactly because the offset was sometimes very small. One could make it out by eye but the standard deviation was not much higher if any than in not affected amplifier columns. A lot more time would have been needed to investigate the exact occurrence of this noise pattern.}

Pattern two shows a sinusoidal noise pattern in every four columns of one amplifier (see second row of Figs. \ref{fig:pattern} \& \ref{fig:patternfull}). The pattern was fitted best with a sine function that added an offset of half a $\pi$ to the next column. This pattern occurs only in amplifiers 14, 16, 18 and 20, counting from left to right and starting with 1.

Both patterns were observed in separate, but also in the same exposure (data file), no superposition was observed. In one data file an amplifier stopped showing a noise pattern after about 1000 rows.

We investigated the last three pixels in every column of the detector. These pixels are control pixels that are not exposed to light but suffer from the same electronic noise as the rest of the detector\footnote{The first and last four pixels in every row and column are control pixels. We used only the last three in every column because in some exposures the first of the last four control pixel in a column seems to be slightly exposed, best seen in the fiber exposure data. In addition, we did not use the first four pixels in every column because the noise offset here was usually different and we could not combine the first and last pixels in every column, also the very first and fourth pixel sometimes suffered from other effects (e.g. slight exposure). The first and last pixels in every row are not interesting since this is a column wise problem.}. From these pixels we were able to determine if there is a noise problem in the amplifier columns by standard and mean deviation methods. We separated the two problems, first correcting for the constant offset pattern, which is easier to determine. The complication here was the automated correction due to positive or negative ADU means combined with negative or positive constant offsets. After that we corrected for the sinusoidal pattern by fitting a two-dimensional sine with 16 columns and 4 rows making up the 64 channels of every amplifier. The noise was detected and corrected in 66 data files (science and calibration data). The correction was successful as shown in figure \ref{fig:patternfull}. Weak noise patterns, which were not selected by our routines, can still make the resulting data files noisier, however, the difference to other noise sources (e.g. photon noise, readout noise) is almost not measurable.

After correcting the unusual detector noise features we were confronted with a more typical problem, the OH emission of the atmosphere. Although we had exceptional weather conditions -- photometric night -- our 150 seconds exposures were too long for the fast changing atmospheric OH line emission. We used the SINFONI pipeline version 2.3.2 for data reduction, except for the OH correction. The usual OH correction performed by the pipeline resulted in strong OH residuals. By activating the higher density OH correction that was implemented following \citet{davies_method_2007}, the correction was better in some parts and worse in others, e.g. 'P-Cygni' profiles in the OH lines were introduced. Comparing the raw target files with their pre/subsequent raw sky files we registered a change in the strength of the OH lines from target to sky by up to 10\%. In addition, a random shift of about 0.04 pixel in the spectral direction of the detector was noticed. We found this by fitting Gaussian profiles to several OH lines. We selected OH lines that had more than 100 peak counts above the continuum and that were isolated enough to assume a reliable profile fit. Furthermore, we selected OH lines from every vibrational transition. All selected lines in one detector column were fitted simultaneously to improve the continuum fit.

We found that all fitted lines in a sky file differed by about the same factor and pixel difference with respect to the corresponding target file, in every detector column. Hence, we took the median of all OH lines (41 in H- and 16 in K-band) and all 32 slits (about 60 of 64 pixels per slit were used) to determine the scaling factor and the shift of the OH lines. The fit did not work well at the edges of the slits due to contamination from the neighboring slit hence we neglected the slit edges. As not to scale the continuum with the OH lines we fitted the continuum in every detector column (spectral direction) and subtracted it before the OH line scaling. In H-band this was done by using a linear function. For the K-band continuum we fitted a black body function, because the grey-body emission of the background becomes prominent at the red end of the K-band. We took the median of the fitted parameters over one slit to determine a robust continuum for the whole slit. After the subtraction of the continuum we scaled the remaining atmospheric emission features added back the continuum and then shifted the full column by the determined subpixel shift. The scaling was done based only on the emission lines because the continuum does not vary whereas the spectral shift has to be done on the full spectrum since the reason for the shift is a shear in the grating rather than some atmospheric effect.

\end{appendix}



\end{document}